\documentclass[twoside,12pt]{article}
\usepackage{epsfig,graphicx,amssymb,amsmath,wasysym,graphicx}

\catcode`@=11
\def\@citex[#1]#2{\if@filesw\immediate\write\@auxout{\string\citation{#2}}\fi
  \def\@citea{}\@cite{\@for\@citeb:=#2\do
    {\@citea\def\@citea{,\penalty\@m}\@ifundefined
      {b@\@citeb}{{\bf ?}\@warning
       {Citation `\@citeb' on page \thepage \space undefined}}%
\hbox{\csname b@\@citeb\endcsname}}}{#1}}

\def\citer{\@ifnextchar
[{\@tempswatrue\@citexr}{\@tempswafalse\@citexr[]}}

\def\@citexr[#1]#2{\if@filesw\immediate\write\@auxout{\string\citation{#2}}\fi
  \def\@citea{}\@cite{\@for\@citeb:=#2\do
    {\@citea\def\@citea{--\penalty\@m}\@ifundefined
       {b@\@citeb}{{\bf ?}\@warning
       {Citation `\@citeb' on page \thepage \space undefined}}%
\hbox{\csname b@\@citeb\endcsname}}}{#1}}
\catcode`@=12

\def\Journal#1#2#3#4{{#1} {#2} (#4) #3}
\def\ARNPS{\em Ann. Rev. Nucl. Part. Sci.}
\def\NCC{{\em Nuovo Cim.} C}
\def\NC{\em Nuovo Cim.}

\def\NP{{\em Nucl. Phys.}}

\def\NPPS{\em Nucl. Phys. Proc. Suppl.}
\def\JPB{{\em J. Phys.} B}
\def\JPG{{\em J. Phys.} G}
\def\NPB{{\em Nucl. Phys.} B}

\def\NPB{{\em Nucl. Phys.} B}
\def\PAN{\em Phys. Atom. Nucl.}  

\def\PLB{{\em Phys. Lett.} B}

\def\PL{{\em Phys. Lett.}}
\def\PPNP{\em Prog. Part. Nucl. Phys.}
\def\PRL{\em Phys. Rev. Lett.}
\def\PREV{\em Phys. Rev.}
\def\PREP{\em Phys. Rep.}
\def\PRA{{\em Phys. Rev.} A}
\def\PRD{{\em Phys. Rev.} D}
\def\PRC{{\em Phys. Rev.} C}

\def\ZPC{{\em Z. Phys.} C}

\def\EPJC{{\em Eur. Phys. J.} C}

\def\RMP{{\em Rev. Mod. Phys.}}

\def\IJMPA{{\em Int. J. Mod. Phys.} A}

\def\msbar{$\overline{\rm MS}$}
\def\drbar{$\overline{\rm DR}$}

\def\seff{\sin^2\theta_{\rm eff}^\ell} 

\def\slash{{\hspace{0pt}\displaystyle\not}\hspace{0pt}}
\def\slashx{{\hspace{-1pt}\displaystyle\not}\hspace{1pt}}
\def\to{\rightarrow}
\def\imply{\Longrightarrow}
\def\ovl{\overline}

\def\bmat{\begin{pmatrix}}
\def\emat{\end{pmatrix}}
\def\ph{\phantom}
\def\cf{{\em cf.}, }
\def\ie{{\em i.e.}, }
\def\vs{{\em vs.}}
\def\etal{{\em et al.}}
\def\ibid{{\em ibid.}}
\def\eg{{\em e.g.}, }
\def\Eg{{\em E.g.}, }
\def\etc{{\em etc}}

\def\rnu{{R_\nu}}
\def\rnubar{{R_{\bar\nu}}}

\newcommand{\be}{\begin{equation}}
\newcommand{\ee}{\end{equation}}
\newcommand{\ba}{\begin{array}}
\newcommand{\ea}{\end{array}}

\begin{document}

\title{\vspace{1cm} The Weak Neutral Current}

\author{Jens Erler$^1$ and Shufang Su$^2$ \\ \\
$^1$Departamento de F\'isica Te\'orica, Instituto de F\'isica, \\
Universidad Nacional Aut\'onoma de M\'exico, 04510 M\'exico D.F., M\'exico \\
$^2$Department of Physics, University of Arizona, \\
P.O. Box 210081, Tucson, AZ 85721, USA}

\maketitle

\begin{abstract} 
This is a review of electroweak precision physics with particular emphasis on low-energy precision 
measurements in the neutral current sector of the electroweak  theory and includes future 
experimental prospects and the theoretical challenges one faces to interpret these observables. 
Within the minimal Standard Model they serve as determinations of the weak mixing angle which 
are competitive with and complementary to those obtained near the $Z$-resonance. 
In the context of new physics beyond the Standard Model these measurements are crucial 
to discriminate between models and to reduce the allowed parameter space within a given model.   
We illustrate this for the minimal supersymmetric Standard Model with or without $R$-parity.
\end{abstract}

\eject

\tableofcontents

\section{Introduction}
\label{intro}

The basic structure of the Standard Model (SM) of the electroweak (EW) interactions was 
established in the 1970s with the help of low-energy (compared to the EW scale, 
$\Lambda_{\rm EW} \equiv 246$~GeV) experiments in neutrino scattering 
and deep inelastic polarized electron-deuteron scattering, 
and later (after some initial confusion) in measurements of atomic parity violation (APV).
The 1980s saw the first precision measurements of the EW mixing angle 
in neutrino scattering and charged lepton scattering in various kinematic regimes.  
The $Z$-pole programs at LEP~1 and the SLC with their high-precision measurements 
of $Z$ boson properties finally established the SM as the correct theory even 
at the level of small quantum corrections.  This permits to view the SM as the low
energy effective theory of a more fundamental theory with a typical energy scale in the TeV
or multi-TeV domain (or conceivably much higher).
The current energy frontier at the Large Hadron Collider (LHC) directly probes this regime by looking 
for new particles or strong deviations from the SM predictions of cross-sections and event rates.  

However, there is an important alternative route.  
One can return to the sub-$Z$-pole regime attaining high precision by performing ultra-high statistics 
experiments at the so-called intensity frontier~\cite{Hewett:2012ns}.
By either using neutrinos or by exploiting the parity-violating nature of the weak interaction one is 
directly sensitive to $\Lambda_{\rm EW}$, so that a measurement of about 1\% precision generally 
probes the multi TeV energy scale.  
Cases in which the SM prediction happens to be suppressed are even more favorable.  
This class of measurements and the associated theoretical challenges are the main subject 
of this review.  
Cases where the SM prediction is even vanishing due to an accidental symmetry such as lepton
number or lepton flavor number, and observables where it is tiny because they violate some 
approximate discrete symmetries like time reversal are covered elsewhere in this volume,
For a previous review discussing low energy neutral current measurements we refer to 
Ref.~\cite{Erler:2004cx}.
The most recent review~\cite{Kumar:2013yoa} covers the weak mixing angle at low momentum 
transfer and parity-violating electron scattering.
It also contains a detailed documentation of the theory issues surrounding the weak charge of 
the proton and the scale dependence of the weak mixing angle,
as well as a discussion of new physics, such as contact interactions, 
extra neutral $Z$ bosons (both visible and dark) and the X parameter (see Section~\ref{STUrho}).

It is difficult to overemphasize the complementarity between the energy and intensity frontiers. 
Measurements of the $W$ and $Z$ boson properties have reached and surpassed 
per mille precision but they may be insensitive to new physics, if mixing and interference effects
are too small.  
New particles can be produced at the LHC, but only if they are either light enough, or
couple sufficiently strong to quarks and gluons, or allow a clean final state 
(preferentially involving charged leptons or photons). 
By contrast, lower energy observables are affected by amplitudes mediated by new physics, 
even if they are hiding below the $Z$ resonance where they might be out of phase.
Also, measurements of EW couplings of the first generation quarks are very difficult at
the energy frontier, a gap that is naturally closed by the intensity frontier.

Of course, the indirect nature of the precision measurements from the intensity frontier makes it 
virtually impossible to identify the type of new physics responsible for a possible deviation from the 
SM.  However, by having at one's disposal a whole array of measurements of comparable precision
reintroduces some discriminatory power, and comparison with the discoveries (or lack thereof)
at the energy frontier strengthens this point further.

This report is structured as follows: 
Section~\ref{SM} summarizes the salient features of the SM\footnote{For a recent 
and much more detailed treatment, see \eg Ref.~\cite{pgl}.}.
In Section~\ref{WZ} we review the very high-precision measurements of various properties 
of the massive gauge bosons and their implications for the mass of the SM Higgs boson.
In Section~\ref{nu} we turn to the intensity frontier and discuss neutrino scattering experiments.
Section~\ref{PAVI} is dedicated to polarized lepton scattering and atomic parity violation.  
As a specific scenario for physics beyond the SM, section~\ref{NP} discusses models of 
supersymmetry which are very well motivated from a theoretical point of view.
Section~\ref{conclusions} concludes with a summary and an outlook.

\section{The Standard Electroweak Theory}
\label{SM}
\subsection{\it Gauge sector and weak mixing angle}

Any relativistic quantum field theory which contains as mediators massless particles of 
helicity $\pm 1$ and which in turn are described in terms of vector fields, $V_\mu$ 
(necessary on dimensional grounds if one wants to avoid extra power suppressions by some large 
mass scale), is subject to a set of exact gauge symmetries~\cite{Weinberg:1965rz}. 
The electroweak SM~\cite{Weinberg:1967tq} is based on the gauge group 
$SU(2)_L \times U(1)_Y$~\cite{Glashow:1961tr} with corresponding
vector fields $W_\mu^i$ ($i = 1, 2, 3$) and $B_\mu$, where $SU(2)_L$ refers to weak isospin
and $Y$ denotes hypercharge.
In terms of the $SU(2)_L$ and $U(1)_Y$ gauge couplings, $g$ and $g'$, 
and the weak mixing angle, $\theta_W$, the linear combinations,
\be
W^\pm_\mu \equiv {W_\mu^1 \mp i W_\mu^2 \over \sqrt{2}}, 
\qquad\qquad
Z_\mu \equiv {g W_\mu^3 - g' B_\mu \over\sqrt{g^2 + g'^2}} \equiv 
\cos\theta_W W_\mu^3 - \sin\theta_W B_\mu,
\ee
\be
A_\mu \equiv {g' W_\mu^3 + g B_\mu \over\sqrt{g^2 + g'^2}} \equiv 
\sin\theta_W W_\mu^3 + \cos\theta_W B_\mu,
\ee
give rise to the charged and neutral gauge bosons of the weak interaction, and the photon,
respectively.  These vector fields enter the SM Lagrangian, 
\be
{\cal L} = {\cal L}_V + {\cal L}_\phi + {\cal L}_f + {\cal L}_Y,
\ee
through the kinetic terms for the vector gauge fields,
\be
{\cal L}_V = - {1\over 4} {W^{\mu\nu}}^i W_{\mu\nu}^i - {1\over 4} {B^{\mu\nu}} B_{\mu\nu},
\ee
where 
\be
W_{\mu\nu}^i = \partial_\mu W_\nu^i  - \partial_\nu W_\mu^i - g \epsilon_{ijk} W_\mu^j W_\nu^k,
\qquad\qquad 
B_{\mu\nu} = \partial_\mu B_\nu - \partial_\nu B_\mu,
\ee
and through the gauge covariant derivatives in ${\cal L}_\phi$ and ${\cal L}_f$ discussed in
Sections~\ref{SMHiggs} and~\ref{SMf}, respectively.

\subsection{\it Higgs sector and the theory after electroweak symmetry breaking}
\label{SMHiggs}

In the minimal model, EW symmetry breaking is introduced with the help 
of a weak iso-doublet, $\phi$~\cite{Djouadi:2005gi}, of complex scalars, with Lagrangian density,
\be\label{Lphi}
{\cal L}_\phi = 
(D^\mu \phi)^\dagger D_\mu\phi - \mu^2 \phi^\dagger\phi - {\lambda^2\over 2} (\phi^\dagger\phi)^2
\equiv (D^\mu \phi)^\dagger D_\mu\phi - V(\phi),
\ee
where $\lambda^2 > 0$ is required for a stable vacuum, and where $\mu^2 < 0$ triggers 
the spontaneous breaking of the EW gauge 
symmetry~\citer{Higgs:1964ia,Kibble:1967sv}.
The Lagrangian~(\ref{Lphi}) is easily seen to have an $SO(4) \simeq SU(2) \times SU(2)$
global (``custodial") symmetry which is, however, broken down to $SU(2)_L \times U(1)_Y$
by the gauge and Yukawa interactions.
The gauge covariant derivative of $\phi$ is,
\be\label{covder}
D_\mu \phi = 
(\partial_\mu + i g T_i^\phi W_\mu^i + i g' Y^\phi B_\mu) \phi =
\left[ \partial_\mu + i {g\over\sqrt{2}} 
(T_+^\phi W_\mu^+ + T_-^\phi W_\mu^- + Q_A^\phi A_\mu + g^\phi Z_\mu) \right] \phi,
\ee
with weak isospin and hypercharge generators given, respectively, 
in terms of the Pauli matrices, $\tau_i$, 
and the identity matrix, $\tau_0 \equiv \mathbb{I}_2$, by,
\be
T_i^\phi = {\tau_i\over 2}, 
\qquad\qquad
Y^\phi = {\tau_0\over 2}.
\ee
For later reference we introduced a common normalization for the generators 
of all four EW gauge bosons in the second form in Eq.~(\ref{covder}), for which we defined, 
\be\label{TQA}
T_\pm^\phi \equiv T_1^\phi \pm i T_2^\phi = {T_\mp^\phi}^\dagger, 
\qquad\qquad
Q_A^\phi \equiv \sqrt{2} \sin\theta_W (T_3^\phi + Y^\phi) \equiv \sqrt{2} \sin\theta_W Q^\phi = 
{Q_A^\phi}^\dagger,
\ee
\be\label{QZ}
g^\phi \equiv \sqrt{2} \cos\theta_W (T_3^\phi - \tan^2\theta_W Y^\phi) = 
{\sqrt{2}\, T_3^\phi - \sin\theta_W Q_A^\phi \over\cos\theta_W} =
\sqrt{2}\, {T_3^\phi - \sin^2\theta_W Q^\phi \over\cos\theta_W}= {g^\phi}^\dagger,
\ee
and where $Q^\phi$ is the electric charge operator normalized in the usual way\footnote{All 
the explicit factors of $\sqrt{2}$ appearing in $Q_A^\phi$, $g^\phi$, and  Eq.~(\ref{covder}) 
are an artifact of the normalization used in the physics literature for the gauge symmetry generators.  
They would be absent in the more natural normalization typically used by mathematicians 
in which, \eg all Dynkin indices are integers.}.
Note the similarity in the relations for $T_\pm^\phi$, $Q_A^\phi$, and $g^\phi$,
in the ideal mixing case $g = g'$ ($\tan\theta_W = 1$) where we also have $Q_A^\phi = Q^\phi$.

Since one wants to quantize the theory around the classical vacuum, 
the physically relevant case, $\mu^2 < 0$, requires a redefinition of the original Higgs field, $\phi$,
so that at the minimum of the potential all fields have vanishing vacuum expectation values (VEVs),
\be
\phi \equiv 
{1\over \sqrt{2}} \bmat \phi_1 + i\phi_2 \\ \phi_3 + i\phi_4 \emat =
{1\over \sqrt{2}} \bmat \phi_1 + i\phi_2 \\ H + v + i\phi_4 \emat \imply
\langle \phi \rangle = {1\over \sqrt{2}} \bmat 0 \\ v \emat.
\ee
The three VEVs, $\langle \phi_i \rangle$ for $i = 1,2,4$, have been arranged to vanish 
by a global $SU(2)_L \times U(1)_Y$ transformation, while for $\langle \phi_3 \rangle = v$
one defines the physical Higgs field, $H \equiv \phi_3 - v$, so that $\langle H \rangle = 0$.
The potential, $V(\phi)$, now gives rise to three massless and non-interacting scalars, 
which in the absence of  the gauge fields would be Goldstone bosons~\cite{Goldstone:1961eq}.
By performing a local $SU(2)_L \times U(1)_Y$ transformation one can finally proceed to
the unitary gauge in which these three bosons disappear from the physical spectrum and
instead provide the longitudinal components of the $W$ and $Z$ bosons.
It remains one massive scalar mode, the physical Higgs boson.
The terms quadratic in $v$ arising from ${\cal L}$ in Eq.~(\ref{Lphi}) are,
$$
{v^2\over 2} \left[ \lambda^2 H^2 + {g^2 \over 2} 
\left( {W^\mu}^- W^+_\mu + (g^\phi)^2 Z^\mu Z_\mu \right) \right],
$$
so that 
\be\label{MHWZ}
M_H = \lambda v = \sqrt{-2\mu^2}, \qquad\qquad 
M_W = {g\over 2} v, \qquad\qquad 
M_Z = {M_W\over \cos\theta_W} = {\sqrt{g^2 + g'^2} \over 2} v.
\ee
The value of $v = 246.22$~GeV is fixed by the Fermi constant, 
\be\label{GF}
G_F = 1.1663787 (6) \times 10^{-5} \mbox{ GeV}^{-2} = {1\over\sqrt{2} v^2},
\ee
and $G_F$ itself can be cleanly extracted from the $\mu$ lifetime, $\tau_\mu$,
which was measured recently by the MuLan Collaboration at the PSI with 
an order of magnitude improved precision,
$\tau_\mu = 2.1969803(22)~\mu\mbox{s}$~\cite{Webber:2010zf}.

\begin{figure}[t]
\begin{center}
\epsfig{file=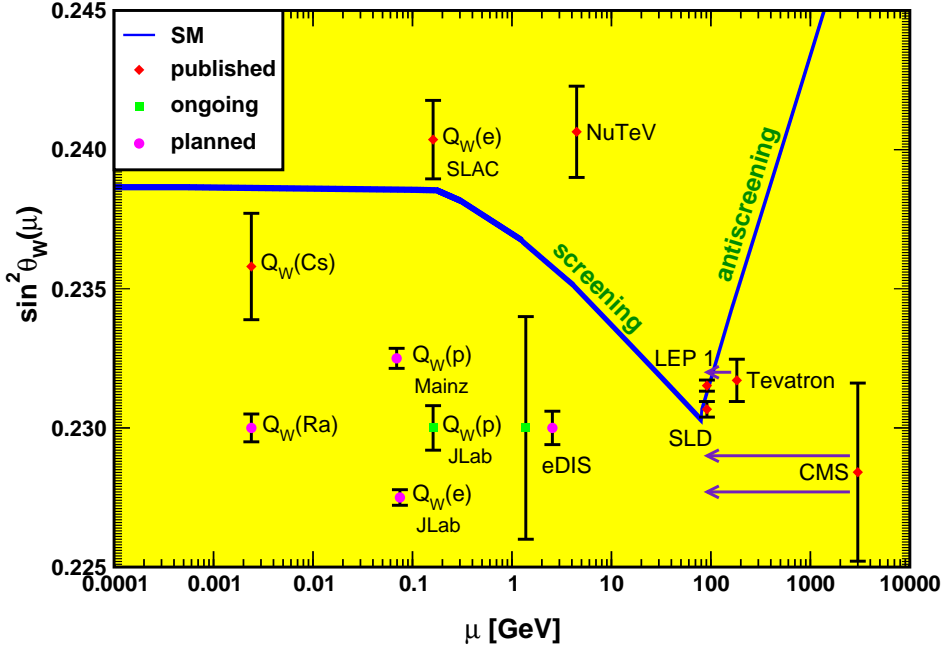,scale=0.49}
\begin{minipage}[t]{16.5 cm}
\caption{Scale dependence of the weak mixing angle in the \msbar-scheme.
At the location of $M_W$ and each fermion mass there 
are discontinuities arising from scheme dependent matching terms which are necessary 
to ensure that the various effective field theories within a given loop order describe the same 
physics. However, in the \msbar-scheme these are very small numerically and barely visible in the 
figure provided one decouples quarks at $\mu = \hat{m}_q (\hat{m}_q)$. 
The width of the curve reflects the theory uncertainty from strong interaction effects 
which at low energies is at the level of $\pm 7\times 10^{-5}$~\cite{Erler:2004in}. 
The various data points are discussed in the subsequent sections. 
The Tevatron and CMS measurements are strongly dominated by invariant masses 
of the final state dilepton pair of ${\cal O}(M_Z)$ and can thus be considered as additional 
$Z$-pole data points, but for clarity we shifted the points horizontally to the right.
\label{s2w}}
\end{minipage}
\end{center}
\end{figure}

Thus, at lowest order one can write,
\be\label{s2theta}
\sin^2\theta_W = 1 - {M_W^2\over M_Z^2} = {{g'}^2\over g^2 + g'^2},
\ee
and one may elevate either form to an exact definition of $\sin^2\theta_W$ to all orders in
perturbation theory.  The first relation defines the on-shell renormalization scheme, and is
manifestly related directly to physically observable particles masses\footnote{We note, 
however, that at higher orders the definition of the mass of an unstable particle becomes
ambiguous, and the whole concept of observability becomes demoted.}. 
The second, coupling-based relation leads to theoretical constructs which depend on details 
of the regularization scheme, the energy scale ($\mu$), the treatment of fermion thresholds, \etc.
This class includes the \msbar-scheme definition, $\sin^2\hat\theta_W(\mu)$, 
and a variant used in supersymmetric theories, which is based on the \drbar-scheme.
These definitions have the advantages that they significantly simplify higher-order calculations,
and the numerical values of different couplings can be directly compared as,
\eg in the discussion of gauge coupling unification.
Moreover, the large Yukawa coupling of the top quark, $Y_t$, and relatedly its heavy mass,
$m_t = Y_t v$, strongly affect
the renormalized value of $M_W$, while its effect on $M_Z$ and $\sin^2\hat\theta_W(\mu)$
is much weaker.  Therefore, using the on-shell definition indiscriminately may distort theoretical
expressions and lead to a poorer convergences of the perturbative series. 
Further definitions and more details can be found in Ref.~\cite{PDGEW2012}.

The scale dependence of the weak mixing angle renormalized in 
the \msbar-scheme~\cite{Erler:2004in} is shown in Figure~\ref{s2w}.  
The minimum of the curve corresponds to $Q = M_W$, below which we switch to an effective 
theory with the $W^\pm$~bosons integrated out, and where the $\beta$-function for the weak 
mixing angle changes sign.
For the scale dependence in a mass-dependent renormalization scheme, 
see Ref.~\cite{Czarnecki:2000ic}, and for a recent review on the low energy measurements of 
the weak mixing angle, see also Ref.~\cite{Kumar:2013yoa}.

\subsection{\it Fermion sector and gauge currents}
\label{SMf}

At the renormalizable level, \ie ignoring, for example, the possibility of Majorana neutrino masses
and other lepton or baryon number violating effects\footnote{Majorana 
neutrino masses from dimension~5 operators~\cite{Weinberg:1979sa} 
and their implications for neutrino oscillations and neutrinoless double beta decay 
are covered elsewhere in this volume.}, 
the parts of ${\cal L}$ containing fermions are given by,
\be
{\cal L}_f = i \sum_{m=1}^3 \left[ 
\ovl{\textsc q}_m \slash{D} {\textsc q}_m +
\ovl{\textsc l}_m \slash{D} {\textsc l}_m + 
\ovl{\textsc u}_m \slash{D} {\textsc u}_m + 
\ovl{\textsc d}_m \slash{D} {\textsc d}_m +
\ovl{\textsc e}_m \slash{D} {\textsc e}_m \right], 
\ee
\be
{\cal L}_Y = - \sqrt{2} \sum_{m,n=1}^3 \left[ 
Y_{mn}^u (i\tau_2 \phi^\dagger)\,  \ovl{\textsc q}_m {\textsc u}_n +
Y_{mn}^d \phi\,  \ovl{\textsc q}_m {\textsc d}_n + 
Y_{mn}^e \phi\, \ovl{\textsc l}_m {\textsc e}_n  \right] + {\rm H.c.},
\ee
where the covariant derivatives are defined in analogy to Eq.~(\ref{covder}) 
and the sums are over fermion families.
The relations~({\ref{TQA}) and~({\ref{QZ}) are general and apply to fermions, as well.
Left-handed fermions, $\psi_L \equiv P_L \psi$, 
form the iso-doublets ${\textsc l} \equiv (\nu_L,e^-_L)^T$ and 
${\textsc q} \equiv (u_L,d_L)^T$, while right-handed fermions, $\psi_R \equiv P_R \psi$, 
are iso-singlets denoted by
${\textsc e} \equiv e_R$, ${\textsc u} \equiv u_R$, and ${\textsc d} \equiv d_R$, and so in summary,
\be
P_{L/R} \equiv {1 \mp \gamma^5 \over 2}, 
\qquad\qquad
T^i_L = {\tau^i \over 2}, 
\qquad\qquad
T^i_R = 0, 
\qquad\qquad
Y_f = Q_f - T_f^3. 
\ee
The $Y_{mn}^f$ ($f = e,u,d$) are arbitrary complex $3\times 3$ Yukawa matrices.
After symmetry breaking one finds,
\be\label{LfY}
{\cal L}_f + {\cal L}_Y = 
\sum_r \ovl\psi_r \left[ i \slashx{\partial} - m_r \left( 1 + {H\over v} \right) \psi_r \right] - {g\over\sqrt{2}} 
\left[ {J_W^\mu}^\dagger W_\mu^+ + J_W^\mu W_\mu^- + J_A^\mu A_\mu + J_Z^\mu Z_\mu \right],
\ee
where $r$ is a double index running over $m$ and $f=\nu,e,u,d$, and where the charged, 
electromagnetic and weak neutral currents are given by,
\be
J_W^\mu = \sum_{m=1}^3 \left[ 
\ovl{d}_m \gamma^\mu V_{CKM}^\dagger P_L u_m +
\ovl{e}_m \gamma^\mu P_L \nu_m \right], 
\ee
\be
J_A^\mu = \sqrt{2} \sin\theta_W \sum_{m=1}^3 \left[ 
{2\over 3} \ovl{u}_m \gamma^\mu u_m -
{1\over 3} \ovl{d}_m \gamma^\mu d_m -
\ovl{e}_m \gamma^\mu e_m \right], 
\ee
\be
J_Z^\mu = 
\sum_f \ovl\psi_f \gamma^\mu \left[ g_L^f P_L + g_R^f P_R \right] \psi_f =
\sum_f \ovl\psi_f \gamma^\mu {g_V^f - g_A^f \gamma^5 \over 2} \psi_f =
\ee
\be
{1\over\sqrt{2}\cos\theta_W} \sum_{m=1}^3 \left[ 
\ovl{u}_m \gamma^\mu P_L u_m -
\ovl{d}_m \gamma^\mu P_L d_m +
\ovl{\nu}_m \gamma^\mu P_L \nu_m -
\ovl{e}_m \gamma^\mu P_L e_m \right] - 
\tan\theta_W J_A^\mu.
\ee
The matrix $V_{CKM} \equiv A_L^{u\dagger} A_L^d$ has entered into $J^W_\mu$ after
unitary field re-definitions, 
\be 
u_L \to A_L^u u_L, 
\qquad\qquad
u_R \to A_R^u u_R, 
\qquad\qquad
d_L \to A_L^u d_L, 
\qquad\qquad
d_R \to A_R^d u_R,
\ee
have been applied in order to write Eq.~(\ref{LfY}) in terms of mass eigenstates.
We also defined vector and axial-vector $Z$ couplings as,
\be\label{vzaz}
g_V^f \equiv g_L^f + g_R^f = \sqrt{2} {T^3_f - 2 \sin^2\theta_W Q_f \over \cos\theta_W}\ , 
\qquad\qquad
g_A^f \equiv g_L^f - g_R^f = \sqrt{2} {T^3_f \over \cos\theta_W}\ .
\ee
At low energies, $Q^2 \equiv -q^2 \ll M_{W,Z}^2$, one finds the effective four-fermion interactions,
\be\label{Leff}
{\cal L}_{\rm CC} = - {2 \over v^2} J^{\mu\dagger}_W J_{W\mu}, 
\qquad\qquad
{\cal L}_{\rm NC} = - {\cos^2\theta_W \over v^2} {J^\mu_Z} J_{Z\mu}.
\ee

\section{Gauge and Higgs Boson Properties
\label{WZ}}
\subsection{\it The effective leptonic weak mixing angle from the Z pole}
\label{sec:seff}

The most precise determinations of the weak mixing angle to date have been obtained at 
the $e^+ e^-$-annihilation $Z$~factories, LEP and SLC, from measurements 
of various $Z$-pole asymmetries in which systematic uncertainties largely cancel.
After correcting for QED (including photon $s$-channel exchange), interference and the tiny box 
graph effects, and where applicable for QCD or $t$-channel $\gamma$ and $Z$ exchanges,
and after extrapolating to $\sqrt{s} = M_Z$ and ideal beam polarizations (0 or 100\%), 
these asymmetries can be expressed in terms of the physics parameters,
\be
A^f \equiv \frac{2 g_V^f g_A^f}{(g_V^f)^2  + (g_A^f)^2}\ .
\ee

In particular, the electron beam at the SLC was longitudinally polarized, 
where the luminosity-weighted average degree of polarization, $P^e$, was about 75\%.
The small polarimetry error of 0.5\% permitted the SLD Collaboration to measure the polarization 
or left-right asymmetry into hadronic final states~\cite{Abe:2000dq}, 
\be\label{alrq}
A_{LR}^q \equiv \frac{\sigma_L - \sigma_R}{\sigma_L + \sigma_R} = A^e
= \frac{1 - 4 \seff}{1 - 4 \seff + 8 \sin^4\theta_{\rm eff}^\ell},
\ee
with high precision, where $\sigma_L$ ($\sigma_R$) is the cross-section 
for left (right)-handed polarized electrons.
$\seff$ is the effective $Z$ pole weak mixing angle entering the vector coupling for charged 
leptons\footnote{We use the symbol $\ell$ for a generic charged lepton when lepton universality 
is assumed.}. Thus it constitutes a coupling-based definition and 
is numerically close to the value in the  \msbar-scheme,
\be\label{seff}
\seff \equiv {1\over 4} \left[ 1 - {g_V^\ell\over g_A^\ell} \right] = \sin^2\hat\theta_W(M_Z) + 0.00029\ .
\ee
As indicated in Eq.~(\ref{alrq}), the initial state polarization asymmetry filters out
the initial state coupling, $A^e$, regardless of the final state. 
This is advantageous because on the one hand $A^e$ is proportional to $1 - 4 \seff$, and on the 
other hand $\seff$ is numerically close to $1/4$, which results in very high sensitivity to $\seff$.
SLD was able to extract the final-state couplings $A^b$, $A^c$~\cite{ALEPH:2005ab}, 
$A^s$~\cite{Abe:2000uc}, $A^\tau$ and $A^\mu$~\cite{Abe:2000hk} 
from combined left-right, forward-backward asymmetries, using
\be
A_{LR,FB}^f 
= \frac{\sigma_{LF}^f - \sigma_{LB}^f - \sigma_{RF}^f + \sigma_{RB}^f}
          {\sigma_{LF}^f + \sigma_{LB}^f + \sigma_{RF}^f + \sigma_{RB}^f} 
= \frac{3}{4} A^f,
\ee
where, for example, $\sigma_{LF}^f$ is the cross-section for left-handed polarized incident
electrons to produce a fermion $f$ traveling in the forward hemisphere. 
Polarized Bhabba scattering represents mostly a measurement of $A_{LR}^\ell$ but it also includes 
information about $A_{LR,FB}^e$, both of which providing the parameter $A_e$~\cite{Abe:2000hk}.
$A^e$ is also proportional to the hadronic asymmetry ratio formed by 
the forward-backward charge asymmetry, $A_{FB}^q$, 
normalized to the left-right forward-backward charge asymmetry, $A_{LR,FB}^q$~\cite{Abe:1996ef}.

\begin{figure}[t]
\begin{center}
\epsfig{file=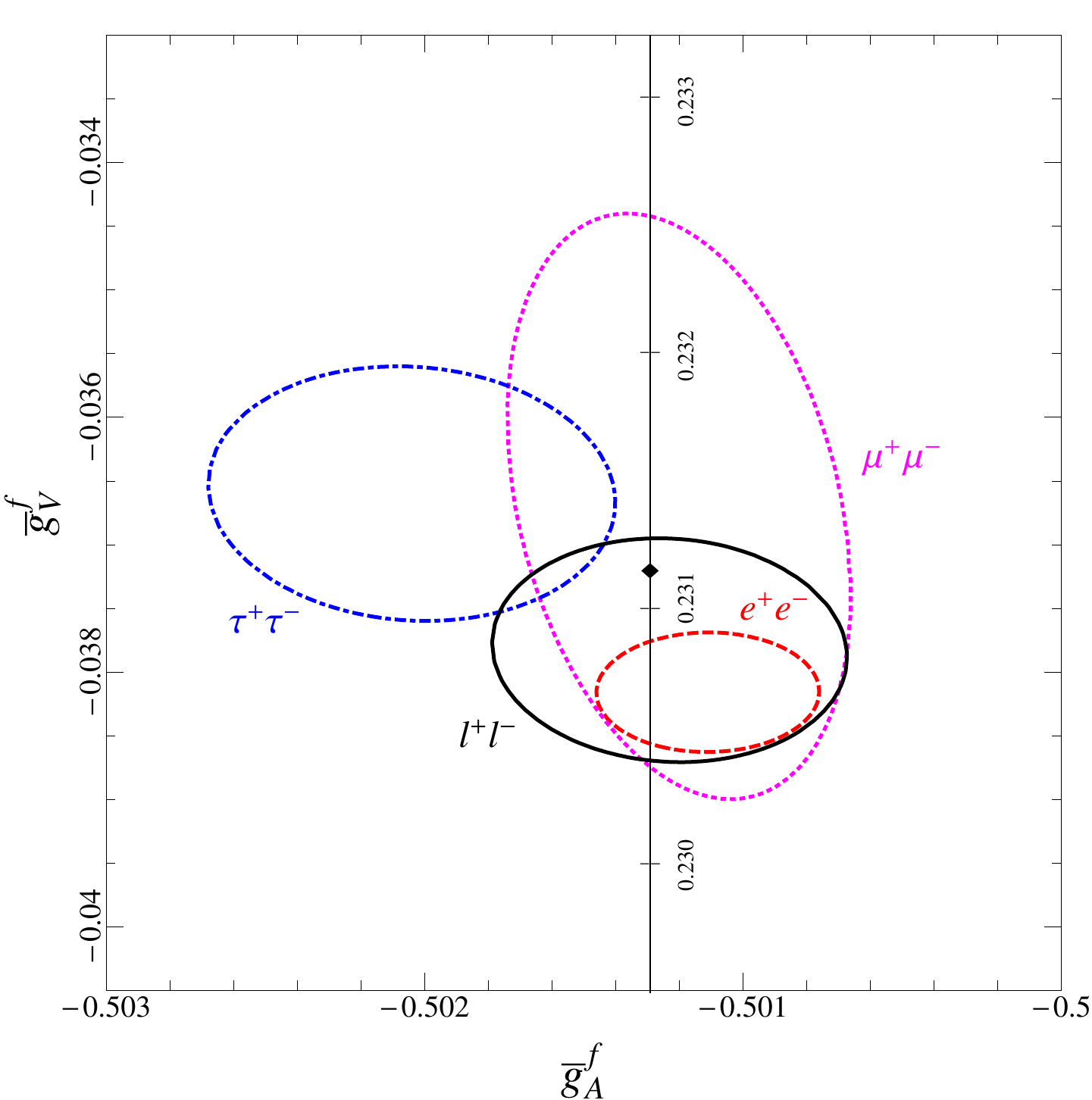,scale=0.7}
\begin{minipage}[t]{16.5 cm}
\caption{1~$\sigma$ (39.35\% CL) regions for the $Z$-pole observables $\bar{g}_V^f$ and 
$\bar{g}_A^f$, $f = e, \mu, \tau$, obtained at LEP and the SLC~\cite{ALEPH:2005ab},
compared to the SM expectation as a function of $\sin^2\hat\theta_W(M_Z)$ 
with the SM best fit value, $\sin^2\hat\theta_W(M_Z) = 0.23116$, indicated. 
Also shown is the 90\%~CL contour in $\bar{g}_{A,V}^\ell$ obtained assuming lepton universality.
(Figure reprinted as permitted according to journal guidelines from
\Journal{\PRD}{86}{010001}{2012}, J.~Erler and P.~Langacker~\cite{PDGEW2012}.)
\label{gae}}
\end{minipage}
\end{center}
\end{figure}

\begin{table}[t]
\begin{center}
\begin{minipage}[t]{16.5 cm}
\caption{$Z$-pole asymmetry measurements from the high-energy frontier (CERN, SLAC and FNAL) 
compared to the SM predictions for $M_H = 125$~GeV.
The corresponding effective weak mixing angles, $\seff$, and pulls (in standard deviations) 
from the SM prediction, $\seff = 0.23158$ (for $M_H = 125$~GeV), 
are shown in the last two columns, respectively.
Note, that the theory (PDF) uncertainty entering the CDF result~\cite{Han:2011vw} is adjusted 
to coincide and treated as fully correlated with the corresponding D\O\ result~\cite{Abazov:2011ws}.
The total average in the last line accounts for further correlations~\cite{ALEPH:2005ab}  
between various individual measurements.}
\label{asymmetries}
\end{minipage}
\begin{tabular}{ccc|c|c|c|c|c}
\\ [-2mm]
\hline\hline
&&&&&&&\\ [-2mm]
Group(s) & collider & Ref. & asymmetry & measurement & SM & $\seff$ & deviation \\ [2mm]
\hline
&&&&&&&\\ [-2mm]
SLD & SLC & \cite{Abe:2000dq} & $A_{LR}^q$ & $0.1514 \pm 0.0022$ & 0.1466
& $0.23097 \mp 0.00027$ & $-2.2$ \\
SLD & SLC & \cite{Abe:2000hk} & $A_{LR}^\ell$ & $0.1544 \pm 0.0060$ & 0.1466 
& $0.23058 \mp 0.00076$ & $-1.3$ \\
SLD & SLC & \cite{Abe:2000hk} & $A_{LR,FB}^\mu$ & $0.142 \pm 0.015$ & 0.1466
& $0.2322 \mp 0.0019$ & $+0.3$ \\
SLD & SLC & \cite{Abe:2000hk} & $A_{LR,FB}^\tau$ & $0.136 \pm 0.015$ & 0.1466
& $0.2329 \mp 0.0019$ & $+0.7$ \\ 
SLD & SLC & \cite{Abe:1996ef} & $A^e$ & $0.162 \pm 0.043$ & 0.1466
& $0.2296 \mp 0.0055$ & $-0.4$ \\ [2mm]
\hline
&&&&&&&\\ [-2mm]
ADLO & LEP~1 & \cite{ALEPH:2005ab} & $A_{FB}^b$ & $0.0992 \pm 0.0016$ & 0.1028
& $0.23221 \mp 0.00029$ & $+2.2$ \\
ADLO & LEP~1 & \cite{ALEPH:2005ab} & $A_{FB}^c$ & $0.0707 \pm 0.0035$ & 0.0734
& $0.23220 \mp 0.00081$ & $+0.8$ \\
ADLO & LEP~1 & \cite{Abe:2000uc} & $A_{FB}^s$ & $0.0976 \pm 0.0114$ & 0.1029
& $0.2325 \mp 0.0021$ & $+0.4$ \\
ADLO & LEP~1 & \cite{ALEPH:2005ab} & $A_{FB}^q$ & $0.0403 \pm 0.0026$ & 0.0421
& $0.2324 \mp 0.0012$ & $+0.7$ \\
ADLO & LEP~1 & \cite{ALEPH:2005ab} & $A_{FB}^e$ & $0.0145 \pm 0.0025$ & 0.0161
& $0.2325 \mp 0.0015$ & $+0.6$ \\
ADLO & LEP~1 & \cite{ALEPH:2005ab} & $A_{FB}^\mu$ & $0.0169 \pm 0.0013$ & 0.0161
& $0.23113 \mp 0.00073$ & $-0.6$ \\
ADLO & LEP~1 & \cite{ALEPH:2005ab} & $A_{FB}^\tau$ & $0.0188 \pm 0.0017$ & 0.0161
& $0.23008 \mp 0.00091$ & $-1.6$ \\
ADLO & LEP~1 & \cite{ALEPH:2005ab} & $-{\cal P}^\tau$ & $0.1439 \pm 0.0043$ & 0.0161
& $0.23192 \mp 0.00055$ & $+0.6$ \\
ADLO & LEP~1 & \cite{ALEPH:2005ab} & $-4/3\ {\cal P}_{FB}^\tau$ & $0.1498 \pm 0.0049$ & 0.0161
& $0.23117 \mp 0.00062$ & $-0.7$ \\ [2mm]
\hline
&&&&&&&\\[-2mm]
D\O & Tevatron & \cite{Abazov:2011ws} & $\sin^2\theta_{\rm eff}^e$ & & 
& $0.2309 \pm 0.0010$ & $-0.7$ \\
CDF & Tevatron & \cite{Han:2011vw} & $\sin^2\theta_{\rm eff}^e$ & & 
& $0.2329 \pm 0.0009$ & $+1.4$ \\
CMS & LHC & \cite{Chatrchyan:2011ya} & $\sin^2\theta_{\rm eff}^\mu$ & & 
& $0.2287 \pm 0.0032$ & $-0.9$ \\ [2mm]
\hline\hline
&&&&&&&\\ [-2mm]
all & all & & $\seff$ & & & $0.23155 \pm 0.00016$ & $-0.2$ \\ [2mm] 
\hline\hline
\end{tabular}
\end{center}
\end{table}     

LEP was not operating with polarized beams, but $A^\tau$ was measured by 
the LEP~1 Collaborations (ALEPH, DELPHI, L3 and OPAL $\equiv$ ADLO)~\cite{ALEPH:2005ab} 
through the total final state $\tau$ polarization, ${\cal P}^\tau$, and $A^e$ was extracted 
from its angular distribution or forward-backward asymmetry,
\be
{\cal P}^\tau = - A^\tau, 
\qquad\qquad\qquad
{\cal P}_{FB}^\tau = - {3\over 4} A^e.
\ee
The $Z$-pole forward-backward asymmetries at LEP~1 are given by 
\be\label{afb}
A_{FB}^f = {3\over 4} A^e A^f,
\ee
where $f = e$, $\mu$, $\tau$, $b$, $c$, $s$~\cite{Abreu:1994pj} and $q$, 
and where as before $A_{FB}^q$ refers to the hadronic charge asymmetry.

As for hadron colliders, the forward-backward asymmetry, $A_{FB}$, for $e^+ e^-$ final states 
(with invariant masses restricted to or dominated by values around $M_Z$) 
in $p\bar{p}$ collisions has been measured by 
the D\O~\cite{Abazov:2011ws} and CDF~\cite{Han:2011vw} Collaborations 
at the Tevatron and values for the weak mixing angle were extracted, which combine to 
$\sin^2\theta_{\rm eff}^e = 0.2320 \pm 0.0008$ (assuming common PDF uncertainties).

This kind of measurement is harder in the $pp$ environment due to the difficulty to assign 
the initial quark and antiquark in the underlying Drell-Yan process to the protons. 
Nevertheless, the CMS Collaboration~\cite{Chatrchyan:2011ya} at the LHC was able to report 
a measurement of $\sin^2\theta_{\rm eff}^\mu$ from their dimuon data based on
an integrated luminosity of 1.1~fb$^{-1}$. 
Given that the event sample size will grow by several orders of magnitude, and further results 
can be anticipated from $e^+ e^-$ final states, as well as from the ATLAS Collaboration, 
there is the potential of much more precise determinations of $\seff$ from the LHC.

The LEP and SLC asymmetries and branching ratios can also be analyzed in terms of 
model-independent couplings, $\bar{g}_V^f$ and $\bar{g}_A^f$, where the bar indicates a different 
normalization and the inclusion of radiative corrections~\cite{Hollik:1993cf}.
Their SM values can be obtained by multiplying Eqs.~(\ref{vzaz}) by $\sqrt{2\rho_f} \cos\theta_W$, 
where $\rho_f \neq 1$ is a radiative correction parameter, and by employing 
$\sin^2\theta_{\rm eff}^f$ in $g_V^f$. 
The resulting $\bar{g}_V^f$ and $\bar{g}_A^f$ for $f = e, \mu, \tau$ and $\ell$ are shown 
in Figure~\ref{gae}.

Table~\ref{asymmetries} summarizes the $Z$-pole asymmetry measurements with enhanced 
sensitivity to $\seff$ and the corresponding extractions.
Notice, that there is a 3.1~$\sigma$ discrepancy between the two most precise determinations,
namely $A_{LR}^q$ from the SLC and $A_{FB}^b$ from LEP.
On the other hand, the average of all measurements coincides exactly with 
the SM prediction for $M_H = 125$~GeV.
Note also, that all other measurements, \ie excluding $A_{LR}^q$ and $A_{FB}^b$,
average to $\seff = 0.23153 \pm 0.00025$, which is also in perfect agreement with the SM prediction
for $M_H = 125$~GeV.
This may be an indication that the $2.2~\sigma$ deviations in $A_{LR}^q$ and $A_{FB}^b$
might be due to fluctuations in opposite directions from the SM.
Also, the $Z \to b\bar{b}$ 
partial decay width normalized to the total hadronic width, $R^b$, and other analogously defined
$R^q$, are generally in reasonable agreement with the SM. 
This makes it difficult to construct plausible new physics models which would shift $A_{FB}^b$ 
away from its SM value by modifying the $A^b$ factor in Eq.~(\ref{afb}).
There is, however, a very recent fermionic two-loop calculation~\cite{Freitas:2012sy} of $R^b$, 
which shifts the SM prediction about $2.3$ below the experimental value. 
This is an interesting development but it should be cautioned that the contribution of purely bosonic 
loops at this order is still unknown.
Also, even if confirmed, the new physics effects proportional to $\sin^2\theta_{\rm eff}^b$ 
would have to be an order of magnitude larger than contributions to $\rho_b$, 
so there would likely be some tuning.

\subsection{\it The on-shell weak mixing angle from the W and Z boson masses}
\label{s2onshell}

As discussed in Section~\ref{SMHiggs}, the weak mixing angle can also be derived directly 
from the $W$ and $Z$ boson masses.  
The $Z$-lineshape scan at LEP~1~\cite{ALEPH:2005ab}  produced a determination of $M_Z$ 
of ultra-high precision,
\be\label{MZ}
M_Z = 91.1876 \pm 0.0021 \mbox{ GeV}.
\ee
Precise values for $M_W$ were obtained in $W$-pair production 
at LEP~2~\cite{Alcaraz:2006mx} and single-$W$ production 
at the Tevatron~\cite{TevatronElectroweakWorkingGroup:2012gb}. 
They are shown in Table~\ref{MW} together with the corresponding values for the 
weak mixing angle in the on-shell scheme, $s^2_W = 1 - c^2_W$, computed using Eq.~(\ref{MZ}). 

\begin{table}[t]
\begin{center}
\begin{minipage}[t]{16.5 cm}
\caption{$W$-boson mass measurements from the high-energy frontier (CERN and FNAL) 
to be compared with the SM prediction, $M_W = 80.369$~GeV (for $M_H = 125$~GeV).
The corresponding on-shell weak mixing angles, $s^2_W$, from the combination with $M_Z$, 
and pulls (in standard deviations) from the SM prediction,
$s^2_W = 0.22320$ (for $M_H = 125$~GeV), are shown in the last two columns, respectively.}
\label{MW}
\end{minipage}
\begin{tabular}{ccc|c|c|c}
\\ [-2mm]
\hline\hline
&&&&&\\ [-2mm]
Group(s) & collider & Ref. & $M_W$~[GeV] & $s^2_W \equiv 1 - {M_W^2\over M_Z^2}$ & 
deviation \\ [2mm]
\hline
&&&&&\\ [-2mm]
ADLO & LEP~2 & \cite{Alcaraz:2006mx} & 
$80.376 \pm 0.033$ & $0.22307 \mp 0.00064$ & $-0.2$ \\ 
CDF \& D\O\ & Tevatron & \cite{TevatronElectroweakWorkingGroup:2012gb} & 
$80.387 \pm 0.016$ & $0.22286 \mp 0.00031$ & $-1.1$ \\ [2mm]
\hline\hline
&&&&&\\ [-2mm]
all & both & & $80.385 \pm 0.015$ & $0.22290 \mp 0.00028$ & $-1.1$ \\ [2mm] 
\hline\hline
\end{tabular}
\end{center}
\end{table}     

The error in the world average of $s^2_W$ in Table~\ref{MW} is significantly larger than
the corresponding error in $\seff$ in Table~\ref{asymmetries}.
However, a fairer comparison would be based on the quantities defined in Eqs.~(\ref{MHWZ}) and 
(\ref{s2theta}) which we relate in various ways to $G_F$ and the fine structure constant,
$\alpha$, through,
\be\label{MWZs2theta}
M_W \sqrt{1 - {M_W^2\over M_Z^2}} = M_W \sin\theta_W = 
M_Z \cos\theta_W \sin\theta_W = \sqrt{\pi\alpha\over \sqrt{2} G_F} \equiv 
A = 37.28038(1) \mbox{ GeV}.
\ee
Since $A$ is practically known exactly (to lowest order) 
we find the following relations between experimental uncertainties,
\be
{\delta M_Z \over M_Z} = \left( 2 - {M_Z^2 \over M_W^2} \right) {\delta M_W \over M_W} =
{\tan^2\theta_W - 1 \over 2}\ {\delta \sin^2\theta_W \over  \sin^2\theta_W}\ ,
\qquad\qquad
{\delta M_W \over M_W} = - {1\over 2}\ {\delta \sin^2\theta_W\over \sin^2\theta_W}\ ,
\ee
so that the $M_W$ error of 15~MeV corresponds rather to measurements of $M_Z$ to 12~MeV
precision or to $\seff$ with an uncertainty of $9 \times 10^{-5}$. 
We conclude that $M_W$ is now the most accurately measured derived quantity in
this sector of the EW theory.
Of course, relations such as in Eqs.~(\ref{MWZs2theta}) receive important radiative corrections
(see the next subsection), making $M_W$ and $\seff$ fundamentally distinct observables 
already in the SM (and even more so in the presence of new physics). 
This is illustrated in Figure~\ref{mwmt} where the direct measurements of $M_W$ and $m_t$ are
compared with all other EW precision data (dominated by $\seff$). 

\begin{figure}[t]
\begin{center}
\epsfig{file=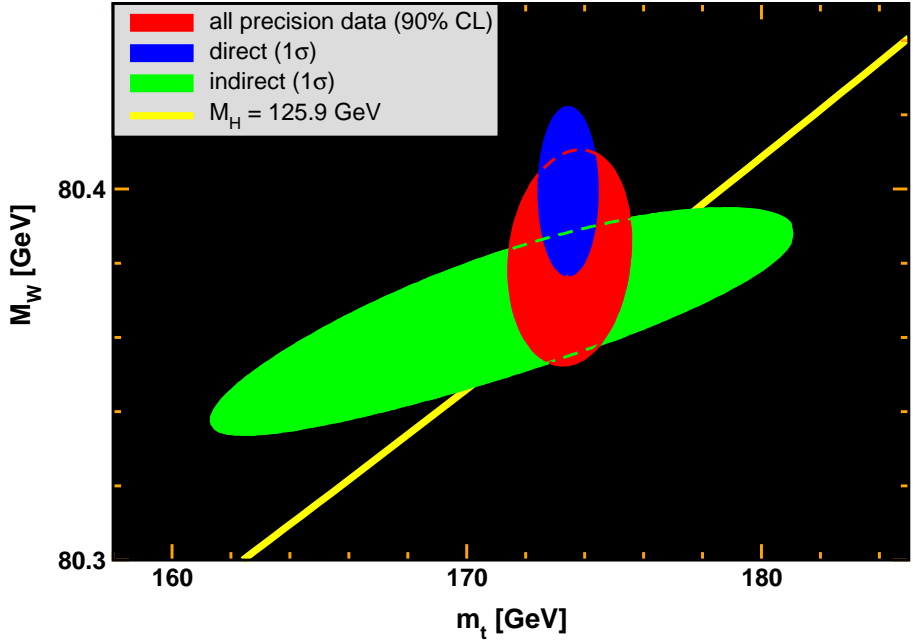,scale=0.49}
\begin{minipage}[t]{16.5 cm}
\caption{$1~\sigma$ (39.35\% CL) region in $M_W$ as a function of $m_t$ for the direct and 
indirect determinations, and the 90\% CL region ($\Delta \chi^2 = 4.605$) allowed by both data sets. 
The SM prediction is indicated as the narrow, bright (yellow) band.
\label{mwmt}}
\end{minipage}
\end{center}
\end{figure}

Most of the LEP~2 measurements and all of those from the Tevatron use direct kinematical 
reconstruction.  
But the LEP Collaborations also performed a $W^+ W^-$ threshold scan of the type 
envisioned with much larger statistics at an International Linear Collider (ILC),
where a 5~MeV uncertainty~\cite{Beneke:2007zg} may be reached within only one year.
Kinematical fitting at the ILC could contribute an independent determination, likewise with 5~MeV
precision, but this would need several years of data taking~\cite{Zanderighi:2007}.

The Tevatron combination~\cite{TevatronElectroweakWorkingGroup:2012gb} is strongly 
dominated by the latest CDF result from Run~II~\cite{Aaltonen:2012bp},
$M_W = 80.387 \pm 0.012_{\rm stat.} \pm 0.010_{\rm syst.} \pm 0.011_{\rm theo.}$~GeV,
even though this is based on only 2.2~fb$^{-1}$ of integrated luminosity.  From past experience 
one anticipates that the systematic error should scale roughly with statistics~\cite{Zeng:2012}.
With the full data set based on 10~fb$^{-1}$ one can therefore reasonably expect the total CDF 
error of 19~MeV to shrink to at least 13~MeV, or even further if there is also progress regarding 
production theory, namely parton distribution functions (PDFs) and QED radiation.
Indeed, while the method was traditionally limited by the lepton energy scale determination, 
the CDF error is now dominated by the PDF uncertainty.
Given these developments, there are also excellent prospects for significant improvements 
by D\O, as well as for the measurements by ATLAS and CMS at the LHC.

\subsection{\it Implications for the Higgs boson mass}
\label{MH}

\begin{figure}[t]
\begin{center}
\epsfig{file=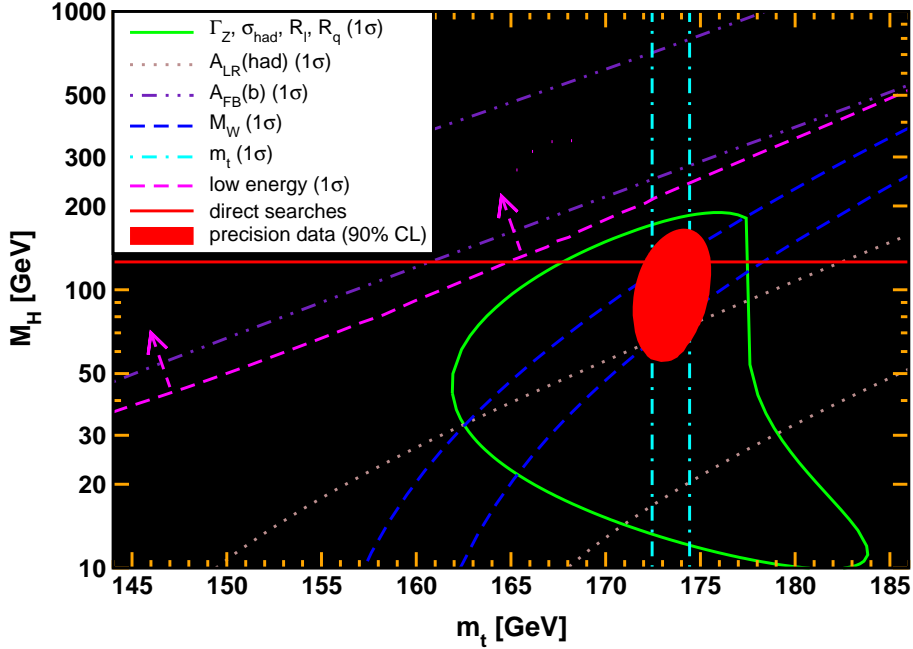,scale=0.49}
\begin{minipage}[t]{16.5 cm}
\caption{$1~\sigma$ (39.35\% CL) contours of $M_H$ as a function of the top quark
pole mass for various inputs, and the 90\% CL region ($\Delta \chi^2 = 4.605$) allowed by all data.
The horizontal red line reflects the Higgs masses indicated by 
the LHC events~\cite{ATLASmh,CMSmh}.
\label{mhmt}}
\end{minipage}
\end{center}
\end{figure}

Other SM parameters, such as $M_H$ and $m_t$, enter into the relations~(\ref{MWZs2theta}) at 
the level of radiative corrections.  
Abbreviating, $\hat{s}^2_Z \equiv \sin^2\hat\theta_W(M_Z)$ and 
$\hat{c}_Z^2 \equiv \cos^2\hat\theta_W(M_Z)$, one can define radiative correction parameters,
$\Delta r$~\cite{Sirlin:1980nh}, $\Delta \hat{r}_W$ and $\Delta \hat{r}$~\cite{Degrassi:1990tu}, 
\be\label{deltar}
M_W^2 s^2_W = M_Z^2\, c^2_W s^2_W \equiv {A^2\over 1 - \Delta r}\ , 
\qquad\qquad
M_W^2 \hat{s}_Z^2  \equiv {A^2\over 1 - \Delta \hat{r}_W}\ , 
\qquad\qquad
M_Z^2\, \hat{c}_Z^2 \hat{s}_Z^2 \equiv {A^2\over 1 - \Delta \hat{r}}\ .
\ee
Note, that all three of these parameters contain a common component $\Delta\hat\alpha(M_Z)$, 
which arises from the renormalization group evolution of $\alpha$ from the Thomson limit to 
$M_Z$ evaluated in the \msbar-scheme~\cite{Erler:1998sy}, 
\be\label{alphat}
\hat\alpha_Z \equiv  \hat\alpha(M_Z) = {\alpha \over 1 - \Delta\hat\alpha(M_Z)}\ , 
\ee
and introduces a theoretical uncertainty of about  $\pm 10^{-4}$~\cite{Davier:2010nc} from 
the hadronic region, as well as the charm and bottom quark thresholds. 
This induces an uncertainty of $\mp 5$~GeV in the extracted $M_H$.
The first of Eqs.~(\ref{deltar}) shows that $M_W = 80.385 \pm 0.015$~GeV can also be interpreted
as a measurement of $\Delta r = 0.03505 \mp 0.00090$.
Similarly, after $\seff$ has been translated with the help of Eq.~(\ref{seff}) into
$\hat{s}^2_Z = 0.23126 \pm 0.00016$, the last of Eqs.~(\ref{deltar}) implies
$\Delta \hat{r} = 0.05982 \pm 0.00045$. Finally, 
combining the results for $M_W$ and $\seff$ gives $\Delta \hat{r}_W = 0.06994 \pm 0.00073$, 
but this is not independent of $\Delta r$ and $\Delta \hat{r}$.

\begin{figure}[t]
\begin{center}
\epsfig{file=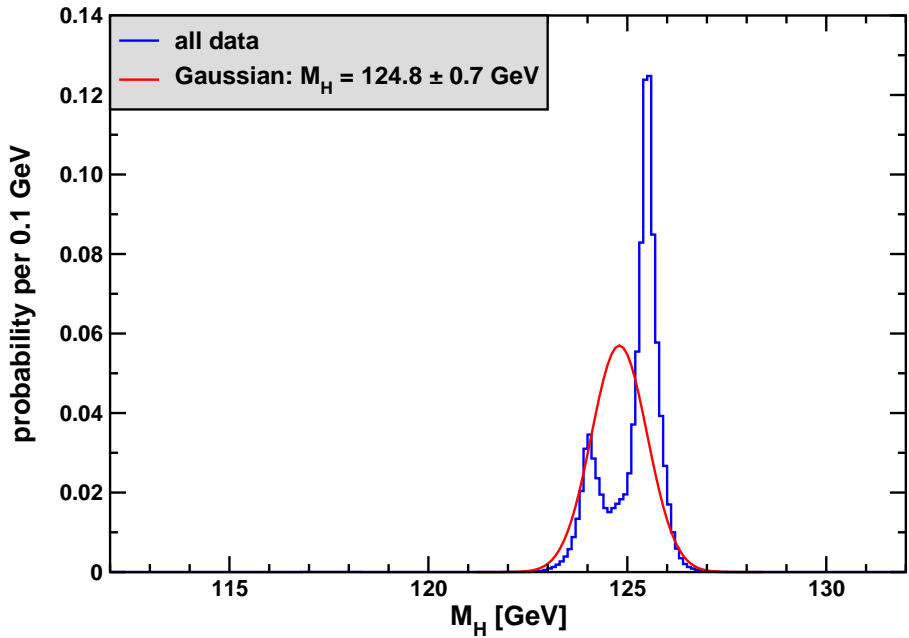,scale=0.49}
\begin{minipage}[t]{16.5 cm}
\caption{The normalized probability distribution (in blue) of $M_H$ based on all (direct and 
indirect) data.
It is highly non-Gaussian, but one can construct a reference bell curve (shown in red) 
providing a $1~\sigma$ estimate of $M_H$, as well as a measure of significance.
\label{mh}}
\end{minipage}
\end{center}
\end{figure}

Both, $\Delta r$ and $\Delta \hat{r}$, are functions of $M_H$ (the leading behavior will be
illustrated in Section~\ref{STUrho}) and can therefore be used to constrain it, 
but both depend quadratically on $m_t$. Denoting, 
\be
\rho_t \equiv {N_C \over 16\pi^2} {m_t^2 \over v^2} = 
0.0094 \left( {m_t \over 173.21 \mbox{ GeV}} \right)^2, 
\ee
where $N_C = 3$ is the color factor, one finds for the leading terms,
$\Delta r \simeq  \Delta\hat\alpha(M_Z) - {\cot^2\theta_W} \rho_t$ and
$\Delta \hat{r} \simeq \Delta\hat\alpha(M_Z) - \rho_t$, 
so that $m_t$ needs to be known to very high precision. 
The various measurements from the Tevatron~\cite{Lancaster:2011wr} and 
the LHC~\cite{LHCmt} (strongly dominated by the CMS $\mu +$jets channel) combine to,
\be
m_t^{\rm Tevatron} = 173.18 \pm 0.56_{\rm stat} \pm 0.75_{\rm syst} \mbox{ GeV}, 
\qquad\qquad
m_t^{\rm LHC} = 173.34 \pm 0.47_{\rm stat} \pm 1.33_{\rm syst}\mbox{ GeV}. 
\ee
Conservatively assuming that the Tevatron systematic error is common to both colliders,
one can form a global average,
\be 
m_t = 173.21 \pm 0.51_{\rm uncorr} \pm 0.75_{\rm corr} \pm 0.5_{\rm theo} \mbox{ GeV} 
= 173.2 \pm 1.0 \mbox{ GeV},
\ee
where we have added a theory uncertainty from the relation~\cite{Chetyrkin:1999qi} 
between the top quark pole mass and the \msbar-mass (the size of the three-loop term).  
The latter is used in the EW library, GAPP~\cite{Erler:1999ug}, to minimize theoretical 
uncertainties. Such a short distance mass definition (unlike the pole mass) is free from 
non-perturbative and renormalon~\cite{Beneke:1998ui} uncertainties. 
We are assuming that the kinematic mass extracted from the collider events corresponds 
within this uncertainty to the pole mass. Constraints on $M_H$ as a function of $m_t$ 
are shown in Figure~\ref{mhmt} for various data sets.

A global fit to all EW precision data including the observables in Tables~\ref{asymmetries} 
and \ref{MW}, as well as further measurements from high and low energies, gives
\be\label{MHfit}
M_H = 101^{+25}_{-20} \mbox{ GeV}.
\ee
The quality of the fit is excellent with a $\chi^2$ of 43.658 for 42 effective degrees of freedom,
which translates into a probability for a higher $\chi^2$ of 41\%.
Reflecting the discussion of the previous paragraph, there is a large (42\%) correlation 
between $M_H$ and $m_t$.
The fit value~(\ref{MHfit}) is slightly lower (by 1.0~$\sigma$) than the
\be\label{MHall}
M_H = 124.8 \pm 0.7 \mbox{ GeV}, 
\ee
suggested~\cite{Erler:2012uu} by the Higgs boson candidates seen at 
the LHC~\cite{ATLASmh,CMSmh}. 
Indeed, if one combines the precision data with the direct Higgs boson search results from
LEP~2~\cite{Barate:2003sz}, the Tevatron~\cite{TEVNPH:2012ab}, and 
the LHC~\cite{ATLASmh,CMSmh}, one can construct 
the proper probability density~\cite{Erler:2012uu} shown in Figure~\ref{mh}.
The distribution shows two peaks (traceable to the two LHC experiments) and is highly 
non-Gaussian not only in the bulk but also (and more importantly) in the tails 
where further local maxima occur. 
By cutting the distribution off where it falls below the density of the highest such local maximum
defines a signal region. 
The integral under this signal region and its center define a reference Gaussian,
which is also shown in the Figure. 
Moreover, one can now unambiguously find the number of standard deviations corresponding 
to the signal region, giving rise to a significance of $3.4~\sigma$. 
This method~\cite{Erler:2012uu} avoids the poorly defined look-elsewhere effect correction
which needs to be applied by the LHC Collaborations.

\subsection{\it Implications for physics beyond the Standard Model
\label{STUrho}} 

The parameters, $\Delta r$, $\Delta \hat{r}_W$ and $\Delta \hat{r}$, introduced in the previous
subsection, are also possibly affected by and can provide constraints on hypothetical particles
beyond the SM which may appear in loop corrections to the transverse parts of the 
current-current correlators (\ie in the so-called oblique corrections),
\be
\hat\Pi^{\rm new}_{\gamma\gamma}(q^2) \equiv {g^2\over 2} \langle J^A J^A \rangle, 
\qquad\qquad
\hat\Pi^{\rm new}_{\gamma Z}(q^2)            \equiv {g^2\over 2} \langle J^A J^Z \rangle,
\ee
\be
\hat\Pi^{\rm new}_{ZZ}(q^2)                         \equiv {g^2\over 2} \langle J^Z J^Z \rangle, 
\qquad\qquad
\hat\Pi^{\rm new}_{WW}(q^2)                      \equiv {g^2\over 2} \langle J^W J^W \rangle.
\ee
Consider first the $WW$ and $ZZ$ self-energies.
Since at present precise measurements are available only at low (basically vanishing) energies 
and at the EW scale, one is left with $\hat\Pi^{\rm new}_{WW}(0)$, 
$\hat\Pi^{\rm new}_{WW}(M_W^2)$, $\hat\Pi^{\rm new}_{ZZ}(0)$ and 
$\hat\Pi^{\rm new}_{ZZ}(M_Z^2)$. 
These are infinite (their UV-divergencies can be absorbed in the counterterm related to $G_F$)
but appropriately chosen differences are finite,
\be\label{piwwzz}
\Delta\hat{r}_W^{\rm new} =
{\hat\Pi_{WW}^{\rm new} (M_W^2) - \hat\Pi_{WW}^{\rm new} (0) \over M_W^2} \equiv 
{\alpha \over 4\hat{s}^2_Z} S_W,
\qquad\qquad
\Delta\hat{r}_Z^{\rm new} \equiv
{\hat\Pi_{ZZ}^{\rm new} (M_Z^2) - \hat\Pi_{ZZ}^{\rm new} (0) \over M_Z^2} \equiv 
{\alpha \over 4 \hat{s}^2_Z\hat{c}^2_Z} S_Z,
\ee
\be\label{rho}
\Delta \hat\rho^{\rm new} = 
{\hat\Pi_{WW}^{\rm new} (M_W^2) - \hat{c}^2_Z \hat\Pi_{ZZ}^{\rm new} (M_Z^2) \over M_W^2}\ ,
\quad\qquad
\Delta \rho^{\rm new} = {\hat\Pi_{WW}^{\rm new} (0) - 
\hat{c}^2_Z \hat\Pi_{ZZ}^{\rm new} (0) \over M_W^2 (1 - \Delta\hat{r}_W)} \equiv 
{\alpha T\over 1-\Delta\hat{r}_W} \approx 
\hat\alpha_Z T,
\ee
\vspace{4pt}
\be\label{deltarz}
\Delta \hat{r}^{\rm new} = 
\hat\rho\, (\Delta \hat{r}_W^{\rm new} - \Delta \hat\rho^{\rm new}) = {1 \over M_Z^2} 
\left[ \hat\Pi_{ZZ}^{\rm new}(M_Z^2) - {\hat\Pi_{WW}^{\rm new}(0) \over \hat{c}_Z^2} \right] =
\alpha \left( {S_Z  \over 4 \hat{s}^2_Z \hat{c}^2_Z} - \hat\rho T \right),
\ee
so that only three are independent~\citer{Marciano:1990dp,Peskin:1991sw}.
For these formulae, we defined the classic $\rho$ parameter~\cite{Veltman:1977kh} 
describing the ratio of neutral-to-charged current interaction strengths
in analogy to the high-energy $\hat\rho$ parameter~\cite{Degrassi:1990tu},
\be\label{rhohat}
\hat\rho \equiv {1\over 1 - \Delta\hat\rho} \equiv {c_W^2\over \hat{c}_Z^2} = 
\frac{1 - \Delta \hat{r}}{1 - \Delta \hat{r}_W}\ ,
\qquad\qquad
\rho \equiv {1\over 1 - \Delta\rho} \equiv {G_{\rm NC}\over G_F} \equiv
\frac{1 - \Delta \hat{r}}{1 - \Delta \hat{r}_Z}\ ,
\ee
and kept track of reducible higher-order effects, so that 
\be
M_Z^2\, \hat{c}_Z^2 \hat{s}_Z^2 = 
{A^2\over 1 - \Delta \hat{r}} = 
{A^2\over \hat\rho (1 - \Delta \hat{r}_W)} =
{A^2\over        \rho( 1 - \Delta \hat{r}_Z)}
\label{eq:deltar}
\ee
exactly. It is understood that these new physics contributions are to be added to the SM ones 
(\ie not to be written in factorized form) and that therefore the self-energies are normalized with 
$\alpha$ rather than $\hat\alpha_Z$.
The dominant SM contributions to the oblique parameters for the cases of a heavy top quark and 
a heavy Higgs boson are given by,
\be
\Delta\hat{r}_W \approx 
\Delta\hat\alpha(M_Z) + {\alpha\over 48\pi\hat{s}_Z^2} \ln {M_H^2\over M_Z^2}\ ,
\qquad\qquad
\Delta\hat{r}_Z \approx
\Delta\hat\alpha(M_Z) + {\alpha\over 48\pi\hat{s}_Z^2 \hat{c}_Z^2}  \ln {M_H^2\over M_Z^2}\ ,
\ee
\be
\Delta r \approx 
\Delta\hat\alpha(M_Z) + 
{11 \alpha\over 48\pi\hat{s}_Z^2} \ln {M_H^2\over M_Z^2}  - {\hat{c}_Z^2 \over \hat{s}_Z^2} \rho_t \ ,
\qquad\qquad
\Delta\rho \approx \rho_t - {3 \hat\alpha_Z \over 16\pi\hat{c}^2_Z} \ln {M_H^2\over M_Z^2} \ .
\ee

\begin{figure}[t]
\begin{center}
\epsfig{file=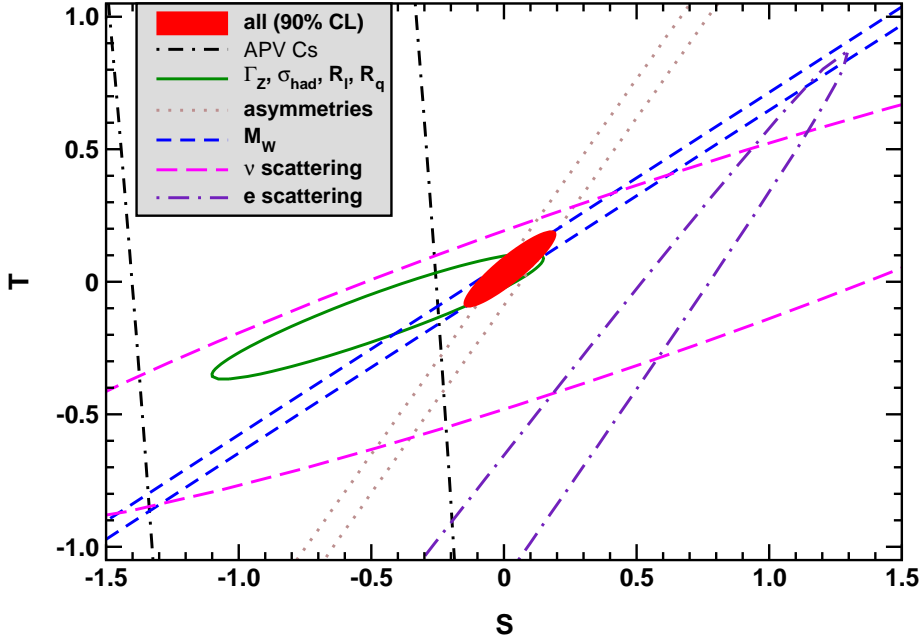,scale=0.49}
\begin{minipage}[t]{16.5 cm}
\caption{1~$\sigma$ regions in $S$ and $T$ from various inputs. 
Data sets not involving $M_W$ or $\Gamma_W$ are insensitive to $U$. 
The strong coupling constant, $\hat\alpha_s(M_Z)$, is constrained by using the $\tau$ lifetime as
additional input. 
This allows $\Gamma_Z$, the hadronic $Z$-peak cross-section, $\sigma_{\rm had}$, 
and the $R_q$, to provide additional new physics probes rather than fixing $\hat\alpha_s$.
The long-dash-dotted (indigo) contour from polarized $e$ 
scattering~\cite{Anthony:2005pm,Young:2007zs} is the upper tip of an elongated ellipse 
centered at $S \approx -14$ and $T \approx -20$.  
It may look as if it is deviating strongly but it is off by only 1.8~$\sigma$,
an illusion arising because $\Delta \chi^2 > 0.77$ throughout the visible part of the contour.
\label{ST}}
\end{minipage}
\end{center}
\end{figure}

A non-vanishing $T$ is associated with new physics disrespecting the custodial symmetry 
mentioned in Section~\ref{SMHiggs}, such as from mass splittings within iso-doublets or 
higher-dimensional Higgs representations.
Experimentally, $T$ can be separated from $S_Z$ in Eq.~(\ref{deltarz}) by low-energy neutral
current processes.
One can also use the $Z$-width, $\Gamma_Z$, which through $Z$-boson wave-function 
renormalization is affected by the derivative of $\hat\Pi^{\rm new}_{ZZ}(q^2)$.
To the extent to which this differs from the difference defining $S_Z$ in Eq.~(\ref{piwwzz}),
there would then be an additional measurable parameter, called $V$~\cite{Maksymyk:1993zm}.
A similar remark applies to $\hat\Pi^{\rm new}_{WW}(q^2)$ where the difference in 
Eq.~(\ref{piwwzz}) and its derivative give rise to the parameters
$S_W$~\cite{Marciano:1990dp} and $W$~\cite{Maksymyk:1993zm}, respectively.
However, new physics at a very high scale, $\Lambda_{\rm new}$, 
will affect $S_Z$ and $V$ (or $S_W$ and $W$)
in very similar ways, and only $S_W$, $S_Z$ and $T$ survive 
when one additionally assumes the oblique new physics to be very heavy.

\begin{table}[t]
\begin{center}
\begin{minipage}[t]{16.5 cm}
\caption{Result of a global fit to the three parameters for heavy oblique new physics 
and their correlations.
The agreement with the SM prediction, $S = T = U = 0$, is remarkable.}
\label{STU}
\end{minipage}
\\ [2mm]
\begin{tabular}{c|c||ccc}
\hline\hline
&&&& \\ [-2mm]
parameter & global fit result & \multicolumn{3}{c}{correlation matrix} \\ [2mm]
\hline
&&&& \\ [-2mm]
$S$ & $0.00 \pm 0.10$ & $\ph{-}1.00$ & $\ph{-}0.91$ &       $- 0.55$ \\
$T$ & $0.02 \pm 0.11$ & $\ph{-}0.91$ & $\ph{-}1.00$ &       $- 0.79$ \\
$U$ & $0.04 \pm 0.09$ & $      - 0.55$ & $       - 0.79$ & $\ph{-}1.00$ \\ [2mm]
\hline\hline
\end{tabular}
\end{center}
\end{table}     

This case also greatly simplifies the analysis for the QED vacuum polarization tensor, 
\be
\hat\Pi^{\rm new}_{\gamma\gamma}(q^2) = 
\hat\Pi^{\rm new}_{\gamma\gamma}(0) 
+ q^2 {d\over d q^2} \hat\Pi^{\rm new}_{\gamma\gamma}(q^2) \big|_{q^2 = 0} 
+ {\cal O} \left( {q^4\over \Lambda^4_{\rm new}} \right),
\ee
where the first term vanishes as a consequence of the QED Ward identity, 
while the remaining term can be absorbed into the \msbar-definition, $\hat{\alpha}_Z$,
of the gauge coupling.
Parallel remarks apply to $\hat\Pi^{\rm new}_{\gamma Z}(q^2)$ whose effects are
absorbed into $\hat{s}^2_Z$, but as was the case before with 
the $V$ and $W$ parameters, relatively light new physics may, at least in principle, separate 
differences from derivatives (or equivalently allow higher orders in $\Lambda^{-1}_{\rm new}$) giving 
rise to what is called the $X$ parameter~\cite{Maksymyk:1993zm}. 

Finally, similar observations can be made about the vector parts of $\hat\Pi^{\rm new}_{ZZ}(q^2)$ 
and $\hat\Pi^{\rm new}_{WW}(q^2)$, so that in the limit $\Lambda_{\rm new} \to \infty$ 
the parameters $S_W$ and $S_Z$ are only affected by new physics breaking axial 
$SU(2)_L$~\cite{Kennedy:1990ib}.
The combination~\cite{Peskin:1991sw} given by $U \equiv S_W - S_Z = \Delta\hat\rho - \Delta\rho$ 
manifestly breaks the vector part, as well, and in this sense $U$ can be regarded as second order 
in the new physics and is expected to be small.
This is borne out in concrete scenarios of dominantly oblique new physics like technicolor,
extra fermion generations, or additional Higgs multiplets, with $U$ often of similar size as the 
usually neglected parameters $V$, $W$, and $X$~\cite{Maksymyk:1993zm}. 
Thus, one frequently reduces the analysis allowing only $S \equiv S_Z$ and $T$, 
as illustrated in Figure~\ref{ST}. The current three parameter fit result is shown in Table~\ref{STU}.
Note, that contributions to $S$ are not decoupling.
\Eg a degenerate extra fermion generation contributes $S = 2/3\pi$ independently 
of the fermion masses and is excluded along with many technicolor models.

In the above discussion, we have implicitly assumed that the new physics is at or above the EW 
scale.  Another possibility is that there are new particles with masses much lighter than $M_Z$,
as long as they are very weakly coupled.  An example of this~\cite{Davoudiasl:2012ag}
would be a ``dark-$Z$" boson with no direct SM fermion couplings, 
but both kinetic and mass mixing effects with the photon and the ordinary $Z$.  
It could affect the predicted value of the weak mixing angle at low energies
in terms of $\hat s_Z^2$ and would therefore contribute to the $X$ parameter.

\section{Neutrino scattering}
\label{nu}

The gauge and Higgs boson properties reviewed in Section~\ref{WZ} strongly
constrain new physics scenarios that may alter them.  This may occur through mixing
or by otherwise modifying the tree-level relations of the SM.  The quantum oblique corrections 
discussed in Section~\ref{STUrho} are another possibility.

On the other hand, there may be new particles beyond the SM which mostly generate new
amplitudes without strongly affecting the masses and couplings of the $W$ and $Z$ bosons.
Then the new physics would not be resonating and one best studies processes away from 
the $Z$-pole.   
This may be at the hadron colliders, Tevatron and LHC, \ie at the current energy frontier,
but lower energy processes typically have the advantage of much higher rates. 
To screen the EW and possibly the new physics from the electromagnetic and 
strong interactions one either probes with neutrinos or utilizes violations of symmetries 
such as parity and CP.
This will be discussed, respectively, in the present and in the following section (previous reviews
can be found in Refs.~\cite{Erler:2004cx,PDGEW2012,Langacker:1996qb}).

High-energy neutrino beams are produced by directing protons of very high energy on
a fixed target from which pions and kaons emerge as secondary beams.
These tend to be positively charged so that it is easier to make high-intensity 
$\nu_\mu$ than $\bar\nu_\mu$ beams
(one needs to correct for small $\nu_e$ and $\bar\nu_e$ contaminations).
If the decaying mesons are energy selected the neutrino energies can be constrained
(narrow band beams) to some extent, but one still needs to measure the final state $\mu^\pm$
momenta to reliably determine $E_\nu$. 
This is not an option, however, for the deep-inelastic scattering (DIS) of neutrinos from hadrons or 
nuclei mediated by the neutral current so that here one focuses on integrated cross-sections and 
their ratios.
A future direction are neutrino factories~\cite{Albright:2000xi} where muons decay in 
primary beams, allowing better knowledge of the $\nu$-spectra and a composition of exactly 50\% 
$\nu_\mu$ and $\bar\nu_e$.

\subsection{\it Neutrino-electron scattering}
\label{sec:nue}

The elastic scattering of $\nu_\mu$ or $\bar\nu_\mu$ from electrons~\cite{Panman:1996} 
is at the leading order mediated by the weak neutral current (NC), which in turn was 
discovered through a single $\bar\nu_\mu$ event of this kind at CERN~\cite{Hasert:1973cr}.
From the second Eq.~(\ref{Leff}) one can glean the relevant interaction\footnote{In order to suppress
some spurious factors of $\sqrt{2}$ we normalize the low-energy Lagrangians in terms of $v$ rather 
than $G_F$, with the understanding that a renormalization scheme is used so that Eq.~(\ref{GF}) 
holds to all orders.} (in the limit of vanishing $\nu$ masses),
$$
{\cal L}_{\rm NC}^{\nu_\mu e} = - 2 {\cos^2\theta_W \over v^2} 
\ovl\nu_\mu \gamma^\rho {g_V^{\nu_\mu} - g_A^{\nu_\mu} \gamma^5 \over 2} \nu_\mu
\ovl{e} \gamma_\rho {g_V^e - g_A^e \gamma^5 \over 2} e = 
$$
\be\label{Lnumue}
- {2 \over v^2} 
\ovl\nu_\mu \gamma^\rho P_L \nu_\mu 
\ovl{e} \gamma_\rho {g_{LV}^{\nu_\mu e} - g_{LA}^{\nu_\mu e} \gamma^5 \over 2} e =
- {2 \over v^2} \ovl{\nu_{\mu L}} \gamma^\rho \nu_{\mu L} 
\left[ g_{LL}^{\nu e}  \ovl{e_L} \gamma_\rho e_L + g_{LR}^{\nu e} \ovl{e_R} \gamma_\rho e_R \right],
\ee
where the SM tree-level relations for the coefficients of these effective four-Fermi operators are 
given by,
\be\label{gLRnumue}
g_{LL}^{\nu_\mu e} \equiv \cos^2\theta_W g_L^{\nu_\mu} g_L^e = - {1\over 2} + \sin^2\theta_W,
\qquad\qquad
g_{LR}^{\nu_\mu e} \equiv \cos^2\theta_W g_L^{\nu_\mu} g_R^e =  \sin^2\theta_W,
\ee
\be\label{gvanumue}
g_{LV}^{\nu_\mu e} \equiv g_{LL}^{\nu_\mu e} + g_{LR}^{\nu_\mu e} = - {1\over 2} + 2 \sin^2\theta_W,
\qquad\qquad
g_{LA}^{\nu_\mu e} \equiv g_{LL}^{\nu_\mu e} - g_{LR}^{\nu_\mu e} = - {1\over 2}\ .
\ee
The first index $L$ is redundant in the absence of right-handed neutrinos
(or when their masses are very large) and therefore it is often dropped.
Experimentally, one can separate these couplings by using 
spectral information or by comparing the $\nu_\mu$ and $\bar\nu_\mu$ rates. 
For example, in the ultra-relativistic regime, $E_\nu \gg m_e$, and for events which scatter strictly in 
the forward direction, helicity conservation implies that only left-handed (right-handed) electrons can 
participate in $\nu$ ($\bar\nu$)-scattering, thus filtering out $g_{LL}^{\nu_\mu e}$ 
($g_{LR}^{\nu_\mu e}$).
In general, the differential neutrino scattering cross-section in the laboratory frame 
reads~\cite{Sarantakos:1982bp},
\be\label{dsigmanue}
{d \sigma(\nu e^- \to \nu e^-)\over d y} =
{m_e E_\nu \over \pi v^4} \left[ (g_{LL}^{\nu e})^2 + 
(g_{LR}^{\nu e})^2 (1 - y)^2 - g_{LL}^{\nu e} g_{LR}^{\nu e} {m_e \over E_\nu} y \right],
\ee
where we neglected $-q^2 $ against $M_W$ in the $W$-propagator, and where the Lorentz-invariant quantity,
\be\label{y}
y \equiv {p_e (p_\nu - p_\nu') \over p_e p_\nu} 
\equiv  {p_e q \over p_e p_\nu} 
= {-q^2 \over 2 p_e p_\nu} 
= 1 - {E_\nu' \over E_\nu}
= {E_e' - m_e \over E_\nu}\ ,
\qquad\qquad
0 \leq y \leq \left( 1 + {m_e\over 2 E_\nu} \right)^{-1},
\ee
is the relative neutrino energy transfer in terms of the initial neutrino and 
electron 4-momenta, $p_\nu$ and $p_e$, and their primed final state counterparts. 
The $\bar\nu$ cross-section has exactly the same form as Eq.~(\ref{dsigmanue}) except for 
the interchange $g_{LL}^{\nu e} \leftrightarrow g_{LR}^{\nu e}$.
In the ultra-relativistic limit, this integrates to
\be\label{sigmanue}
\sigma(\nu e^- \to \nu e^-) = {m_e E_\nu \over \pi v^4} 
\left[ (g_{LL}^{\nu e})^2 + {(g_{LR}^{\nu e})^2 \over 3} \right],
\qquad\qquad
\sigma(\bar\nu e^- \to \bar\nu e^-) = {m_e E_{\bar\nu} \over \pi v^4} 
\left[ {(g_{LL}^{\nu e})^2 \over 3}+ (g_{LR}^{\nu e})^2 \right],
\ee
Thus, the small $m_e$ strongly suppresses the cross-sections which do not exceed half a fb 
even for 300~GeV incident neutrinos, and event rates of only a few thousand have been achieved.
With future neutrino beams of about a factor of 100 higher intensity, so-called {\em superbeams},
the errors would decrease by an oder of magnitude and very competitive results would become
possible.
Yet higher precision could be obtained at a neutrino factory~\cite{Albright:2000xi}.

The most precise measurements (shown in Table~\ref{nue}) are from
the CHARM~\cite{Dorenbosch:1988is} and CHARM~II~\cite{Vilain:1994qy} Collaborations 
at CERN and the BNL--734 (CALO) experiment~\cite{Ahrens:1990fp}. 
In addition to these NC results, the CC {\em inverse muon-decay\/} cross-section, 
$\sigma(\nu_\mu e^- \to \nu_e \mu^-)$,
is needed in conjunction with ordinary muon-decay 
to establish the $V-A$ structure of the weak charged current unambiguously.
It is given by Eq.~(\ref{sigmanue}) with the square bracket dropped.

\begin{table}[t]
\begin{center}
\begin{minipage}[t]{16.5 cm}
\caption{Results of the most precise $\nu_\mu e$ and $\bar{\nu}_\mu e$ scattering experiments
and their combinations.  The SM predictions are given by 
$g_{LV}^{\nu_\mu e} = -0.0396$ and $g_{LA}^{\nu_\mu e} = -0.5064$.}
\label{nue}
\end{minipage}
\\ [2mm]
\begin{tabular}{cc|c|c|c|c}
\hline\hline
&&&&& \\ [-2mm]
Group & Ref. & Laboratory & Accelerator & $g_{LV}^{\nu_\mu e}$ & $g_{LA}^{\nu_\mu e}$ \\ [2mm]
\hline
&&&&& \\ [-2mm]
CHARM & \cite{Dorenbosch:1988is} & CERN & SpS & $-0.06 \pm 0.07 \pm 0.02$ & 
$-0.54 \pm 0.04 \pm 0.06$ \\ [2mm]
CHARM~II & \cite{Vilain:1994qy} & CERN & SpS & $-0.035 \pm 0.017$ & $-0.503 \pm 0.017$ 
\\ [2mm]
CALO & \cite{Ahrens:1990fp} & BNL & AGS & $-0.107 \pm 0.035 \pm 0.028$ & 
$-0.514 \pm 0.023 \pm 0.028$ \\ [2mm]
\hline
&&&&& \\ [-2mm]
all &  & both & both & $-0.045 \pm 0.016$ & $-0.507 \pm 0.015$ \\ [2mm]
\hline\hline
\end{tabular}
\end{center}
\end{table}     

For $\nu_e e$ and $\bar\nu_e e$ elastic scattering there is an extra contribution from the weak charged current (CC),
\be\label{Lnuee}
{\cal L}_{\rm NC + CC}^{\nu_e e} = - 2 {\cos^2\theta_W\over v^2} 
\ovl\nu_e \gamma^\rho {g_V^{\nu_e} - g_A^{\nu_e} \gamma^5 \over 2} \nu_e 
\ovl{e} \gamma_\rho {g_V^e - g_A^e \gamma^5 \over 2}  e - {2 \over v^2} 
\ovl\nu_e \gamma^\rho P_L e 
\ovl{e} \gamma_\rho P_L \nu_e,
\ee
which in the SM appears naturally in charge-changing order, but can be brought into 
the charge-retention form of Eq.~(\ref{Lnumue}) by means of a Fierz re-ordering
(in fact, the second term in Eq.~(\ref{Lnuee}) is Fierz invariant),
and one finds at the SM tree-level,
\be\label{gvanuee}
g_{LV}^{\nu_e e} \equiv \cos^2\theta_W g_L^{\nu_e} g_V^e + 1 = {1\over 2} + 2 \sin^2\theta_W, 
\qquad\qquad
g_{LA}^{\nu_e e} \equiv \cos^2\theta_W g_L^{\nu_e} g_A^e + 1= {1\over 2}.
\ee
The $\nu_e e$ process has been measured by the CNTR (LAMPF)~\cite{Allen:1992qe} and 
LSND (LANSCE)~\cite{Auerbach:2001wg} experiments at LANL,
while $\bar\nu_e e$ scattering was studied by the TEXONO Collaboration~\cite{Deniz:2009mu} at 
the Kuo-Sheng Nuclear Power Reactor in Taiwan. 
These experiments are generally less precise than those using $\nu_\mu$ or 
$\bar{\nu}_\mu$-beams, but as shown in Figure~\ref{numue}, they are very useful to reduce 
the four-fold ambiguity from $g_{V,A} \to -g_{V,A}$ and $g_{V,A} \to g_{A,V}$ to only two solutions.
The interference term between the neutral and charged current amplitudes could also be
extracted~\cite{Auerbach:2001wg}.

The tree-level relations~(\ref{gLRnumue}), (\ref{gvanumue}) and (\ref{gvanuee}) 
are subject to the EW radiative corrections discussed in Section~\ref{radcorrnu}.
The remaining QED corrections have been obtained in the ultra-relativistic limit 
in Ref.~\cite{Sarantakos:1982bp} (which also gives corrections to the spectrum), 
and are assumed to be removed beforehand. 

\begin{figure}[t]
\begin{center}
\epsfig{file=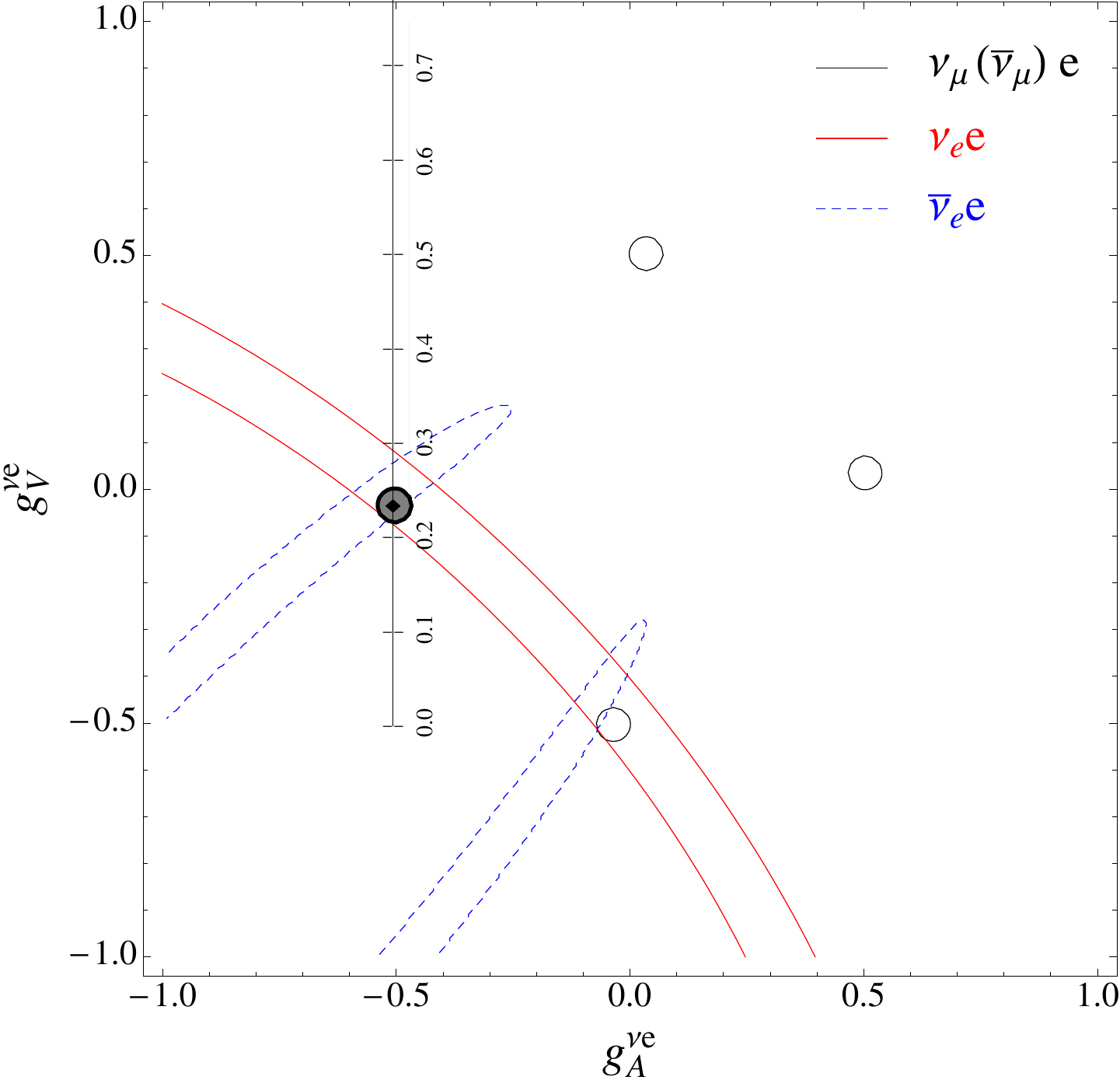,scale=0.7}
\begin{minipage}[t]{16.5 cm}
\caption{Allowed contours in $g_V^{\nu e}\ vs.\ g_A^{\nu e}$ from neutrino-electron scattering and
the SM prediction as a function of the weak mixing angle $\hat s\,^2_Z$
(the SM best fit value $\hat s\,^2_Z = 0.23116$ is also indicated).
The $\nu_e e$~\cite{Allen:1992qe,Auerbach:2001wg} and $\bar{\nu}_e e$~\cite{Deniz:2009mu}
constraints are at  1~$\sigma$ while each of the four equivalent $\nu_\mu (\bar{\nu}_\mu) e$ 
solutions ($g_{V,A} \to -g_{V,A}$ and $g_{V,A} \to g_{A,V}$) are at 90\% CL. 
The global best fit region (shaded) almost exactly coincides with the corresponding 
$\nu_\mu (\bar{\nu}_\mu) e$ region. 
The solution near $g_A= 0$, $g_V = -0.5$ is eliminated by $e^+ e^- \to \ell^+ \ell^-$ data under 
the weak additional assumption that the neutral current is dominated by the exchange of a single 
$Z$ boson.
(Figure reprinted as permitted according to journal guidelines from
\Journal{\PRD}{86}{010001}{2012}, J.~Erler and P.~Langacker~\cite{PDGEW2012}.)
\label{numue}}
\end{minipage}
\end{center}
\end{figure}

\subsection{\it Deep-inelastic neutrino-nucleus scattering and related processes}
\label{nuDIS}

The most direct way to access the $\nu$-quark sector is in the deep-inelastic kinematic 
regime~\cite{Perrier:1996qg,Conrad:1997ne},
where neutrinos scatter in a first approximation incoherently from individual quarks.
The use of heavy nuclei such as iron increases interaction rates compared to 
the leptonic processes covered in the previous section where lower density materials such as glass
are preferred to facilitate better angular resolution of the showers produced by the recoil electrons.
Even more importantly, the cross-sections will be seen 
to be proportional to the nucleon mass rather than $m_e$.

On the other hand, while CC events can be reconstructed from the recoiling muons and the hadronic 
energy, NC events cannot. 
Moreover, there are many complications due to hadronic and nuclear structure effects,
and one resorts to cross-section ratios involving both neutral and charged currents
in which many of the associated uncertainties cancel.
Suppressing family indices, the relevant quark-level effective Lagrangians are given by,
\be\label{LuqCC}
{\cal L}_{\rm CC}^{\nu q} = - {2 \over v^2} \left[ \ovl{e} \gamma^\mu {1 - \gamma_5\over 2} \nu \,
\ovl{u}  \gamma_\mu {1 - \gamma^5 \over 2} V_{\rm CKM}\, d + {\rm H. c.}  \right],
\ee
$$
{\cal L}_{\rm NC}^{\nu q} = - {2 \over v^2} \ovl\nu \gamma^\mu {1 - \gamma_5\over 2} \nu 
\left[ \ovl{u} \gamma_\mu {g_{LV}^{\nu u} - g_{LA}^{\nu u} \gamma^5 \over 2} u +
\ovl{d} \gamma_\mu {g_{LV}^{\nu d} - g_{LA}^{\nu d} \gamma^5 \over 2} d \right] =
$$
\be\label{LnuqNC}
- {2 \over v^2} \ovl{\nu_L} \gamma^\mu \nu_L 
\left[ g_{LL}^{\nu u}  \ovl{u_L} \gamma_\mu u_L + g_{LR}^{\nu u} \ovl{u_R} \gamma_\mu u_R +
        g_{LL}^{\nu d}  \ovl{d_L} \gamma_\mu d_L + g_{LR}^{\nu d} \ovl{u_R} \gamma_\mu d_R \right],
\ee
where the SM relations for the real-valued coefficients in Eq.~(\ref{LnuqNC}) are,
\be\label{epsilonu}
g_{LL}^{\nu u} \equiv {g_{LV}^{\nu u} + g_{LA}^{\nu u} \over 2} = 
\ph{-} {1\over 2} - {2\over 3} \sin^2\theta_W,
\qquad\qquad
g_{LR}^{\nu u} \equiv {g_{LV}^{\nu u} - g_{LA}^{\nu u} \over 2} = - {2\over 3} \sin^2\theta_W,
\ee
\be\label{epsilond}
g_{LL}^{\nu d} \equiv {g_{LV}^{\nu d} + g_{LA}^{\nu d} \over 2} = 
- {1\over 2} + {1\over 3}  \sin^2\theta_W,
\qquad\qquad
g_{LR}^{\nu d} \equiv {g_{LV}^{\nu d} - g_{LA}^{\nu d} \over 2} = \ph{-} {1\over 3} \sin^2\theta_W.
\ee

One can generalize Eq.~(\ref{dsigmanue}) to $\nu$-nucleon scattering by introducing proton PDFs, 
$q(x)$, which are functions of the Bjorken scaling variable ($p_N$ and $m_N$ are the nucleon 
momenta and mass, respectively),
\be
x \equiv {-q^2 \over 2 p_N q} =
{-q^2 \over 2 m_N (E_{\rm had} - m_N)}\ ,
\qquad\qquad
0 \leq x \leq 1.
\ee
One also defines the {\em inelasticity\/} parameter,
\be\label{inelasticity}
y \equiv {p_N q \over p_N p_\nu} = 
{- q^2 \over 2 x p_N p_\nu} = 
1 - {E_\nu' \over E_\nu} =
{E_{\rm had} - m_N \over E_\nu}\ ,
\qquad\qquad
0 \leq y \leq \left( 1 + {x m_N\over 2 E_\nu} \right)^{-1},
\ee
which is to be compared with the parameter defined in Eq.~(\ref{y}).  
One then has for $\nu p$-scattering,
\be \label{dsigmanup}
{d^2 \sigma(\nu p \to \nu X)\over dx dy} = {m_p E_\nu \over \pi v^4} \sum_q x \left\{ 
\left[ (g_{LL}^{\nu q})^2 + (g_{LR}^{\nu q})^2 (1 - y)^2 \right] q(x) +
\left[ (g_{LR}^{\nu q})^2 + (g_{LL}^{\nu q})^2 (1 - y)^2 \right] \bar{q}(x) \right\}.
\ee
The $\bar\nu p$ cross-section is again implied by interchanging $g_{LL}^{\nu q}$ and 
$g_{LR}^{\nu q}$, while the neutron case is obtained from these under 
the assumption of {\em charge symmetry} by exchanging the $u$ and $d$ quark PDFs,
and likewise for the anti-quarks.
Top and bottom quarks can safely be neglected, and when one also ignores the second generation 
quarks, then one can write on average per nucleon, $N$, 
in an isoscalar target~\cite{Llewellyn Smith:1983ie},
$$
{d^2 \sigma(\nu N \to \nu X)\over dx dy} = {m_N E_\nu \over 2 \pi v^4} \left\{ 
\left[ g_L^2 + g_R^2 (1 - y)^2 \right] [x u(x) + x d(x)] + 
\left[ g_R^2 + g_L^2 (1 - y)^2 \right] [x \bar{u}(x) + x \bar{d}(x)] \right\}
$$
\be\label{dsigmanuN}
= g_L^2 {d \sigma(\nu N \to \mu^- X)\over dx dy} 
+ g_R^2 {d \sigma(\bar\nu N \to \mu^+ X)\over dx dy}\ ,
\ee
where it was used that (anti)-neutrinos can only emit $W^+$ ($W^-$) bosons and thus can only be 
absorbed by negatively (positively) charged quarks and anti-quarks.
In Eq.~(\ref{dsigmanuN}) we abbreviated,
\be 
g_L^2 \equiv (g_{LL}^{\nu u})^2 + (g_{LL}^{\nu d})^2,
\qquad\qquad
g_R^2 \equiv (g_{LR}^{\nu u})^2 + (g_{LR}^{\nu d})^2,
\ee
\be 
h_L^2 \equiv (g_{LL}^{\nu u})^2 - (g_{LL}^{\nu d})^2,
\qquad\qquad
h_R^2 \equiv (g_{LR}^{\nu u})^2 - (g_{LR}^{\nu d})^2,
\ee
and with the analogous result for $\bar\nu N$ scattering, 
one arrives at the Llevellyn Smith relations~\cite{Llewellyn Smith:1983ie},
\be\label{LSrelations}
R_\nu \equiv {\sigma(\nu N \to \nu X) \over \sigma(\nu N \to \mu^- X)} = g_L^2 + r g_R^2,
\qquad\qquad
R_{\bar\nu} \equiv {\sigma(\bar\nu N \to \bar\nu X) \over \sigma(\bar\nu N \to \mu^+ X)} = 
{g_R^2 \over r} + g_L^2,
\ee
in terms of the the CC cross-section ratio, $r$, which 
cancels in the Paschos-Wolfenstein ratios, $R_\pm$~\cite{Paschos:1972kj},
\be
r \equiv {\sigma(\bar\nu N \to \mu^+ X) \over \sigma(\nu N \to \mu^- X)}\ ,
\qquad\qquad
R_\pm \equiv {\sigma(\nu N \to \nu X) \pm \sigma(\bar\nu N \to \bar\nu X) 
\over \sigma(\nu N \to \mu^- X) \pm \sigma(\bar\nu N \to \mu^+ X)} = 
{R_\nu \pm r R_{\bar\nu} \over 1 \pm r} = g_L^2 \pm g_R^2.
\ee
In the absence of sea quarks and for an ideal experiment with full acceptance, $r = 1/3$
(\cf Eq.~(\ref{sigmanue})), but $r$ grows almost linearly with the ratio of the fraction of the nucleon's 
momentum carried by anti-quarks to that carried by quarks,
and typically decreases with realistic acceptances. 
In practice, $r \sim 0.4$ is determined experimentally.
The ratio $R_-$ is particularly clean, since any effect shifting $\sigma(\nu N \to \nu X)$ and 
$\sigma(\bar\nu N \to \bar\nu X)$ equally, drops out.  

As for $R_\nu$ and $R_{\bar\nu}$ individually, the $s$ and $c$ quarks cause effects due to 
$V_{\rm CKM} \neq 1$, induce the leading theoretical uncertainty through the non-vanishing charm
quark mass, and require knowledge of the strange sea and its asymmetry, $x s(x) - x \bar{s}(x)$. 
Furthermore. it is necessary to account for the imperfect cancellation of the $M_W$ and $M_Z$ 
propagator effects, as well as for $m_\mu \neq 0$, $m_N \neq 0$, non-isoscalarity due to 
the neutron excess in heavy nuclei, and the $x$ and $Q^2$ dependences of the PDFs.
In the approximation of charge symmetry, higher-order QCD corrections are suppressed by
$\sin^4\theta_W$~\cite{Dobrescu:2003ta}.
Non-perturbative (higher-twist) QCD effects cause, \eg a violation of 
the Callan-Gross relation~\cite{Callan:1969uq} and therefore non-vanishing longitudinal 
structure functions, but their effects on Eqs.~(\ref{LSrelations}) are suppressed for large $Q^2$.
Most difficult to quantify are charge symmetry violations 
(from $Q_d \neq Q_u$~\cite{Gluck:2005xh} and from $m_d \neq m_u$~\cite{Londergan:2003ij}) 
and nuclear effects (including medium modification of the nucleon PDFs~\cite{Cloet:2009qs}). 
EW and QED radiative corrections~\cite{Diener:2003ss,Arbuzov:2004zr}, beyond those treated
in Section~\ref{radcorrnu}, are large and kinematics dependent, and have to be applied by
the experimentalists.

\begin{table}[t]
\begin{center}
\begin{minipage}[t]{16.5 cm}
\caption{Results of the most precise isoscalar $\nu$DIS experiments as quoted in the original 
publications and the SM predictions adjusted to the applicable $Q^2$ values as explained in
Section~\ref{radcorrnu}.
The CCFR experiment was mostly sensitive to $R_\nu$ and quotes the combination
$\kappa = 1.7897\, g_L^2 + 1.1479\, g_R^2 - 0.0916\, h_L^2 - 0.0782\, h_R^2$.
In all cases, additional theory corrections and uncertainties may have to be applied as discussed in 
the text.}
\label{numuDIS}
\end{minipage}
\\ [2mm]
\begin{tabular}{cc|c|c|c|c|c|c}
\hline\hline
&&&&&& \\ [-2mm]
Group & Ref. & Laboratory & Accelerator & Target & Quantity & Measurement & SM \\ [2mm]
\hline
&&&&&& \\ [-2mm]
CHARM & \cite{Allaby:1987vr} & CERN & SpS & CaCO$_3$ & $R_\nu$ & 
$0.3093 \pm 0.0031$ & 0.3156 \\ [2mm]
CDHS & \cite{Blondel:1989ev} & CERN & SpS & Fe & $R_\nu$ & 
$0.3072 \pm 0.0033$ & 0.3091 \\ [2mm]
CCFR & \cite{McFarland:1997wx} & FNAL & Tevatron & BeO & $\kappa$ & $0.5820 \pm 0.0041$ &
0.5830 \\ [2mm] 
NuTeV & \cite{Zeller:2001hh} & FNAL & Tevatron & BeO & $g_L^2$ & $0.30005 \pm 0.00137$ &
0.3039 \\ [2mm]
NuTeV & \cite{Zeller:2001hh} & FNAL & Tevatron & BeO & $g_R^2$ & $0.03076 \pm 0.00110$ & 
0.0300 \\ [2mm]
\hline\hline
\end{tabular}
\end{center}
\end{table}     

The results of the most precise $\nu$DIS experiments on isoscalar targets are summarized in
Table~\ref{numuDIS}. 
Note, that there is a 2.7~$\sigma$ deviation in the NuTeV determination of $g_L^2$. 
This corresponds to the result of the original NuTeV publication~\cite{Zeller:2001hh}.
Corrections of the kind discussed in the previous paragraph that go beyond those already 
considered in the original analysis have not been applied in Table~\ref{numuDIS}.

Determinations of the isovector combinations are more difficult since the experiments are
harder to interpret theoretically.
They can be constrained by using $\nu$DIS from non-isoscalar targets 
(see \eg Ref.~\cite{Amaldi:1987fu}), but there is more model dependence.
 
Another possibility is in principle the elastic scattering of neutrinos from protons~\cite{Mann:1996qh}.
At very low $Q^2$ these experiments are sensitive to the combination,
$(g_{LV}^{\nu p})^2 + g_A^2 (g_{LA}^{\nu p})^2 (1 + \Delta S_s)^2$, where
\be
g_{LV}^{\nu p} \equiv 2 g_{LV}^{\nu u} + g_{LV}^{\nu d} = {1\over 2} - 2 \sin^2\theta_W,
\qquad\qquad
g_{LA}^{\nu p} \equiv 2 g_{LA}^{\nu u} + g_{LA}^{\nu d} = {1\over 2},
\ee
and where $g_A = 1.27$ is the axial-vector coupling constant as measured in neutron decays
(the right hand sides show the SM tree-level).
$\Delta S_s$ is the strange quark contribution to the nucleon spin which is poorly known. 
One can try to separate $(g_{LV}^{\nu p})^2$ from the axial contributions kinematically 
by exploiting the different behavior at $Q^2 \neq 0$.
Unfortunately, this induces further parameters, such as the nucleon axial dipole mass, $M_A$,
and the strange quark electric and magnetic form factors, even though lattice 
calculations~\cite{Leinweber:2004tc,Leinweber:2006ug} indicate that the latter are rather 
small\footnote{This is consistent with the experimental results of the HAPPEX Collaboration at 
JLab~\cite{Ahmed:2011vp}, but since it was necessary to assume the SM values for the EW 
couplings their analysis cannot be used directly for model independent EW fits.
Similar remarks apply to the results by the PVA4 Collaboration at Mainz~\cite{Baunack:2009gy} and
the G\O\ Collaboration at JLab~\cite{Androic:2009aa}.
For recent reviews on parity violation in elastic electron-nucleon scattering with focus on 
the strangeness content of the nucleon, including more comprehensive lists of references,
see Refs.~\cite{GonzalezJimenez:2011fq,Armstrong:2012bi}.}.
At this level, one also faces the neutron combinations,
\be
g_{LV}^{\nu n} \equiv g_{LV}^{\nu u} + 2 g_{LV}^{\nu d} = - {1\over 2},
\qquad\qquad
g_{LA}^{\nu n} \equiv g_{LA}^{\nu u} + 2 g_{LA}^{\nu d} = - {1\over 2},
\ee
where $g_{LV}^{\nu n}$ enters proportional to the neutron magnetic moment.
$g_{LA}^{\nu n}$ multiplies the CP-violating electric dipole moment operator and can safely be 
ignored, but the induced pseudoscalar form factor, $G_P$ (which is often neglected), 
survives even the additional assumption of vanishing second class currents~\cite{Weinberg:1958ut}.
This separation also needs full kinematic information, ideally on an event-by-event basis.
Ignoring strange quark complications, one could reinterpret the combined value of $\sin^2\theta_W$
as extracted by the two most precise experiments~\cite{Horstkotte:1981ne,Ahrens:1986xe} as 
the constraint, $(g_{LV}^{\nu p})^2 + g_A^2 (g_{LA}^{\nu p})^2 = 0.4 \pm 0.1$, 
but this is only to illustrate the sensitivity. 
A global analysis including differential cross-sections from $\nu$, $\bar\nu$ and $e^-$ elastic and 
quasi-elastic scattering from both protons and neutrons, and NC and CC channels, may be in order.
A very promising future opportunity may be provided by so-called 
$\beta$-beams~\cite{Zucchelli:2002sa}, in which relatively low-energy radioactive nuclei serve as 
primary neutrino beams.
These would allow much better control of the neutrino spectra and one may be able to separate
the various form factors kinematically~\cite{Antonelli:2007eb}.

The isovector axial-vector combination, 
$\beta \equiv g_{LA}^{\nu u} - g_{LA}^{\nu d} = g_{LA}^{\nu p} - g_{LA}^{\nu n} = 1$ (at the SM 
tree-level), can be accessed in neutrino induced coherent neutral pion production from nuclei.
In the most recent experiment, NOMAD~\cite{Kullenberg:2009pu} normalized its data to 
the inclusive CC cross-section, $\sigma(\nu_\mu A \to \mu^- X)$, which was then taken from their 
earlier measurement~\cite{Wu:2007ab}.
Comparing the resulting cross-section, $\sigma(\nu_\mu A \to \nu_\mu A \pi^0)$, to the prediction 
(based on the PCAC hypothesis) from Ref.~\cite{Rein:1982pf}, we find $\beta^2 = 0.93 \pm 0.14$.
The SKAT Collaboration~\cite{Grabosch:1985mt} relied only on isospin symmetry when
using the NC to CC cross-section ratio,
\be
\beta^2 = 2 |V_{ud}|^2 
{\sigma(\nu_\mu A \to \nu_\mu A \pi^0) \over \sigma(\nu_\mu A \to \mu^- A \pi^+)} = 0.93 \pm 0.37\ .
\ee
Similarly, the CHARM Collaboration~\cite{Bergsma:1985qy} extracted both 
$\sigma(\nu_\mu A \to \nu_\mu A \pi^0)$ and $\sigma(\bar\nu_\mu A \to \bar\nu_\mu A \pi^0)$,
and later measured~\cite{Vilain:1993sf} $\sigma(\nu_\mu A \to \mu^- A \pi^+)$ and
$\sigma(\bar\nu_\mu A \to \mu^+ A \pi^-)$, which we use to derive $\beta^2 = 1.08 \pm 0.54$ and 
$\beta^2 = 0.93 \pm 0.38$, respectively. 
Disregarding some older experiments which relied on the same model~\cite{Rein:1982pf} as 
NOMAD, we combine these results to obtain $\beta^2 = 0.94 \pm 0.12$.

\subsection{\it Radiative corrections}
\label{radcorrnu}

The leptonic and semileptonic neutrino scattering processes discussed in Sections~\ref{sec:nue} 
and~\ref{nuDIS} are modified by radiative corrections, which in general depend on energies, 
experimental cuts, \etc.  
This is also true of the $Z$-pole observables mentioned in Section~\ref{WZ}.
For those it is conventional to divide the radiative corrections into two classes~\cite{PDGEW2012}:
one of them consists of QED graphs involving the emission of real photons combined with 
certain diagrams of virtual photons in loops to form finite and gauge-invariant sets.
Photon exchange diagrams, except for vacuum polarization effects, also belong to this class
which then needs to be calculated and removed individually for each experiment.
On the other hand, purely EW diagrams and associated photonic and gluonic corrections enter EW
parameters such as the $\bar{g}_V^f$ and $\bar{g}_A^f$ in Section~\ref{sec:seff}, and are
considered part of so-called {\em pseudo-observables} containing the interesting physics.
Deviating from these rules, final-state QED and QCD effects (but not initial-final state interference) 
contributing to the partial and total decay widths of the $W$ and $Z$ bosons are kept.

We now propose a similar strategy where purely EW diagrams and certain photonic loops and 
$\gamma$-exchange graphs are absorbed in the definitions of the low-energy (pseudo-observable)
EW couplings, appearing, \eg in Eqs.~(\ref{Lnumue}) and (\ref{LnuqNC}). 
The remaining corrections are assumed to be applied individually for each experiment. 
However, while the $Z$-pole pseudo-observables are naturally defined at the scale $\mu = M_Z$, 
the effective four-Fermi couplings are obtained at momentum transfers $Q^2 \ll M_Z^2$ (even for
experiments in the deep-inelastic regime) and the genuine EW radiative corrections will in general
depend on the specific kinematic points or ranges at which the low-energy experiments are 
performed.
Thus, one needs to introduce idealized EW coupling parameters defined at some common reference 
scale $\mu$ (we choose $\mu = 0$), and have the experimental collaborations correct for effects due 
to $Q^2 \neq 0$.
The effective NC couplings are modified by the following radiative corrections:
\begin{description}
\item[\em W and Z boson self-energies:]
Fermion and $W$ boson bubbles inserted into the $Z$ boson propagator are to be evaluated at 
$Q^2 = 0$ and subtracted from the analogous corrections to the $W$ propagator. 
This results in the most important contribution to the universal low-energy $\rho$ parameter defined
in the second Eq.~(\ref{rhohat}).
$\rho$ also contains vertex and $WZ$ box\footnote{The $W\gamma$ box has been removed as it is 
part of the traditional QED correction to $\mu$-decay within the $V-A$ theory~\cite{Sirlin:1980nh}.} 
corrections to the $\mu$-lifetime from which $G_F$ in Eq.~(\ref{GF}) was obtained and in terms of 
which the low-energy Lagrangians are normalized.

\item[\em $\gamma$-Z mixing:]
Vacuum polarization diagrams of $\gamma$-$Z$ mixing type give rise to 
a scale-dependence~\cite{Czarnecki:2000ic} of the weak mixing angle (see Figure~\ref{s2w}).
In the perturbative QCD domain, $\mu \gtrsim 1$~GeV, these effects can be 
re-summed~\cite{Erler:2004in}, while the non-perturbative region introduces a hadronic uncertainty 
which is, however, small compared to the current and foreseeable experimental 
precision~\cite{Kumar:2013yoa,Erler:2004in}.
Here we employ the weak mixing angle at the reference scale $\mu = 0$ and abbreviate, 
$\hat s_0^2 \equiv \sin^2\theta_W(0)$.

\item[\em Neutrino charge radius:]
The charge radius of the $\ell$-neutrino,  $\diameter_{\nu_\ell W}$,  is generated by a loop insertion 
into the $\nu_\ell$-line consisting of a $W$ boson and the charged lepton $\ell$ 
(in most cases $\nu_\ell = \nu_\mu$).
Attaching the photon to the $W$ boson produces a simple universal correction.
By contrast, attaching it to the lepton $\ell$ generates a large EW logarithm,
which is regulated at $m_\ell$.
This allows photon exchange diagrams connecting $\diameter_{\nu_\ell W}$ with the target fermion.

\item[\em WW and ZZ box diagrams:]
The left-handed couplings $g_{LL}^{\nu e}$ and $g_{LL}^{\nu d}$ receive a contribution from
the $WW$ box, $\Box_{WW}$, and $g_{LL}^{\nu u}$ from the $WW$ crossed-box,
$\hspace{3pt}{\raisebox{-1pt}{\rotatebox{90}{\Bowtie}}}_{WW}$.
In addition, the effective couplings of either chirality, $g_{LX}^{\nu f}$, get a correction term 
proportional to $(g_{LX}^{\nu f})^2$ from the sum of both types of $ZZ$ box diagrams, 
$\boxtimes_{ZZ}$, in which $(g_{LX}^{\nu f})^2$ is evaluated at lowest order, but replacing 
$\hat s_0^2$ by $\hat s_Z^2$.
\end{description} 
Collecting these corrections~\cite{Marciano:1980pb} one obtains,
\be\label{gLLnuu} 
g_{LL}^{\nu_\ell f} = \rho \left[ {1 \over 2} - Q_f \hat s_0^2 + \boxtimes_{ZZ} \right] - 
Q_f \diameter_{\nu_\ell W} + \hspace{3pt}{\raisebox{-1pt}{\rotatebox{90}{\Bowtie}}}_{WW}
\qquad\qquad (f = u),
\ee
\be\label{gLLnud} 
g_{LL}^{\nu_\ell f} = \rho \left[ - {1 \over 2} - Q_f \hat s_0^2 + \boxtimes_{ZZ} \right] - 
Q_f \diameter_{\nu_\ell W} +  \Box_{WW} 
\qquad\qquad (f = d,e),
\ee
\be\label{gLRnu}
g_{LR}^{\nu_\ell f} = - \rho \left[ Q_f \hat s_0^2 + \boxtimes_{ZZ} \right] - Q_f \diameter_{\nu_\ell W} 
\qquad\qquad (f = u,d,e),
\ee
where 
\be
\diameter_{\nu_\ell W} = - {\alpha \over 6 \pi} \left( \ln {M_W^2 \over m_\ell^2} + {3 \over 2} \right),
\ee
\be\label{WWbox}
\Box_{WW} = - {\hat\alpha_Z \over 2 \pi \hat s_Z^2}\left[ 1 - {\hat\alpha_s(M_W) \over 2 \pi} \right], 
\qquad\qquad
\hspace{3pt}{\raisebox{-1pt}{\rotatebox{90}{\Bowtie}}}_{WW} = 
{\hat\alpha_Z \over 8 \pi \hat s_Z^2} \left[ 1 + {\hat\alpha_s(M_W) \over \pi} \right], 
\ee
\be
\boxtimes_{ZZ} = - {3 \hat\alpha_Z \over 8 \pi \hat s_Z^2 \hat c_Z^2} (g_{LX}^{\nu_\ell f})^2
\left[ 1 - {\hat\alpha_s(M_Z) \over\pi} \right].
\ee
For the numerical SM evaluation we assume $M_H = 125.5$~GeV (see Section~\ref{MH}) yielding,
\be\label{inputs}
\rho = 1.00064, \qquad\qquad
\hat{s}^2_0 = 0.23865, \qquad\qquad
\hat{s}^2_Z = 0.23126, \qquad\qquad
\hat\alpha_Z^{-1} = 127.94,
\ee
and with that we find the SM values in Table~\ref{gnu}. 
We also give the results for the combinations, 
\be
g_L^2 = 0.3034, \qquad\qquad
h_L^2 = - 0.0643, \qquad\qquad
g_R^2 = 0.0302, \qquad\qquad
h_R^2 = 0.0181,
\ee
\be
\tan \theta_L \equiv {g_{LL}^{\nu_\mu u} \over g_{LL}^{\nu_\mu d}} = - 0.8062, 
\qquad\qquad
\tan \theta_R \equiv {g_{LR}^{\nu_\mu u} \over g_{LR}^{\nu_\mu d}} =
-2 +  {3\, \hat\alpha_Z\over 4 \pi \hat{c}^2_Z} \left( 1 - {\hat\alpha_s\over\pi} \right) = - 1.9977,
\ee
where the explicit expression for $\tan \theta_R$ is exact to one-loop order,
with $\tan \theta_R \neq -2 $ entirely due to the $ZZ$ box diagrams.
Similarly, $\beta^2 = 1.0151$, where $\beta^2 \neq 1$ is mostly due to the $WW$ box diagrams.
The parameters $\theta_{L,R}$ are useful because together with $g_{L,R}^2$ they form a parameter 
set with small correlations when it is extracted from the current data, 
while the $g_{LL}^{\nu_\mu q}$ and $g_{LR}^{\nu_\mu q}$ have non-Gaussian errors. 

\begin{table}[t]
\begin{center}
\begin{minipage}[t]{16.5 cm}
\caption{SM values of the one-loop and leading two-loop corrected 
effective NC $\nu$ couplings for the charged SM fermions and the nucleons.
Note, that the small $g_{LR}^{\nu_\mu n}$ arises solely from the $ZZ$ box diagrams.}
\label{gnu}
\end{minipage}
\\ [2mm]
\begin{tabular}{c|c|c|c|c|c}
\hline\hline
&&&& \\ [-2mm]
$f$ & $e$ & $u$ & $d$ & $p$ & $n$ \\ [2mm]
\hline
&&&& \\ [-2mm]
$g_{LL}^{\nu_\mu f}$ & $-0.2730$ & $\ph-$0.3457 & $-0.4288$ & $\ph-$0.2626 & $-0.5119$ \\ [2mm]
$g_{LR}^{\nu_\mu f}$&$\ph-$0.2334&$-0.1553$& $\ph-$0.0777 & $-0.2328$ & $\ph-$0.0002\\ [2mm]
$g_{LV}^{\nu_\mu f}$ & $-0.0396$ & $\ph-$0.1905 & $-0.3511$ & $\ph-$0.0298 & $-0.5117$ \\ [2mm]
$g_{LA}^{\nu_\mu f}$ & $-0.5064$ & $\ph-$0.5010 & $-0.5065$ & $\ph-$0.4955 & $-0.5121$ \\ [2mm]
\hline\hline
\end{tabular}
\end{center}
\end{table}

Under realistic conditions one has to correct the $g_{LV}^{\nu f}$ for $Q^2 \neq 0$
(the $g_{LA}^{\nu f}$ are $Q^2$-independent).
In the case of $\nu_\mu$-$e$ scattering with energies $E_{\nu_\mu} \lesssim 100$~GeV, the 
resulting values of $Q^2 \lesssim m_\mu^2$ hardly affect $\diameter_{\nu_\mu W}$, and the 
correction to $\hat{s}^2$ is suppressed by a factor $(1 - 4 \hat{s}_0^2) \ll 1$, so that the residual 
$Q^2$-correction is two orders of magnitude below the experimental uncertainty in Table~\ref{nue}.  
The average $Q^2$ of most $\nu$DIS experiments is around the bottom quark and $\tau$ lepton 
thresholds so that we can write,
\be\label{q2nu}
g_{LV}^{\nu_\mu q} \to g_{LV}^{\nu_\mu q} - 2 Q_q \left[ \hat{s}_Q^2 - \hat{s}_0^2 + 
{\alpha \over 6 \pi} \left( \ln {Q^2 \over m_\mu^2} + 5 - {160 \over 9}\hat{s}_0^2 + 
\Delta_Q^{\nu_\mu q} \right) \right]
\approx g_{LV}^{\nu_\mu q} + 0.0005\, Q_q \ln {Q^2 \over 2.2 \mbox{ GeV}^2}\ ,
\ee
where $\hat{s}_Q^2 \equiv \sin^2\hat\theta_W(Q^2)$, and where significant cancellations occur 
between the $Q^2$-variations of $\sin^2\hat\theta_W$ and $\diameter_{\nu_\mu W}$.
The small $Q^2$-dependent correction term $\Delta_Q^{\nu_\mu q}$ accounts for non-decoupling 
bottom and $\tau$ effects, the non-vanishing masses of the lighter fermions, 
and some reducible higher-order contributions.

\section{Parity Violation}
\label{PAVI}
\subsection{\it Parity-violating M\o ller scattering}
\label{qweake}

The parity-violating part of the electron-electron interaction is to leading order a purely weak 
NC process.  From the second Eq.~(\ref{Leff}) one finds,
\be\label{Lee}
{\cal L}_{\rm NC}^{e e} = - {\cos^2\theta_W \over v^2} 
\ovl e \gamma^\mu {g_V^e - g_A^e \gamma^5 \over 2} e\,
\ovl{e} \gamma_\mu {g_V^e - g_A^e \gamma^5 \over 2} e = 
- {1 \over v^2} \ovl e \gamma^\mu {g_{VV}^{\, ee} e \ovl{e} 
- g_{AA}^{\, ee} \gamma^5 e \ovl{e} \gamma^5 
+ 2 g_{VA}^{\, ee} e \ovl{e} \gamma^5 \over 4} \gamma_\mu e,
\ee
where the SM tree-level relations for the coefficients multiplying parity-conserving 
four-Fermi operators are,
\be
g_{VV}^{\, ee} \equiv (\cos\theta_W g_V^e)^2 = {(1 - 4 \sin^2\theta_W)^2 \over 2}\ , \qquad\qquad
g_{AA}^{\, ee} \equiv (\cos\theta_W g_A^e)^2 = {1 \over 2}\ ,
\ee
while for parity-violating processes one has,
\be
g_{VA}^{\, ee} = \cos^2\theta_W g_V^e g_A^e  \equiv 
\cos^2\theta_W ({g_L^e}^2 - {g_R^e}^2) = {1\over 2} - 2 \sin^2\theta_W.
\ee
$g_{VA}^{\, ee}$ can be measured in fixed target polarized M\o ller scattering, 
${\vec{e}}^{\, -} e^- \rightarrow e^- e^-$, by observing the left-right cross-section 
asymmetry~\cite{Derman:1979zc}, $A_{LR}^{ee}$, which --- to an excellent approximation --- 
reduces to the interference term of the parity-violating part of the EW amplitude 
with the parity-conserving QED amplitude.
For large incident electron energies, $E_e \gg m_e$, it can be written as,
\be\label{ALRee}
A_{LR}^{ee} \equiv
\frac{d\sigma_L - d\sigma_R}{d\sigma_L + d\sigma_R} =
2 {s \over M_Z^2} {{g_L^e}^2 - {g_R^e}^2 \over (2 Q_A^e)^2} {\cal F}^{ee} =
{2 m_e E_e \over v^2} {g_{VA}^{\, ee} \over 4 \pi \alpha} {\cal F}^{ee} \approx
{s \over M_Z^2} \left[ {1\over \sin^2 2\theta_W} - {1\over \cos^2\theta_W} \right] {{\cal F}^{ee} \over 2},
\ee
where $s = 2 m_e E_e$ is the square of the center-of-mass energy, 
$y$ is again the relative energy transfer, and
\be
{\cal F}^{ee} \equiv {\cal F}^{ee} (Q^2,y) = 
{2 y (1 - y) \over (1 - y + y^2)^2} {\cal F}^{ee}_{\rm QED} (Q^2,y) =
{4 y (1 - y) \over 1 + y^4 + (1 - y)^4} {\cal F}^{ee}_{\rm QED} (Q^2,y).
\ee
The QED radiative correction factor, ${\cal F}_{\rm QED} (Q^2,y)$, includes kinematically weighted 
hard initial and final state radiation effects, the $\gamma\gamma$ box graphs, and 
the non-logarithmic contributions from the charge radius and the $\gamma Z$ box 
diagrams~\cite{Zykunov:2004nk,Zykunov:2005md}.
$A_{LR}^{ee}$ has been obtained at $y = 0.6$ and at low $Q^2 = y s = 0.026$~GeV$^2$ in the 
SLAC--E158 experiment~\cite{Anthony:2005pm} using the 89\% polarized $e^-$ beam of the SLC, 
with the result,
\be
A_{LR}^{ee} = (1.31 \pm 0.14_{\rm \, stat.} \pm 0.10_{\rm \, syst.}) \times 10^{-7}.
\ee
For the conditions\footnote{The numerical values for $y$ and $Q^2$ are effective
quantities defined to match the analyzing power obtained from a Monte Carlo simulation which 
accounts for energy losses in the target, \etc, and are not identical to the nominal ones.
Note, that even though the analyzing power is proportional to 
$\alpha^{-1}(Q^2)$~\cite{Anthony:2005pm} we used instead $\alpha^{-1}$ in Eq.~(\ref{ALRee}) 
because the $Q^2$-dependence due to lepton loops cancels exactly between $\sin^2\theta_W$ in 
$g_{VA}^{\, ee}$ and $\alpha$, while hadronic loops are negligible~\cite{Czarnecki:1995fw}.} of the 
experiment at the SLC one obtains ${\cal F}_{\rm QED} (0.026 \mbox{ GeV}^2,0.6) = 1.01 \pm 0.01$ 
and $F^{ee} (0.026 \mbox{ GeV}^2,0.6) \approx 0.84$, and anticipating the first 
correction~(\ref{MOLLERcorr}) in Section~\ref{radcorre} we can extract, 
\be\label{gVAee}
g_{VA}^{\, ee} = 0.0190 \pm 0.0027,
\ee
which is 1.3~$\sigma$ below the SM prediction of $g_{VA}^{\, ee} = 0.0225$.
Expressed in terms of the weak mixing angle in the \msbar-scheme, Eq.~(\ref{gVAee}) yields,
\be
\hat s^2 (0.16 \mbox{ GeV}) = 0.2403 \pm 0.0013,
\ee
and establishes the scale dependence of the weak mixing angle (see Figure~\ref{s2w}) at the level 
of 6.4~standard deviations. 
The implications for physics beyond the SM are discussed in 
Refs.~\cite{Kumar:2013yoa,Czarnecki:2000ic,Erler:2003yk}.

A new and five times more precise experiment~\cite{Mammei:2012ph} of this type is planned at 
Jefferson Laboratory (JLab) at the 11~GeV upgraded CEBAF. 
The kinematics will be at $y \approx 0.57$ and $Q^2 \approx 0.0056$~GeV$^2$, giving an even 
smaller asymmetry.
As shown in Figure~\ref{KK}, this experiment will provide one of the most precise determinations of
$\sin^2\theta_W$ and the most precise one off the $Z$ pole, 
and will have important implications for the indirect determination of $M_H$.

\begin{figure}[t]
\begin{center}
\epsfig{file=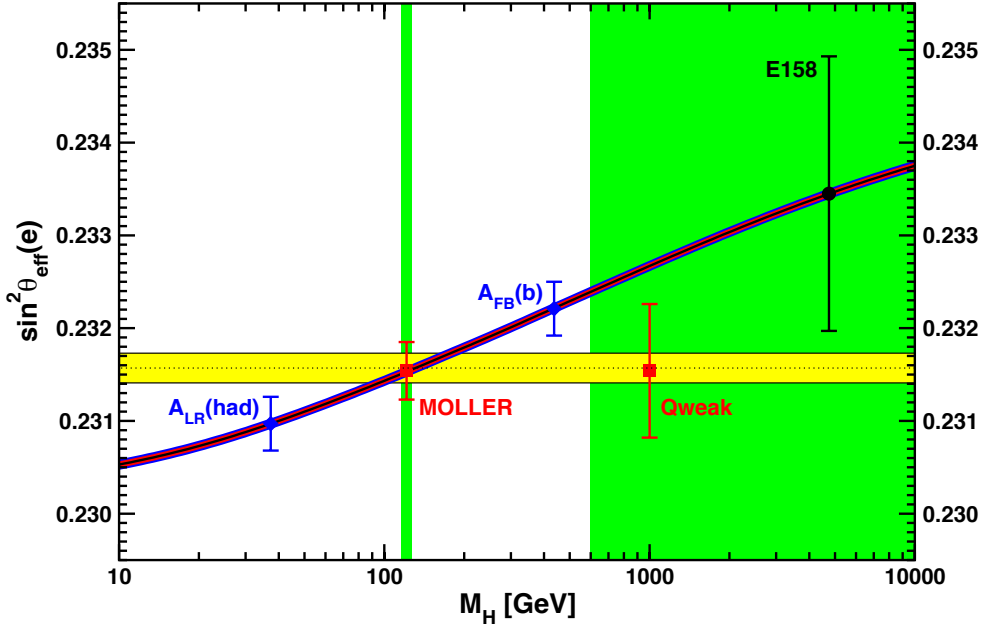,scale=0.49}
\begin{minipage}[t]{16.5 cm}
\caption{High precision measurements of $\sin^2\theta^\ell_{\rm eff}$.  
Shown in blue are the most precise $Z$-pole values and in black the presently most precise
low-energy determination~\cite{Anthony:2005pm}.
Projections for two of the future experiments are shown in red.  
See the body of the text for more details.
\label{KK}}
\end{minipage}
\end{center}
\end{figure}

\subsection{\it Parity non-conservation in atoms and ions}
\label{APV}

The parity-violating part of the NC electron-quark interactions is described by the Lagrangian,
\be\label{LeqNC}
{\cal L}_{\rm NC}^{\, e q} = - {2 \over v^2} {\ovl e \gamma^5\gamma^\mu e \over 2}
\left[ g_{AV}^{\, e u} {\ovl{u} \gamma_\mu u \over 2} + 
g_{AV}^{\, e d} {\ovl d \gamma_\mu d \over 2} \right] - 
{2 \over v^2} {\ovl e \gamma^\mu e \over 2} 
\left[ g_{VA}^{\, e u} {\ovl{u} \gamma^5 \gamma_\mu u \over 2} + 
g_{VA}^{\, e d} {\ovl d \gamma^5 \gamma_\mu d \over 2} \right],
\ee
where the SM tree-level relations for the real-valued coefficients $g_{AV}^{\, e q}$ and 
$g_{VA}^{\, e q}$,
\be\label{gAVeud}
g_{AV}^{\, e u} \equiv \cos^2\theta_W g_A^e g_V^u = - {1\over 2} + {4\over 3} \sin^2\theta_W,
\qquad\qquad
g_{AV}^{\, e d} \equiv \cos^2\theta_W g_A^e g_V^d = {1\over 2} - {2\over 3}  \sin^2\theta_W,
\ee
\be\label{gVAeud}
g_{VA}^{\, e u}  \equiv \cos^2\theta_W g_V^e g_A^u = - {1\over 2} + 2 \sin^2\theta_W,
\qquad\qquad
g_{VA}^{\, e d}  \equiv \cos^2\theta_W g_V^e g_A^d ={1\over 2} - 2 \sin^2\theta_W,
\ee
coincide with the similarly defined effective couplings, $C_{1q}$ and $C_{2q}$, respectively.
These interactions induce extremely small parity-violating effects in atomic 
physics~\cite{Bouchiat:1986gc,Masterson:1996qi} which grow roughly with the third power of atomic 
number, $Z$, and atomic parity violation (APV) has been observed only in heavy atoms, such as
cesium~\cite{Wood:1997zq,Guena:2004sq} and thallium~\cite{Edwards:1995zz,Vetter:1995vf}. 
In order to interpret these effects in terms of the interactions~(\ref{LeqNC}) one also needs a good
understanding of atomic structure~\cite{Blundell:1996qj,Ginges:2003qt}. 
This has been achieved for $^{133}$Cs~\cite{Porsev:2009pr,Dzuba:2012kx}, 
the nucleus where also the greatest experimental precision was obtained~\cite{Wood:1997zq}.
The effective couplings in Eq.~(\ref{gAVeud}) add up coherently across the nucleus and give rise to 
the nuclear spin-independent interaction. 
It can be isolated from the much smaller spin-dependent interaction by measuring different hyperfine
transitions.
However, the couplings in Eq.~(\ref{gVAeud}) are clouded by the nuclear anapole 
moment~\cite{Zeldovich:1958,Haxton:2001ay}, 
which grows as $Z^{2/3}$ and dominates in heavy nuclei.

The EW physics is contained in the nuclear weak charges which are defined by,
\be\label{qwzn}
Q_W^{Z,N} \equiv
- 2 \left[ Z (g_{AV}^{\, e p} + 0.00005) + N (g_{AV}^{\, e n} + 0.00006) \right]
\left( 1 - {\alpha\over 2 \pi} \right) \approx 
Z (1 - 4 \sin^2 \theta_W) - N,
\ee
where $N$ is the number of neutrons in the nucleus,
and where the nucleon couplings are given by,
\be
g_{AV}^{\, e p} \equiv 2 g_{AV}^{\, e u} + g_{AV}^{\, e d} = - {1\over 2} + 2 \sin^2\theta_W,
\qquad\qquad
g_{AV}^{\, e n} \equiv g_{AV}^{\, e u} + 2 g_{AV}^{\, e d} = {1\over 2}.
\ee
The small numerical adjustments in Eq.~(\ref{qwzn}) are discussed in Section~\ref{radcorre}.
{\it E.g.}, the weak charge of $^{133}$Cs, $Q_W ^{55,78}$, is extracted by measuring experimentally 
the ratio of the parity-violating amplitude, $E_{\rm PNC}$, to the Stark vector transition polarizability,
$\beta$, and by calculating $E_{\rm PNC}$ theoretically in terms of $Q_W ^{55,78}$,
\be\label{QW}
Q_W^{Z,N} = N \left( {{\rm Im}\, E_{\rm PNC}\over\beta} \right)_{\rm exp.} 
\left( {Q_W^{Z,N} \over N\, {\rm Im}\, E_{\rm PNC}} \right)_{\rm th.} \beta_{\rm exp.+th.}
\approx 
2 (N - Z) \left[ {1\over 2} + {2 Z \over N - Z} \sin^2\theta_W \right].
\ee
Notice, the reduced sensitivity to $\sin^2\theta_W$.
The ratio of the off-diagonal hyperfine amplitude to the polarizability was measured directly by the 
Boulder group~\cite{Bennett:1999pd}.
Combined with the precisely known hyperfine amplitude~\cite{Bouchiat:1988} one extracts the 
value $\beta = (26.991 \pm 0.046) a_B^3$, where $a_B$ is the Bohr radius.
A recent state-of-the-art many body calculation~\cite{Porsev:2009pr} yields, 
\be\label{EPNC}
{\rm Im}\, E_{\rm PNC} = (0.8906 \pm 0.0026) \times 10^{-11} |e|\, a_B {Q_W^{Z,N} \over N}\ ,
\ee
while the two measurements~\cite{Wood:1997zq,Guena:2004sq} combine to give 
${\rm Im}\, E_{\rm PNC}/\beta = - (1.5924 \pm 0.0055)$~mV/cm, or if $\beta$ is gven in atomic units 
as is adequate for Eq.~(\ref{QW}), 
${\rm Im}\, E_{\rm PNC}/\beta = - (3.0967 \pm 0.0107) \times 10^{-13} |e|/a_B^2$.
We finally obtain $Q_W ^{55,78} = - 73.20 \pm 0.35$, and by virtue of Eq.~(\ref{qwzn}),
\be\label{cs}
55 g_{AV}^{\, e p} + 78 g_{AV}^{\, e n} = 36.64 \pm 0.18,
\ee
in excellent agreement with the SM prediction, $55 g_{AV}^{\, e p} + 78 g_{AV}^{\, e n} = 36.66$.
However, a very recent atomic structure calculation~\cite{Dzuba:2012kx} found significant
corrections to two non-dominating terms, shifting the numerical coefficient in Eq.~(\ref{EPNC}) 
to $(0.8977 \pm 0.0040) \times 10^{-11}$, and yielding in place of Eq.~(\ref{cs}) the constraint,
$55 g_{AV}^{\, e p} + 78 g_{AV}^{\, e n} = 36.35 \pm 0.21$ ($Q_W ^{55,78} = - 72.62 \pm 0.43$),
a $1.5~\sigma$ SM deviation~\cite{Dzuba:2012kx}.

The theoretical uncertainties are 3\% for thallium~\cite{Dzuba:1987px} but larger for other atoms. 
In the future it could be possible to reduce the theoretical wave function uncertainties by 
taking ratios of parity violation in different isotopes~\cite{Bouchiat:1986gc,Rosner:1995aj}.
There would still be some residual uncertainties~\cite{Pollock:1992mv,Chen:1993fw} from 
differences in the {\em neutron skin\/} (the excess of the root-mean-square radii of the neutron over 
the proton distributions), however. 
This is because the atomic wave function for $s$-states is maximal at the origin so that a broader
neutron distribution results in a smaller overall effect (the neutron weak charge 
dominates over that of the proton).
Incidentally, the neutron skin may also affect APV in single isotopes~\cite{Brown:2008ib}.
It has recently been observed in polarized electron scattering from $^{208}$Pb by the
PREX Collaboration~\cite{Abrahamyan:2012gp}.
Note also, that unlike single isotopes, the isotope ratios constrain mostly new physics contributions
to $g_{AV}^{\, e p}$~\cite{RamseyMusolf:1999qk}.
Experiments in hydrogen and deuterium are another possibility for reducing the atomic theory 
uncertainties~\cite{Dunford:2007df}, while measurements of trapped radioactive 
atoms~\cite{Behr:2008at} (most notably francium) and single trapped radium ions are 
promising~\cite{Wansbeek:2008} because of the much larger parity-violating effects.

\subsection{\it Parity-violating deep-inelastic scattering and related processes}
\label{PVDIS}

In an experiment similar to the process discussed in Section~\ref{qweake} and at about the same 
$Q^2$, the Qweak Collaboration at JLab~\cite{Armstrong:2012ps} has completed data taking to 
determine,
\be
A_{LR}^{ep} \equiv
\frac{d\sigma_L - d\sigma_R}{d\sigma_L + d\sigma_R} =
2 {s \over M_Z^2} {g_A^e (2 g_V^u + g_V^d) \over 4 {Q_A^e} (2 Q_A^u + Q_A^d)} {\cal F}^{ep} =
- {m_p (2 E_e + m_p) \over v^2} {g_{AV}^{\, ep} \over 4 \pi \alpha} {\cal F}^{ep},
\ee
in elastic 85\% polarized $ep$ scattering, ${\vec{e}}^{\, -} p \rightarrow e^- p$,
where $s = m_p (2 E_e + m_p)$ for $m_e \ll E_e$, and
\be\label{fqweak}
{\cal F}^{ep} \equiv {\cal F}^{ep} (Q^2,y) =
\left[ y + {\cal O}(y^2) \right] {\cal F}^{ep}_{\rm QED} (Q^2,y).
\ee
A beam energy of $1.165$~GeV at a nominal scattering angle of $\theta_{\rm lab} = 7.9^\circ$ 
keeps both $Q^2 = 0.026 \mbox{ GeV}^2$ and $y \approx 0.0085$ perturbatively small. 
This is necessary because the ${\cal O}(y^2)$-term in Eq.~(\ref{fqweak}) is plagued by large 
hadronic uncertainties and must be kept below the experimental error. 
It is precisely the option to restrict to forward angles which makes this kind of measurement
possible in elastic ${\vec{e}}^{\, -}$ but not $\nu$ scattering (see the discussion in 
Section~\ref{nuDIS}).
In practice, an extrapolation to $y \to 0$ will be performed using other asymmetry 
measurements~\cite{GonzalezJimenez:2011fq} in
parity-violating electron scattering (PVES) at higher $Q^2$-values (see Figure~\ref{C1ud}).
Overall, the SM prediction, $A_{LR}^{ep} = 265$~ppb, at Qweak is comparable to $A_{LR}^{ee}$ at 
E158.
The additional experiment specific corrections~(\ref{gammaZqweak}) and (\ref{cradqweak}) to 
$g_{AV}^{\, ep}$ discussed in Section~\ref{radcorre} are sometimes absorbed together with 
the axial current renormalization into the weak charge of the proton,
\be
Q_W^p \equiv - 2 \left( g_{AV}^{\, ep} - 0.0021 - 0.0004 \right) \left( 1 - {\alpha\over 2 \pi} \right) 
\approx 1 - 4 \sin^2 \theta_W.
\ee
The anticipated experimental precision is $2.5\%$ for $A_{LR}^{ep}$, but 
due to the hadronic dilution (the ${\cal O}(y^2)$-term contributes about one third to $A_{LR}^{ep}$), 
as well as the uncertainty ($1.5\%$) from the proton structure, it is larger ($4.1\%$) for $Q_W^p$.
Including the current estimate of the additional hadronic $\gamma Z$ box dilution and 
its uncertainty in Eq.~(\ref{gammaZqweak}) results in a $4.5\%$ or $\pm 0.0016$ error in 
$g_{AV}^{\, ep}$, which in turn would allow for a $\pm 0.0008$ determination of $\sin^2 \theta_W$.
We note that following the appearance of Ref.~\cite{Gorchtein:2008px} the $\gamma Z$ box 
contribution has been subjected to considerable theoretical scrutiny (see the discussion 
and references in Section~\ref{radcorre}).
It will be important to perform additional experiments, in particular in parity-violating
structure functions at low $Q^2$, such as those planned at JLab,
which can constrain the associated dispersion integrals.
The implications for new physics are discussed in Ref.~\cite{Erler:2003yk}.

\begin{figure}[t]
\begin{center}
\epsfig{file=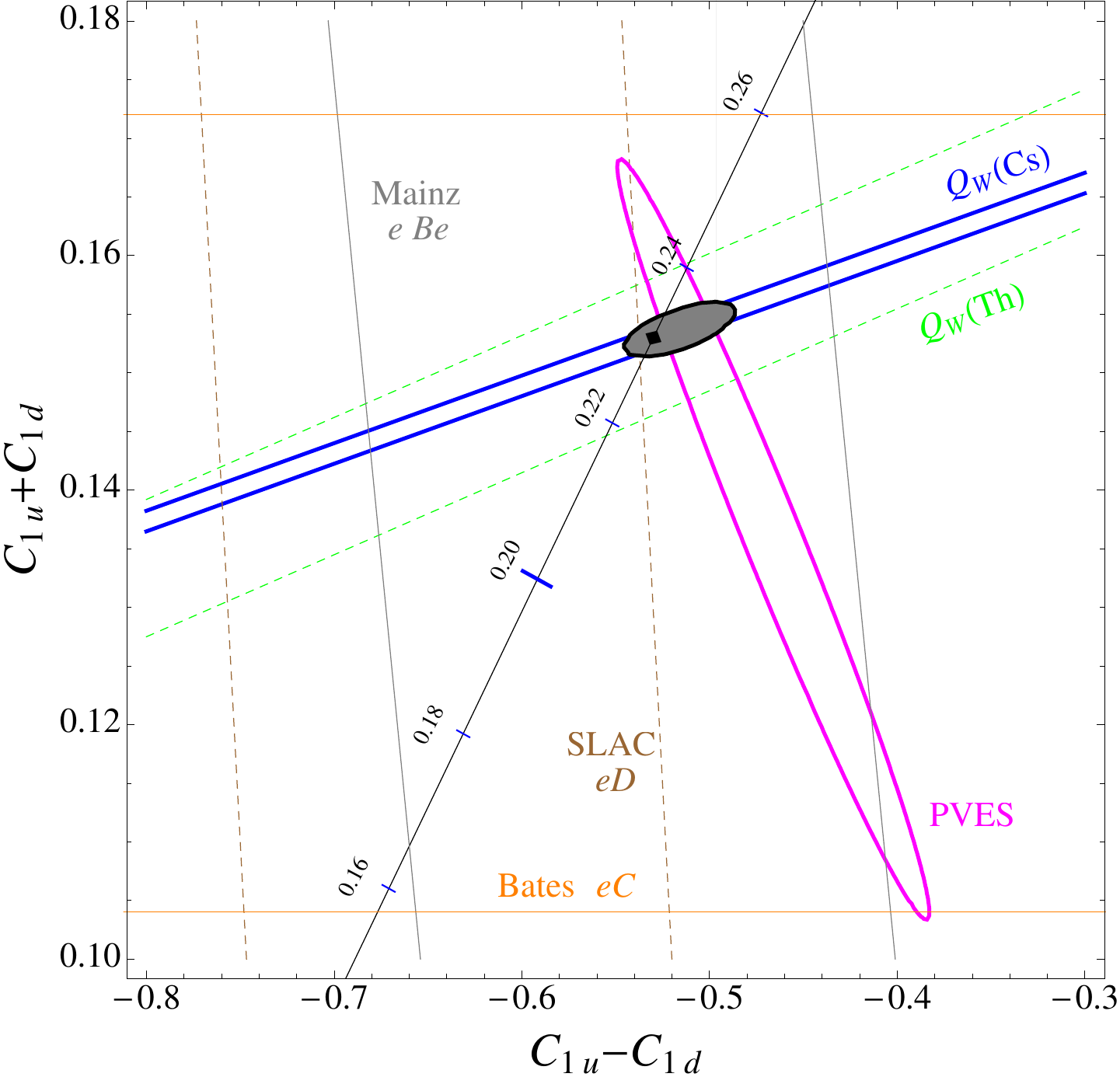,scale=0.7}
\begin{minipage}[t]{16.5 cm}
\caption{Constraints on the effective NC couplings, $C_{1u}$ and $C_{1d}$, from recent (PVES) 
and older parity-violating electron scattering, and from APV at 1~$\sigma$, as well as the 90\% CL 
global best fit (shaded) and the SM prediction as a function of the weak mixing angle $\hat s\,^2_Z$. 
The SM best fit value $\hat s\,^2_Z = 0.23116$ is also indicated.
(Figure reprinted as permitted according to journal guidelines from
\Journal{\PRD}{86}{010001}{2012}, J.~Erler and P.~Langacker~\cite{PDGEW2012}.)
\label{C1ud}}
\end{minipage}
\end{center}
\end{figure}

Both types of hadronic dilutions and uncertainties will be significantly smaller at a similar experiment 
proposed at Mainz~\cite{Baunack:2012} with a lower beam energy of 200~MeV and 
$Q^2 = 0.0048 \mbox{ GeV}^2$ (at  $\theta_{\rm lab} = 20^\circ$).
With these parameters a precision of $1.7\%$ in $A_{LR}^{ep}$ and $2\%$ in $g_{AV}^{\, ep}$ may 
be feasible, \ie an extraction of $\sin^2 \theta_W$ with $\pm 0.00036$ accuracy.
A precise measurement of the model-independent coupling $g_{AV}^{\, ep}$ would also greatly 
reduce the allowed parameter space in the $C_{1q}$  ($g_{AV}^{\, eq}$) plane shown in 
Figure~\ref{C1ud}. 

The coherent sum of couplings, $g_{AV}^{\, eu} + g_{AV}^{\, ed}$, can be extracted by elastic 
scattering off isoscalar nuclei. 
The asymmetry has been measured at MIT-Bates to $25\%$ using a $^{12}$C 
target~\cite{Souder:1990ia}.
A much more precise determination would be very interesting as it would over-constrain
the parameter space in Figure~\ref{C1ud}. 

On the other hand, the $C_{2q}$ ($g_{VA}^{\, eq}$) are harder to come by. 
Experiments in parity-violating deep-inelastic scattering~\cite{Souder:1996qk} are sensitive to 
the interference of the quark-level amplitudes corresponding to ${\cal L}_{\rm NC}^{\, e q}$ with 
the QED amplitudes.
Scattering from an isoscalar target provides information on the charge weighted combinations, 
$2 g_{AV}^{\, eu} - g_{AV}^{\, ed}$ and $2 g_{VA}^{\, eu} - g_{VA}^{\, ed}$.
Specifically, in the simple quark model and in the limit of vanishing $m_p$,  
$$
A_{LR}^{e{\rm DIS}} \equiv
2 {s y \over M_Z^2} {g_A^e \sum_q Q_A^q g_V^q [q(x) + \ovl q(x)] [1 + (1 - y)^2] + 
g_V^e \sum_q Q_A^q g_A^q  [q(x) - \ovl q(x)] [1 - (1 - y)^2] \over 
Q_A^e \sum_q (2 Q_A^q)^2 [q(x) + \ovl q(x)] [1 + (1 - y)^2]}
$$
\be\label{eDIS}
\approx - {9 \over 20 \pi \alpha(Q)} {Q^2 \over v^2} \left[ 
\left( {2 \over 3} g_{AV}^{eu} - {1 \over 3} g_{AV}^{ed} \right) + 
\left( {2 \over 3} g_{VA}^{eu} - {1 \over 3} g_{VA}^{ed} \right) {1 - (1 - y)^2 \over 1 + (1 - y)^2} \right],
\ee
where the last expression is the valence quark approximation, which should be reasonable at larger 
values of $x \gtrsim 0.4$.
One has to correct for higher twist effects (especially at low $Q^2$ and high $x \gtrsim 0.5$),
charge symmetry violations (expected to grow with $x$), 
quark-quark correlations, sea quark contributions, target mass effects, 
longitudinal structure functions and nuclear effects~\cite{Kumar:2013yoa} 
(see also Section~\ref{nuDIS}).
These effects should not exclusively be seen as limitations, but are of considerable interest in
their own right.

Because of the larger values of $Q^2 \gtrsim 1$~GeV in eDIS experiments, and also due to 
the absence of the $1 - 4 \sin^2\theta_W$ suppression, the asymmetry in Eq.~(\ref{eDIS}) is much 
larger ($\gtrsim 10^{-4}$) than the asymmetries measured in elastic scattering.
It has been obtained to about 10\% precision at SLAC~\cite{Prescott:1979dh} in a $Q^2$ range
between 0.92 and 1.96~GeV$^2$, and $0.15 \leq y \leq 0.36$, providing a 7\% determination of 
$\sin^2\theta_W$ and clarifying the confused situation regarding the SM which prevailed at the time. 
Two further data points have been collected~\cite{Zheng:2012vf} at $Q^2 = 1.1$~GeV$^2$ and 
$Q^2 = 1.9$~GeV$^2$ at the 6~GeV CEBAF, each with about 2.5\% precision.
The Collaboration is currently analyzing the data.
A large array of data points will be taken~\cite{Souder:2011zz} after the CEBAF upgrade to 11~GeV.
The kinematic ranges will be between 1.9 and 9.5~GeV$^2$ and $0.2 \lesssim x \lesssim 0.7$
(mostly around $y \approx 0.7$).
These broad ranges will allow to separate the EW physics from the strong interaction issues and it is
hoped to extract the bracketed linear combination of couplings on the r.h.s.\ of Eq.~(\ref{eDIS}) to 
0.5\% accuracy, and $\sin^2\theta_W$ with an uncertainty of $\pm 0.0006$.

An experiment at CERN~\cite{Argento:1982tq} obtained a different kind of DIS asymmetry by 
reversing the charge of projectile muons simultaneously with the helicity 
(the reversals occurred every 6 days rather than in fractions of seconds as in eDIS). 
Since reversing the helicity of muons is much more difficult, the precision ($\approx 25\%$) was 
rather poor.
But this is the only experiment in which the $P$-even, $C$-odd couplings,
\be\label{gAAeud}
g_{AA}^{\, e u} \equiv \cos^2\theta_W g_A^e g_A^u = - {1\over 2}\ ,
\qquad\qquad
g_{AA}^{\, e d} \equiv \cos^2\theta_W g_A^e g_A^d = {1\over 2}\ , 
\ee
entered into the equations (in the combination $2 g_{AA}^{\, eu} - g_{AA}^{\, ed}$)
in addition to $2 g_{VA}^{\, eu} - g_{VA}^{\, ed}$.

Another possibility to find information on the $g_{VA}^{\, eq}$ couplings is through elastic or 
quasi-elastic scattering experiments at backward angles.
They enter through the axial-vector form factor, $G_A^e$ (see, \eg Ref.~\cite{Zhu:2000gn}), 
which becomes dominant in the backward direction. 
Unfortunately, somewhat paralleling the discussion in Section~\ref{nuDIS}, the strange quark 
contribution, $\Delta \mu_s$, to the magnetic moment of the nucleon, as well as the nucleon anapole 
moment (a weak interaction effect between two different quarks in the nucleon, in analogy to 
the nuclear anapole moment mentioned in Section~\ref{APV}) are obstructions to a clean 
determination.
Moreover, according to Eqs.~(\ref{gVAeud}) the isoscalar coupling, $g_{VA}^{\, eu} + g_{VA}^{\, ed}$, 
vanishes to lowest order and is usually set to zero, 
while the focus is on the isovector combination, $g_{VA}^{\, eu} - g_{VA}^{\, ed}$.
Experimental results are available for scattering from hydrogen~\cite{Spayde:1999qg}, 
deuterium~\cite{Hasty:2001ep}, and $^9$B~\cite{Heil:1989dz}.

\subsection{\it Radiative corrections}
\label{radcorre}

The inclusion of EW radiative corrections to the NC effective couplings accessible in parity-violating
low-energy observables follows closely the discussion in Section~\ref{radcorrnu}.
In particular, the $W$ and $Z$ boson self-energy corrections and the related contributions to 
the $\rho$ parameter, as well as the $\gamma$-$Z$ mixing renormalization 
effects~\cite{Erler:2004in,Czarnecki:2000ic} are the same.
The results from the $WW$ and $ZZ$ boxes also carry over.
However, the following comments are in order:

\begin{description}
\item[\em Electron and quark charge radii:]
Both $W$ and $Z$ bosons contribute to the EW charge radii of charged fermions.
The $g_{AV}^{\, \ell f}$ couplings receive large EW logarithms from the $Z$ loop contribution to the 
charge radii of charged leptons, $\diameter_{\ell Z}$.
The logarithms entering the quark charge radii, $\diameter_{q W}$ and $\diameter_{q Z}$, 
are regulated at the strong interaction scale, introducing a hadronic theory uncertainty into 
the $g_{VA}^{\, \ell q}$ unless they are extracted in the perturbative QCD regime of DIS.
For definiteness we choose for the quark mass $m_q = m_p$ in the low-energy parton model 
expression.
The $\diameter_{f Z}$ are proportional to the vector couplings of fermion $f$ which introduce
additional dependences on $\sin\hat\theta_W(\mu)$, and so $\mu$ needs to be chosen 
appropriately.
We take $\mu^2 = m_f M_Z$ for the low-energy couplings, and in the absence of a two-loop 
calculation, we postulate that more generally $\mu^2 = \sqrt{Q^2} M_Z$ will provide a good 
approximation at least when $Q^2 \gg m_f^2$.
Analogous remarks apply to the vector couplings entering through the $\gamma Z$ box graphs
discussed next.

\item[\em $\gamma Z$ box diagrams:]
A new feature compared to the $\nu$ case is the appearance of $\gamma Z$ box,
$\boxtimes_{\gamma Z}$, graphs\footnote{The interference of $\gamma\gamma$ box diagrams
with single $\gamma$ or $Z$ exchanges also enters at this perturbative order but this does not affect
the NC amplitudes or effective Lagrangians such as in Eq.~(\ref{Lee}), so we do not consider them 
here.} generating large logarithms in both $g_{VA}^{\, \ell q}$ and $g_{AV}^{\, \ell q}$.
As before we regulate the parton model result at $m_q = m_p$, while the full effect depends 
on kinematical details. 
Indeed, the $\boxtimes_{\gamma Z}$ terms entering the $g_{AV}^{\, \ell q}$ are suppressed by 
a factor $(1 - 4 \sin^2\hat\theta_W)$ when they are extracted from APV~\cite{Marciano:1982mm} 
but for the conditions of polarized electron scattering~\cite{Gorchtein:2008px} there is 
an admixture of the $\boxtimes_{\gamma Z}^\prime$ structure (see below) where this suppression
is lifted. 
Again we propose to correct for these experiment dependent effects relative to the choice
$m_q = m_p$.

\item[\em Two-loop QCD corrections:]
$g_{AV}^{\, ep}$ is suppressed by a factor $(1-4 \hat{s}_0^2)$ but the $WW$ box contributions are
not. Rather they are further enhanced by an additional factor of 7 compared to $g_{AV}^{\, ee}$, 
resulting in a loop effect similar in size as the tree-level result and perturbative QCD 
corrections to the EW box~\cite{Erler:2003yk} should be included.

\item[\em Axial current renormalization:]
In the original work on radiative corrections to APV~\cite{Marciano:1982mm} the authors included 
QED renormalization terms at $q^2 = 0$, multiplying each axial current vertex, 
$Z_\mu \ovl{f} \gamma^\mu \gamma^5 f$, by a factor $(1 - Q_f^2 \alpha/2 \pi)$.
However, the analogous QCD renormalization of the axial vertices of quarks, 
which cannot be computed perturbatively at small $q^2$, has been omitted. 
Even the term at the free electron vertex should in principle be recalculated for bound state electrons
in heavy atoms~\cite{Marciano:1993ep}.
In any case, consistent with the general strategy proposed in Section~\ref{radcorrnu}, 
QED and QCD corrections to external lines are not considered as part of the EW couplings,
and in fact have not been added in the neutrino scattering case~\cite{Marciano:1980pb}.
We therefore {\em remove\/} these small terms ($ \lesssim 0.1\permil$) 
from the  effective NC couplings in the present case as well, 
with the understanding that they are being accounted for together with the other remaining radiative 
corrections. 
\end{description} 
Collecting these corrections~\cite{Czarnecki:1995fw,Erler:2003yk,Marciano:1982mm,Marciano:1983ss},
\be\label{gAVeu} 
g_{AV}^{\, \ell f} = \rho \left[ - {1 \over 2} + 2\, Q_f \hat s_0^2 -  2\, Q_f \diameter_{\ell Z} + 
\boxtimes_{ZZ} + \boxtimes_{\gamma Z} \right]  - 2\, Q_f \diameter_{\ell W} + \Box_{WW} \ph{e,}
\qquad\qquad (f = u),
\ee
\be\label{gAVed} 
g_{AV}^{\, \ell f} = \rho \left[ \ph{-} {1 \over 2} + 2\, Q_f \hat s_0^2 - 2\, Q_f \diameter_{\ell Z} + 
\boxtimes_{ZZ} + \boxtimes_{\gamma Z} \right] - 2\, Q_f \diameter_{\ell W} +
\hspace{3pt}{\raisebox{-1pt}{\rotatebox{90}{\Bowtie}}}_{WW}
\qquad\qquad (f = d,e),
\ee
\be\label{gVAeu} 
g_{VA}^{\, \ell u} = \rho \left[ - {1 \over 2} + 2 \hat s_0^2 + 2 \diameter_{uZ} + \boxtimes_{ZZ}^\prime + 
\boxtimes_{\gamma Z}^\prime \right] + 2\, \diameter_{uW} + \Box_{WW},
\ee
\be\label{gVAed} 
g_{VA}^{\, \ell d} = \rho \left[ \ph{-} {1 \over 2} - 2 \hat s_0^2 + 2 \diameter_{dZ} + 
\boxtimes_{ZZ}^\prime + \boxtimes_{\gamma Z}^\prime \right] + 2\, \diameter_{dW} +
\hspace{3pt}{\raisebox{-1pt}{\rotatebox{90}{\Bowtie}}}_{WW},
\ee
where using the abbreviation, $\hat\alpha_{ij} \equiv \hat\alpha(\sqrt{m_i M_j})$, we defined,
\be
\diameter_{\ell W} = {2 \alpha \over 9 \pi}\ ,
\qquad\qquad
\diameter_{uW} = - {\alpha \over 18 \pi} \left( \ln {M_W^2 \over m_p^2} + {25 \over 6} \right),
\qquad\qquad
\diameter_{dW} = {\alpha \over 9 \pi} \left( \ln {M_W^2 \over m_p^2} + {13 \over 6} \right),
\ee
\be
\diameter_{fZ} = 
{\alpha \over 6 \pi} Q_f\, g_{VA}^{\, ff} \left( \ln {M_Z^2 \over m_f^2} + {1\over 6} \right),
\ee
and,
\be
\boxtimes_{ZZ} = - {3 \hat\alpha_Z \over 16 \pi \hat s_Z^2 \hat c_Z^2} 
\left( g_{VA}^{\, \ell f} g_{VV}^{\, \ell f} + g_{AV}^{\, \ell f} g_{AA}^{\, \ell f} \right)
\left[ 1 - {\hat\alpha_s(M_Z) \over\pi} \right],
\ee
while $\boxtimes_{ZZ}^\prime$ is given by $\boxtimes_{ZZ}$ with 
$g_{VA}^{\, \ell f} \leftrightarrow g_{AV}^{\, \ell f}$, and $\Box_{WW}$ and 
$\hspace{3pt}{\raisebox{-1pt}{\rotatebox{90}{\Bowtie}}}_{WW}$ are defined in Eqs.~(\ref{WWbox}).  
Furthermore, 
\be\label{gammaZ}
\boxtimes_{\gamma Z} = 
{3 \hat\alpha_{fZ} \over 2 \pi} Q_f\, g_{VA}^{\, \ell f} \left( \ln {M_Z^2 \over m_f^2} + {3 \over 2} \right),
\qquad\qquad
\boxtimes_{\gamma Z}^\prime = 
{3 \hat\alpha_{pZ} \over 2 \pi} Q_f\, g_{AV}^{\, \ell f} \left( \ln {M_Z^2 \over m_p^2} + {5 \over 6} \right).
\ee
The numerical values of these NC couplings are computed using Eqs.~(\ref{inputs}) and are given in 
Table~\ref{ge}.

As in Section~\ref{radcorrnu} one has to correct for $Q^2 \neq 0$.
The only new issue for the $g_{AV}^{e f}$ are the $\gamma Z$ box graphs
which still need to be computed for the relevant $Q^2$-values and electron beam energies.
We expect that in the DIS regime this should be feasible with sufficient accuracy, but we suppose 
that the numerical answer will not differ very strongly from $\boxtimes_{\gamma Z}$ in
Eq.~(\ref{gammaZ}). Ignoring this issue, we find for the SLAC~\cite{Prescott:1979dh}
and Jefferson Lab DIS experiments, all with $Q^2$ values around the charm quark threshold,
$$
g_{AV}^{\, eq} \to g_{AV}^{\, eq} + 2 Q_q \left[ \hat{s}_Q^2- \hat{s}_0^2 + 
{\alpha \over 6 \pi} \left( - g_{VA}^{\, ee} \ln {Q^2 \over m_e^2} + 
{15 \over 2} - {190 \over 9}\hat{s}_0^2 + \Delta_Q^{e q} \right) \right]
$$
\be
\approx g_{AV}^{\, eq}  - 0.0011\, Q_q \ln {Q^2 \over 0.14 \mbox{ GeV}^2}\ .
\ee
The small $Q^2$-dependent correction term $\Delta_Q^{e q}$ accounts for non-decoupling 
$b$ quark and $\tau$ lepton effects, the non-vanishing masses of the lighter fermions, 
and some reducible higher-order effects.
Similarly,
\be
g_{VA}^{\, eu} \to g_{VA}^{\, eu}  - 0.0009 \ln {Q^2 \over 0.078 \mbox{ GeV}^2}\ ,
\qquad\qquad
g_{VA}^{\, ed} \to g_{VA}^{\, ed}  + 0.0007 \ln {Q^2 \over 0.021 \mbox{ GeV}^2}\ .
\ee
The $\boxtimes_{\gamma Z}$ and $\boxtimes_{\gamma Z}^\prime$ contributions induce 
extra $Q^2$-dependences and may change the $A_{LR}^{e{\rm DIS}}$ by several $\permil$.

\begin{table}[t]
\begin{center}
\begin{minipage}[t]{16.5 cm}
\caption{SM values of the one-loop and leading two-loop corrected 
effective NC $e$ couplings for the charged SM fermions and the nucleons.}
\label{ge}
\end{minipage}
\\ [2mm]
\begin{tabular}{c|c|c|c|c|c}
\hline\hline
&&&& \\ [-2mm]
$f$& $e$ & $u$ & $d$ & $p$ & $n$ \\ [2mm]
\hline
&&&& \\ [-2mm]
$g_{AV}^{\ e f}$ & $\ph-$ 0.0225 & $-0.1887$ & $\ph-$ 0.3419 & $-0.0355$ & $\ph-$ 0.4950 \\ [2mm]
$g_{VA}^{\ e f}$ & $\ph-$ 0.0225 & $-0.0351$ & $\ph-$ 0.0247 & $-0.0454$ & $\ph-$ 0.0144 \\ [2mm]
\hline\hline
\end{tabular}
\end{center}
\end{table}

Ref.~\cite{Blunden:2012ty} obtained $\boxtimes_{\gamma Z}$ for the case of APV, which we
represent as a correction relative to Eq.~(\ref{gammaZ}),
\be
g_{AV}^{\, eq} \to g_{AV}^{\, eq} + 
{3 \hat\alpha_{pZ} \over 2 \pi} Q_q\, g_{VA}^{\, e q} \ln {m_p^2 \over m_q^2} \approx \left\{ \ba{c} 
g_{AV}^{\, eu} + 0.00002(4), \\ 
g_{AV}^{\, ed} + 0.00002(2), \\ 
g_{AV}^{\, ep} + 0.00005(9), \\ 
g_{AV}^{\, en} + 0.00006(7), \ea \right.
\ee
choosing $m_u = 1.07$~GeV and $m_d = 1.14$~GeV to reproduce the result~\cite{Blunden:2012ty} 
relevant for bound nucleons.

Likewise, $\boxtimes_{\gamma Z}$ was computed in Ref.~\cite{Blunden:2011rd} for polarized 
electron scattering at $Q^2 = 0$, resulting in
\be\label{gammaZqweak}
g_{AV}^{\, ep} \to g_{AV}^{\, ep} - 0.0021^{+0.0003}_{-0.0006} \mbox{ (CEBAF)}, \qquad\qquad
g_{AV}^{\, ep} \to g_{AV}^{\, ep} - 0.0007^{+0.0002}_{-0.0003} \mbox{ (MESA)},
\ee
for $E_e = 1.165$~GeV and $E_e = 200$~MeV, respectively.  
These shifts are due to the sum of both chirality structures, $\boxtimes_{\gamma Z}$ and
$\boxtimes_{\gamma Z}^\prime$, where the latter~\cite{Sibirtsev:2010zg} (which is dominant
here but irrelevant for APV) agrees well within the quoted uncertainties with the findings of 
Refs.~\cite{Rislow:2010vi,Gorchtein:2011mz}.
The error estimates themselves are currently under discussion, and are expected to improve when 
more experimental data entering the theoretical dispersion integrals will become available.
The effects due to $Q^2 \neq 0$ in the shifts~(\ref{gammaZqweak}) is about 
$-3.5\times 10^{-5}$~\cite{Gorchtein:2011mz} for Qweak and negligible, and the $Q^2$-dependence
of the weak mixing angle can be ignored if $A_{LR}^{ep}$ is normalized using the fine structure
constant in the Thomson limit (see the footnote to Section~\ref{qweake}), but the electron charge 
radius induces the additional shift,
\be\label{cradqweak}
g_{AV}^{\, ep} \to g_{AV}^{\, ep}  - 0.00008 \ln {Q^2 \over 0.00021 \mbox{ GeV}^2}\ .
\ee

Finally, the experiment specific adjustments due to $\boxtimes_{\gamma Z}$ and the electron 
charge radius to be applied to $g_{VA}^{ee}$ for the conditions of the E158 (with a beam energy
of about 48~GeV) and MOLLER (11~GeV) experiments are, respectively,
\be\label{MOLLERcorr}
g_{VA}^{\, ee} \to g_{VA}^{\, ee} + 0.0010 \pm 0.0004 \mbox{ (SLC)}, \qquad\qquad
g_{VA}^{\, ee} \to g_{VA}^{\, ee} + 0.0008 \pm 0.0005 \mbox{ (CEBAF)}.
\ee
Note, that some contributions have been merged together with other radiative corrections into 
${\cal F}_{\rm QED}$~\cite{Zykunov:2005md}.

\section{Constraints on Supersymmetry
\label{NP}}

The low energy NC measurements discussed in the previous sections give
complementary information on possible physics beyond the SM, especially when compared to 
$Z$-pole precision observables as well as direct searches for new particles at high energy colliders.  
In this section, we illustrate the sensitivity to new physics via a few NC measurements
(the weak charges of the electron, the proton and of cesium, NuTeV, and eDIS) using 
the Minimal Supersymmetric Standard Model (MSSM) as a specific example.  
For a review of low energy precision tests of supersymmetry, see Ref.~\cite{RamseyMusolf:2006vr}.

\subsection{\it Minimal Supersymmetric Standard Model}

The SM has been very successful in describing the strong, weak and electromagnetic interactions 
and has been confirmed to high precision by a wide variety of experiments.  
As reviewed in Section~\ref{SMHiggs}, the Higgs mechanism is introduced to spontaneously break 
$SU(2)_L \times U(1)_Y$ down to the electromagnetic gauge group, $U(1)_Q$.
Being a fundamental scalar particle, the Higgs boson receives large corrections to its mass from 
quantum loop effects, which are quadratic in terms of  the cut-off scale $\Lambda_{\rm UV}$.  
If $\Lambda_{\rm UV}$ is of the order of Planck mass scale,  a precise cancellation of 32 orders of 
magnitude between the tree-level bare Higgs mass squared and the radiative corrections is needed 
to obtain a physical $M_H$ around the EW scale.
If not helped by a symmetry~\cite{'tHooft:1979bh}, 
such a high level of fine tuning is referred to as a naturalness problem.
Finding a solution to the naturalness problem of the SM Higgs sector (the hierarchy problem) points 
to physics beyond the SM, such as 
supersymmetry (SUSY)~\cite{Haber:1984rc,Martin:1997ns},  
large extra dimensions~\cite{ArkaniHamed:1998,Antoniadis:1998},
warped extra dimensions~\cite{Randall:1999}, 
little Higgs theories~\cite{ArkaniHamed:2001nc,ArkaniHamed:2002},
composite Higgs models~\cite{Harnik:2003rs}, 
Higgs-less models~\cite{Csaki:2003zu}, \etc.

Supersymmetry is a symmetry under the interchange of  bosonic and fermionic degrees of freedom  
and is considered to be one of the most  promising new physics scenarios among various proposals.  
For each particle in a supersymmetric theory, there exists a superpartner with spin differing by half 
a unit.   
\Eg the fermionic superpartners of Higgs bosons are called {\it Higgsinos}, while
the scalar superpartners of quarks and leptons, are called {\it squarks} and {\it sleptons}, respectively.  
The spin-$1/2$ superpartners of the gauge bosons are called {\it gluinos, winos} and the {\it bino}.
When SUSY is exact, the masses and the gauge quantum numbers of the superpartners are the 
same, and the couplings are related by the symmetry.  
These features protect the Higgs mass from receiving the problematic quadratic dependence on 
$\Lambda_{\rm UV}$ as these contributions from fermionic and bosonic superpartners cancel.  
SUSY has to be broken, however, since no scalar-electron with the same mass and 
coupling of that of the electron has been observed.  
Soft SUSY breaking by superrenormalizable terms in the Lagrangian, \ie terms with coefficients of 
strictly positive mass dimension, is needed to retain the cancellation of the quadratic
$\Lambda_{\rm UV}$ dependence of the  Higgs mass corrections. 

The remaining logarithmic corrections to $M_H$ are proportional to the soft SUSY breaking 
masses  $\tilde{m}$,
\be\label{eq:mhsoft}
\Delta {M_H^2} \propto 
- \frac{\tilde{m}^2}{8 \pi^2} \ln {\Lambda_{\rm UV}^2 \over \tilde{m}^2} + \cdots,
\end{equation}
where $\tilde{m}$ should not exceed a few TeV to avoid reintroduction of the naturalness problem.   
Besides providing an elegant solution to the hierarchy problem, SUSY offers other  
attractive features, such as approximate gauge coupling unification, radiatively generated EW  
symmetry breaking, a dark matter candidate, and (in MSSM extensions) the possibility to 
generate the baryon asymmetry of the universe.

By definition, the MSSM  is the SUSY model with the minimal particle content and provides 
a useful framework for discussing the phenomenology of low energy SUSY.   
Note, that the introduction of two Higgs doublets with opposite hypercharge, $H_u$ 
and $H_d$, is dictated by the requirement of anomaly cancellation among the Higgsinos, 
and independently by the holomorphicity of the superpotential.  
In contrast to the SM, where the same Higgs doublet gives masses to both 
the up and down type quarks, here they receive their masses from the VEVs 
of the neutral $H_u$ and $H_d$, respectively.  
  
The most general MSSM superpotential also includes baryon number ($B$) and 
lepton number ($L$) violating interactions, which lead to rapid proton decay already at 
the renormalizable level, which is in sharp conflict with bounds on the proton lifetime.  
One way to eliminate  such terms is to introduce  a new symmetry called $R$-parity, 
defined by conservation of the quantum number,
\be
P_R = (-1)^{ 3 (B - L) + 2 S}\ ,
\ee
where $S$ is the spin of the particle.  
All SM particles are assigned $P_R = + 1$, while all the superpartners have $P_R = - 1$.  
Exact $R$-parity has two important phenomenological consequences:
(i) The lightest supersymmetric particle (LSP) is absolutely stable.
(ii) SM particles are coupled to even numbers of superpartners (usually two).
If the LSP is colorless and electrically neutral, it may be a viable candidate to constitute cold dark 
matter. 
For low-energy processes involving only SM particles in the initial and final states, such as those of 
interest in this section, supersymmetric contributions appear only at the loop-level via virtual pair 
production of superpartners.
However, one may relax the constraint of $R$-parity conservation while preserving proton 
stability, \eg by forbidding only the baryon number violating terms. 
In this case, the LSP is no longer stable and tree-level SUSY contributions to low energy processes 
occur through $R$-parity violating interactions. 
In what follows, we will consider the implications of both $R$-parity conserving (RPC) 
and $R$-parity violating (RPV) supersymmetry.

The MSSM soft terms contribute masses to the Bino ($M_1$), Winos ($M_2$), Gluinos ($M_3$), 
the squarks and sleptons, the Higgs bosons, as well as a bilinear Higgs mixing ``$B$-term",
and trilinear ``$A$-terms" that couple Higgs scalars with the superpartners of left- and right-handed 
quarks and leptons.
The $A$-terms and the soft SUSY breaking squark and slepton masses are in general non-diagonal 
in the flavor basis which could lead to large flavor-changing neutral current (FCNC) effects.
They might also have additional phases that cannot be eliminated by field redefinitions, thus 
inducing new sources of CP-violation.
For generic parameter values, the predicted effects are considerably larger than allowed by 
the experimental limits (the SUSY flavor and CP problems).  
For illustration, we consider one approach to the flavor problem in which it is assumed that all soft 
masses of squarks and sleptons are diagonal in the flavor basis and that $A$-terms are 
proportional to the corresponding Yukawa matrices.   
Under these simple assumptions, the mass matrix of the superpartners of the left- and right-handed 
SM fermions reduces to block-diagonal form of $2 \times 2$ sub-matrices for each flavor, governed 
by mixing angles $\theta_{\tilde{f}}$. 
 
After EW symmetry breaking, the neutral gauginos and Higgsinos, $\tilde{B}$, $\tilde{W}^0$, 
$\tilde{H}_d^0$, $\tilde{H}_u^0$, mix with each other.  
The mass eigenstates are called neutralinos, $\chi_i^0$ ($i=1 \cdots 4$), with
$m_{\chi_1^0} < m_{\chi_2^0} < m_{\chi_3^0} < m_{\chi_4^0}$ by definition.    
In the limit, $M_Z \ll M_1, M_2, |\mu|$, each $\chi_i^0$ is a pure gaugino or Higgsino state,  
but in general the $\chi_i^0$ are mixtures.  
Similarly, charged gauginos and Higgsinos, $\tilde{W}^+, \tilde{H}_u^+, \tilde{W}^-, \tilde{H}_d^-$,
mix to form chargino states, $\chi_i^{\pm}$ ($i=1,2$), with $m_{\chi_1^\pm} < m_{\chi_2^\pm}$.
The ${ SU}(3)_C$ gluinos are color octet fermions, and cannot mix with other particles in the MSSM.
In phenomenological studies one usually assumes the mass unification relation at the EW scale 
$M_3 : M_2 : M_1= \alpha_3 : \alpha_2 : \alpha_1 \approx 7 : 2 : 1$.
However, such a  relation need not hold in general and $M_3$, $M_2$ and $M_1$ could 
be completely independent of  each other. 
 
The gauge interactions involving superpartners in the MSSM can be obtained from the usual SM 
gauge interactions by  replacing two of the SM particles with their superpartners.    
Similarly, Higgs-squark-squark and Higgsino-quark-squark couplings can be obtained through 
Yukawa interactions, which also give rise to additional  Higgs-Higgs-squark-squark couplings.  
The $A$-terms give rise to additional Higgs-squark-squark couplings.  
Analogous remarks apply to the leptonic sector.

\subsection{\it R-parity conserving MSSM contributions to neutral current processes}

The precise measurements of the electron and proton weak charges could probe both 
the supersymmetric loop effects as well as the tree-level RPV contributions. 
For RPC SUSY, contributions to $Q_W^e$ and $Q_W^p$, or equivalently to $g_{AV}^{ee}$ and 
$g_{AV}^{ep}$, appear at the loop-level and have been analyzed in detail in 
Ref.~\cite{Kurylov:2003zh}.
The results are summarized and updated in Figure~\ref{fig:mssm-vs-parity},  
taking into account the latest limits from superparticle searches 
at the LHC~\cite{Aad:2012pxa,Chatrchyan:2013lya}, 
as well as tighter constraints on the oblique parameters~\cite{PDGEW2012}.
We plotted the contributions to $g_{AV}^{ep}$ \vs\ those to $g_{AV}^{ee}$, 
normalized to the respective SM values.  
The dots show the results of a random scan over a range of MSSM parameters.
The loop corrections in the RPC case are nearly always positive and can reach 
2--3\% for $g_{AV}^{ep}$.  
For $g_{AV}^{ee}$ it can be even larger but this is not supported by the result of 
E158~\cite{Anthony:2005pm} which observed a suppressed value compared to the SM.
The planned PVES experiments~\cite{Mammei:2012ph,Baunack:2012} would reach factors of five 
and two better than E158 and Qweak, respectively, and would provide very significant 
constraints on the loop effects in RPC SUSY.

The corrections to $g_{AV}^{ee,ep}$ are dominated by the SUSY contribution proportional to 
$\hat s^2_Z$, which is identical for $ee$ and $ep$ scattering.  
Such dominance produces a nearly linear correlation between these two couplings. 
This universal correction is almost always negative, predicting an apparent reduction in 
the extracted value of $\hat s^2_Z$ from PVES experiments.

\begin{figure}[t]
\begin{center}
\epsfig{file=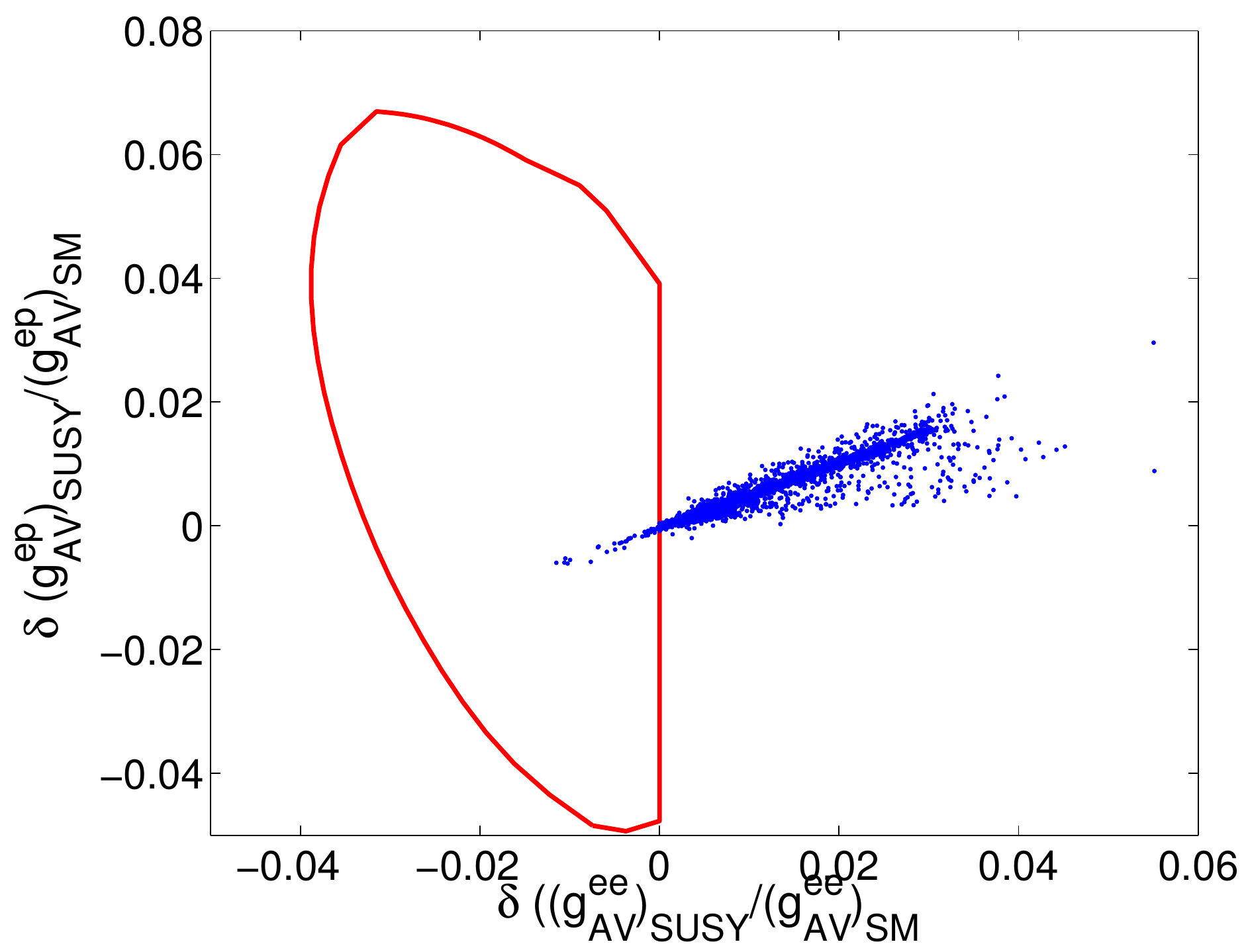,scale=0.7}
\begin{minipage}[t]{16.5 cm}
\caption{Relative shifts in $g_{AV}^{ee}$ and $g_{AV}^{ep}$ 
(normalized to the respective SM values) due to SUSY effects.
The dots indicate the RPC corrections for $\sim 3000$ randomly generated SUSY-breaking 
parameters. 
The interior of the truncated elliptical region gives the possible shifts due to the RPV SUSY 
interactions at the 95\% CL.
(Figure updated from Ref.~\cite{Kurylov:2003zh}.) }
\label{fig:mssm-vs-parity}
\end{minipage}
\end{center}
\end{figure}

The possible effects of new physics on the $Q_W^{Z,N}$, or equivalently, 
$Z g_{AV}^{ep} + N g_{AV}^{en}$, can be written as a sum of the corresponding effects on 
$g_{AV}^{eu, ed}$. 
Since the sign of $\delta g_{AV}^f/g_{AV}^f$ due to superpartner loops is nearly always the same, 
and since $g_{AV}^{eu}<0$ and $g_{AV}^{ed}>0$ in the SM, a strong cancellation between 
$\delta g_{AV}^{eu}$ and $\delta g_{AV}^{ed}$ occurs in heavy nuclei. 
This means that the magnitude of superpartner loop contributions to the weak charge of cesium,
$Q_W^{\rm Cs}$, is generally less than about 0.2\%, and is equally likely to have either sign. 
Since the total uncertainty is presently about 0.6\%~\cite{Dzuba:2012kx}, it does not substantially 
constrain the RPC SUSY parameter space.   
This contrasts with a class of models with extra $Z$ bosons, where sizable shifts in 
$g_{AV}^{ee,ep}$ would also imply observable deviations in $Q_W^{\rm Cs}$  \cite{Erler:2003yk}.
In the case where such models and the MSSM have similar effects on $g_{AV}^{ee}$ and 
$g_{AV}^{ep}$, the determination of $Q_W^{\rm Cs}$ may tell these two apart.  

In addition to the $g_{AV}^{eq}$ ($C_{1q}$), the RPC version of the MSSM also affects 
the $g_{VA}^{eq}$ ($C_{2q}$) which can be probed in eDIS~\cite{Zheng:2012vf,Souder:2011zz} 
as discussed in Section~\ref{PVDIS}.
The shifts in $A_{LR}^{e {\rm DIS}}$ are correlated with those in $g_{AV}^{ee}$ and 
$g_{AV}^{ep}$ and may reach 0.7\%~\cite{Kurylov:2003xa}. 

The RPC contributions to $R_{\nu}$ and $R_{\bar\nu}$ are highly correlated and can reach
$1.5\times 10^{-3}$~\cite{Kurylov:2003by,Davidson:2001ji}.
Their sign is almost always positive, which is in conflict with the sign of the NuTeV anomaly.
Contributions from gluino loop with nearly degenerate first 
generation up-type squark and down-type squarks and close to maximal left-right mixing could admit 
a negative loop contribution.  However, this corner of parameter space is disfavored theoretically
and has tensions with other precision electroweak inputs.
In any case, these negative gluino contributions are too small to fully account for NuTeV.

\subsection{\it R-parity violating supersymmetry}

Additional $B$- and $L$-violating interactions may appear in the MSSM if RPV is allowed.
Rapid proton decay can still be avoided if we only turn on $B$ or $L$ violating terms, 
but not both simultaneously.      
For low energy processes where light quarks are present in the initial and final states, the RPV 
terms that are of interest are the Yukawa-type interactions,
\be\label{eq:rpv-super}
{\cal L}_{\rm RPV}^{\Delta L = 1} = \lambda_{ijk} \left( {1 \over 2} L_i L_j \tilde{\bar{e}}^\dagger_k + 
\tilde{L}_i L_j\bar{e}^{\dagger}_k \right) + \lambda_{ijk}^{\prime} \big( L_i Q_j \tilde{\bar{d}}^\dagger_k
+ \tilde{L}_i Q_j \bar{d}^{\dagger}_k + L_i \tilde{Q}_j \bar{d}^{\dagger}_k \big),
\ee
\be
{\cal L}_{\rm RPV}^{\Delta B = 1} = \lambda_{ijk}^{\prime\prime} \big(
\bar{u}^{\dagger}_i \bar{d}^{\dagger}_j \tilde{\bar{d}}^\dagger_k +
\tilde{\bar{u}}^\dagger_i \bar{d}^{\dagger}_j \bar{d}^{\dagger}_k \big),
\ee
which will contribute via the exchange of heavy squarks or sleptons.  
For small momentum transfers, $q^2 \ll m_{\tilde f}^2$, 
the corrections can be parametrized in terms of
\be\label{eq:deltas}
\Delta_{ijk}(\tilde f) \equiv {|\lambda_{ijk}|^2\over 4} {v^2 \over m_{\tilde f}^2} \ge 0,
\ee 
and similarly for the primed and doubly primed quantities.    
The shifts in the couplings are then~\cite{RamseyMusolf:2000qn},
\be\label{eq:rpv-weak1}
\delta g_{AV}^{\, ee} \approx - \left[ g_{AV}^{\, ee} + 2 \lambda_x \right] \Delta_{12k}({\tilde e}_R^k), 
\ee
\be\label{eq:rpv-weak2}
\delta g_{AV}^{ep} \approx - \left[ g_{AV}^{\, ep} - 2\lambda_x \right] \Delta_{12k}({\tilde e}_R^k) - 
2 \Delta_{11k}^\prime({\tilde d}_R^k) + \Delta_{1k1}^\prime({\tilde q}_L^k),
\ee
\be\label{eq:rpv-weak3}
\delta g_{AV}^{\, eu} = - \left[ g_{AV}^{\, eu} - \frac{4}{3} \lambda_x \right]
\Delta_{12k}({\tilde e}^k_R) - \Delta^\prime_{11k}({\tilde d}^k_R),
\qquad
\delta g_{AV}^{\, ed} = - \left[ g_{AV}^{\, ed} + \frac{2}{3} \lambda_x \right]
\Delta_{12k}({\tilde e}^k_R) + \Delta^\prime_{1k1}({\tilde q}^k_L),
\ee
\be
\delta g_{VA}^{\, eu} = - \left[ g_{VA}^{\, eu} - 2 \lambda_x \right]
\Delta_{12k}({\tilde e}^k_R) - \Delta^\prime_{11k}({\tilde d}^k_R),
\qquad\qquad
\delta g_{VA}^{\, ed} = - \left[ g_{VA}^{\, ed} + 2 \lambda_x \right]
\Delta_{12k}({\tilde e}^k_R) - \Delta^\prime_{1k1}({\tilde q}^k_L),
\ee
where,
\be\label{eq:lambda}
\lambda_x = \frac{\hat s_Z^2 \hat c_Z^2}{1-2{\hat s}^2} \frac{1}{1-\Delta {\hat r_Z^{\rm SM}}}\ .
\ee
Since the $\Delta_{ijk}^{(\prime)}$ are non-negative, Eq.~(\ref{eq:rpv-weak1}) shows that the shifts
in the $g_{AV}^{ee}$ are negative semidefinite, while according to Eq.~(\ref{eq:rpv-weak2}) 
those in the $g_{AV}^{ep}$ can have either sign depending on the relative magnitudes of 
$\Delta_{12k}$, $\Delta_{11k}^\prime$, and $\Delta_{1k1}^\prime$.


Quantities such as the $\Delta_{ijk}$ are also constrained by other precision measurements and rare 
decays.  
A summary of the current experimental bounds is shown in Table~\ref{tab:rpv-constrain}.
It includes CKM unitarity tests obtained from superallowed nuclear $\beta$-decays 
(constraining $|V_{ud}|$~\cite{Hardy:2008gy}) and the kaon-decay determination of $V_{us}$, 
the APV measurement of $Q_W^{\rm Cs}$~\cite{Dzuba:2012kx,Bennett:1999pd},
the ratio $R_{e/\mu}$~\cite{Britton:1992pg,Czapek:1993kc} of $\pi_{l2}$
decays~\cite{Cirigliano:2007ga,Bauman:2012fx}, 
and the allowed range for $\Delta \hat{r}$ defined in Eq.~(\ref{eq:deltar})].
We also indicate the sensitivity to the various $\Delta_{ijk}^{(\prime)}(\tilde f)$.
  
\begin{table}[t]
\begin{center}
\begin{minipage}[t]{16.5 cm}
\caption{RPV contributions to $|V_{ud}|^2$, $Q_W^{\rm Cs}$, $R_{e/\mu}$ and $\Delta \hat{r}$.
$\delta |V_{ud}|^2/|V_{ud}|^2$ are the corrections to $|V_{ud}|^2$ 
extracted from beta-decay that are allowed by first row CKM unitarity tests.   
The mid columns display the coefficients of the various corrections from $\Delta_{ijk}^{\prime}$ and 
$\Delta_{12k}$.  
The last column gives the values extracted from experiment assuming only SM contributions.}
\label{tab:rpv-constrain}
\end{minipage}
\\ [2mm]
\begin{tabular}{c|llll|c}   
\hline \hline 
&&&&&\\ [-2mm]
Quantity & $\Delta_{11k}^{\prime}(\tilde{d}_R^k)$ & $\Delta_{1k1}^{\prime}(\tilde{q}_L^k)$ & 
$\Delta_{12k}(\tilde{e}_R^k)$ & $\Delta_{21k}^{\prime}(\tilde{d}_R^k)$ & Value \\ [2mm]
\hline
&&&&&\\ [-2mm]
$\delta |V_{ud}|^2/|V_{ud}|^2$ & $\ph-2$ & $\ph-0$ & $-2$ & $\ph-0$ & 
$-0.0001 \pm 0.0006$ \\ [2mm]
$\delta Q_W^{ Cs}/Q_W^{ Cs}$ & $-4.82$ & $\ph-5.41$ & $\ph- 0.05$ & $\ph-0$ & 
$-0.0089 \pm 0.0059$ \\ [2mm]
$\delta R_{e/\mu}$ & $\ph-2$ & $\ph-0$ & $\ph-0$ & $-2$ & 
$-0.0034 \pm 0.0030$ \\ [2mm]
$\Delta\hat{r}$ & $\ph-0$ & $\ph-0$ & $\ph-1$ & $\ph-0$ & 
$\ph- -0.00002 \pm 0.00045$ \\ [2mm]
\hline\hline
\end{tabular}
\end{center}
\end{table}
 
The 95\% CL allowed region in the $\delta g_{AV}^{ep}/g_{AV}^{ep}$ \vs\
$\delta g_{AV}^{ee}/g_{AV}^{ee}$ plane is shown by the closed curve in 
Figure~\ref{fig:mssm-vs-parity}.
Note, that the truncation of the initially elliptical curve is due to the inequality in Eq.~(\ref{eq:deltas}).  
The corrections to $g_{AV}^{ee}$ from RPV SUSY are always negative and less than about 4\% in 
magnitude, while those to $g_{AV}^{ep}$ vary in the range of $-5\%$ to 7\%.  
In addition, the prospective effects of RPV SUSY where $\delta g_{AV}^{ep}/g_{AV}^{ep}$ can have 
either sign, are quite distinct from SUSY loops where it is non-negative.
Thus, a comparison of the results for the two PVES experiments could 
help determine whether this extension of the MSSM is favored over other new physics 
scenarios (see also Ref.~\cite{Erler:2003yk}).
 
Given the 95\% CL region for the RPV coefficients, the maximum RPV correction to 
$A_{LR}^{e \rm{DIS}}$ is about $\pm 0.4\%$, close to the precision proposed in 
Ref.~\cite{Souder:2011zz}.
The RPV effects would induce opposite shifts in $A_{LR}^{e \rm {DIS}}$ and $g_{AV}^{ee}$, 
whereas the loop corrections are positive in both cases.  
A sizable positive shift in $g_{AV}^{ep}$  due to the RPV contributions 
could correspond to a tiny effect in $A_{LR}^{e\rm {DIS}}$.  
The addition of an eDIS measurement would provide a useful complement to the PVES $ee$ 
and $ep$ measurements, assuming it can be performed with $\sim$ 0.4\% precision or better. 

Similarly, the shifts in $R_{\nu(\bar\nu)}$ introduced by the tree-level RPV interactions are given by,
\be\label{eq:rnurpv}
\delta R_{\nu (\bar\nu)} = {1 + r \over r} 
\left[ - {4 \over 3} g_{LL}^{\nu u} + {2 \over 3} g_{LL}^{\nu d} \right] \lambda_x 
\Delta_{12k}(\tilde{e}_R^k) - 2 \left[ R_{\nu (\bar\nu)}^{\rm SM} + g_{LL}^{\nu d} \right] 
\Delta_{21k}^{\prime}(\tilde{d}_{R}^k) + {2 \over r} g_{LR}^{\nu d} 
\Delta_{2k1}^{\prime}(\tilde{d}_{L}^k). 
\end{equation}
While $\Delta_{12k}({\tilde e}^k_R)$ and $\Delta^{\prime}_{21k}({\tilde d}^k_R)$ are constrained by 
other precision EW data, $\Delta^{\prime}_{2k1}({\tilde d}^k_L)$ is relatively unconstrained. 
The present constraints on $\Delta_{12k}({\tilde e}^k_R)$ from other EW observables, 
as listed in Table~\ref{tab:rpv-constrain}, however, are fairly stringent.  
The possible effects on $\rnu$ and $\rnubar$ from RPV interactions are by and large positive. 
Negative corrections are also possible, but they are too small to be interesting~\cite{Kurylov:2003by}.

We have used SUSY as an example of how new physics may affect the low energy NC precision 
observables. 
In addition, there are various studies in the literature exploring other types of new physics, such as 
models with extra $Z^\prime$ bosons~\citer{Chang:2009yw,GonzalezAlonso:2012jb} and 
leptoquarks~\cite{RamseyMusolf:1999qk,Herczeg:2003ag}.
The bottom line is that the combination of various NC experiments can distinguish between different 
new physics scenarios, as illustrated in 
Refs.~\cite{Kurylov:2003zh,Chang:2009yw,Li:2009xh,Diener:2011jt}.   
\Eg while both the superpartner loops and leptoquark exchange give positive contributions
to the proton weak charge, only the MSSM gives rise to a sizable effect on the electron weak 
charge~\cite{Erler:2003yk,RamseyMusolf:1999qk}.
For the class of $Z^{\prime}$ theories based on the $E_6$ gauge group with $Z^{\prime}$ masses 
$\lesssim 1$~TeV, the effects on $Q_W^p$ and $Q_W^e$ also correlate, 
but $\delta Q_W^{e,p}/Q_W^{e,p}$ can have either sign~\cite{Erler:2003yk,RamseyMusolf:1999qk}. 
Finally, in the case where $E_6~Z^\prime$ models and the MSSM have similar effects on 
$g_{AV}^{ep}$ and $g_{AV}^{ee}$, the determination of $Q_W^{\rm Cs}$ can further tell these two 
apart.

\section{Conclusions
\label{conclusions}}
 
In this review, we surveyed low energy neutral current measurements, including neutrino 
scattering, parity-violating electron scattering, and atomic parity violation.  
We reviewed the experimental status and the theoretical challenges of these observables, 
and pointed to future experiments that are either planned or proposed.  
We also explored the sensitivity of those measurement to new physics, using the minimal 
supersymmetric Standard Model as a specific example.  
Furthermore, we illustrated how the interplay between the various NC observables 
provides extra discriminating power between different types of physics beyond the SM.
 
While the direct measurements of $W$ and $Z$ properties have reached per mille precision, and 
the LHC at its design energy of 14~TeV could produce strong interacting particles with masses in the 
few TeV range, the low energy precision measurements provide an alternative probe of new physics.  
\Eg they are sensitive to new physics that does not mix with $W$ and $Z$ bosons.  
With the high statistics achieved at the intensity frontier and the advances in both the 
theoretical and experimental sides, they serve as indispensable complements to the energy frontier.   
If a significant deviation from the SM prediction is observed in a low energy neutral current 
observable, comparison with other precision measurements and the collider results will sharpen our 
understanding of physics beyond the SM.

\section*{Acknowledgements}
It is a pleasure to thank Krishna Kumar and Paul Langacker for a careful reading of 
the manuscript and stimulating discussions.
The work of J.E. was supported by CONACyT (M\'exico) projects 82291--F and 15 1234
and by PAPIIT (DGAPA--UNAM) project IN106913.  
The work of S.S. was supported by the Department of Energy under 
Grant~DE--FG02--04ER--41298.


\begin{thebibliography}{199}
\itemsep -2pt 

\bibitem{Hewett:2012ns} J.~L.~Hewett \etal, arXiv:1205.2671 [hep-ex]

\bibitem{Erler:2004cx} J.~Erler and M.~J.~Ramsey-Musolf, \Journal{\PPNP}{54}{351}{2005}
    
\bibitem{Kumar:2013yoa} 
K.~S.~Kumar, S.~Mantry, W.~J.~Marciano and P.~A.~Souder,  arXiv:1302.6263 [hep-ex]
  
\bibitem{pgl} P.~Langacker, {\it The Standard Model and Beyond} 
(Taylor \& Francis, Boca Raton, FL, 2010)

\bibitem{Weinberg:1965rz} S.~Weinberg, \Journal{\PREV}{138}{B988}{1965}

\bibitem{Weinberg:1967tq} S.~Weinberg, \Journal{\PRL}{19}{1264}{1967}

\bibitem{Glashow:1961tr} S.~L.~Glashow, \Journal{\NP}{22}{579}{1961}

\bibitem{Djouadi:2005gi} For a review, see A.~Djouadi, \Journal{\PREP}{457}{1}{2008}

\bibitem{Higgs:1964ia} P.~W.~Higgs, \Journal{\PL}{12}{132}{1964} 
and \Journal{\PRL}{13}{508}{1964}

\bibitem{Guralnik:1964eu} G.~S.~Guralnik, C.~R.~Hagen and T.~W.~B.~Kibble,
\Journal{\PRL}{13}{585}{1964}

\bibitem{Englert:1964et} F.~Englert and R.~Brout, \Journal{\PRL}{13}{321}{1964}

\bibitem{Kibble:1967sv} T.~W.~B.~Kibble,  \Journal{\PREV}{155}{1554}{1967}

\bibitem{Goldstone:1961eq} J.~Goldstone, \Journal{\NC}{19}{154}{1961}

\bibitem{Webber:2010zf} D.~M.~Webber \etal\ (MuLan Collaboration), 
\Journal{\PRL}{106}{041803}{2011}and \ibid\ 079901

\bibitem{PDGEW2012} J.~Erler and P.~Langacker, 
{\em Electroweak Model and Constraints on New Physics}, pp.~136--156 of Ref.~\cite{PDG2012}

\bibitem{PDG2012} J.~Beringer \etal\ (Particle Data Group), \Journal{\PRD}{86}{010001}{2012}

\bibitem{Erler:2004in} J.~Erler and M.~J.~Ramsey-Musolf, \Journal{\PRD}{72}{073003}{2005}

\bibitem{Czarnecki:2000ic} A.~Czarnecki and W.~J.~Marciano, 
\Journal{\IJMPA}{15}{2365}{2000}

\bibitem{Weinberg:1979sa} S.~Weinberg, \Journal{\PRL}{43}{1566}{1979}

\bibitem{Abe:2000dq} K.~Abe \etal\ (SLD Collaboration), \Journal{\PRL}{84}{5945}{2000}

\bibitem{ALEPH:2005ab} The ALEPH, DELPHI, L3, OPAL and SLD Collaborations, 
the LEP Electroweak Working Group and the SLD Electroweak and Heavy Flavour Groups,
\Journal{\PREP}{427}{257}{2006}

\bibitem{Abe:2000uc} K.~Abe \etal\ (SLD Collaboration), \Journal{\PRL}{85}{5059}{2000}

\bibitem{Abe:2000hk} K.~Abe \etal\ (SLD Collaboration), \Journal{\PRL}{86}{1162}{2001}

\bibitem{Abe:1996ef} K.~Abe \etal\ (SLD Collaboration), \Journal{\PRL}{78}{17}{1997}

\bibitem{Abreu:1994pj} P.~Abreu \etal\ (DELPHI Collaboration), \Journal{\ZPC}{67}{1}{1995}

\bibitem{Abazov:2011ws} V.~M.~Abazov \etal\ (D\O\ Collaboration), 
\Journal{\PRD}{84}{012007}{2011}

\bibitem{Han:2011vw} J.~Han (CDF Collaboration), 
in the Proceedings of DPF--2011, arXiv:1110.0153 [hep-ex]

\bibitem{Chatrchyan:2011ya} S.~Chatrchyan \etal\ (CMS Collaboration), 
\Journal{\PRD}{84}{112002}{2011}

\bibitem{Hollik:1993cf} W.~Hollik, {\em Predictions for $e^+ e^-$ processes}, 
pp.~117--169 of Ref.~\cite{Langacker:1996qb}

\bibitem{Langacker:1996qb} P.~Langacker (ed.), 
{\em Precision tests of the standard electroweak model},
Advanced series on directions in high energy physics: 14 
(World Scientific, Singapore, Singapore, 1995)

\bibitem{Freitas:2012sy} A.~Freitas and Y.~C.~Huang, \Journal{\em JHEP}{1208}{050}{2012}


\bibitem{Alcaraz:2006mx} J.~Alcaraz \etal\ (ALEPH, DELPHI, L3 and OPAL Collaborations and 
LEP Electroweak Working Group), arXiv:hep-ex/0612034

\bibitem{TevatronElectroweakWorkingGroup:2012gb} 
Tevatron Electroweak Working Group (CDF and D\O\ Collaborations), arXiv:1204.0042 [hep-ex]

\bibitem{Beneke:2007zg} M.~Beneke, P.~Falgari, C.~Schwinn, A.~Signer and G.~Zanderighi,
\Journal{\NPB}{792}{89}{2008}

\bibitem{Zanderighi:2007} G.~Zanderighi,
{\tt http://www.ggi.fi.infn.it/talks/talk484.pdf}

\bibitem{Aaltonen:2012bp} T.~Aaltonen \etal\ (CDF Collaboration), 
\Journal{\PRL}{108}{151803}{2012}

\bibitem{Zeng:2012} Y.~Zeng,
{\tt http://www.phy.duke.edu/\~{}yz38/wmw\_meeting/CollabMeeting\_Wmass\_2012Mar29.pdf}

\bibitem{Sirlin:1980nh} A.~Sirlin, \Journal{\PRD}{22}{971}{1980}

\bibitem{Degrassi:1990tu} G.~Degrassi, S.~Fanchiotti and A.~Sirlin, \Journal{\NPB}{351}{49}{1991}

\bibitem{Erler:1998sy} J.~Erler, \Journal{\PRD}{59}{054008}{1999}

\bibitem{Davier:2010nc} M.~Davier, A.~Hoecker, B.~Malaescu and Z.~Zhang,
\Journal{\EPJC}{71}{1515}{2011}

\bibitem{Lancaster:2011wr} Tevatron Electroweak Working Group and 
CDF and D\O\ Collaborations, arXiv:1107.5255 [hep-ex]

\bibitem{LHCmt} ATLAS and CMS Collaborations, \\
{\tt https://twiki.cern.ch/twiki/bin/view/CMSPublic/PhysicsResultsTOP12001}

\bibitem{Chetyrkin:1999qi} K.~G.~Chetyrkin and M.~Steinhauser, \Journal{\NPB}{573}{617}{2000}

\bibitem{Erler:1999ug} J.~Erler, in the Proceedings of the Physics at Run~II Workshop at FNAL, 
arXiv:hep-ph/0005084

\bibitem{Beneke:1998ui} M.~Beneke, \Journal{\PREP}{317}{1}{1999}
  
\bibitem{ATLASmh} G.~Aad \etal\ (ATLAS Collaboration), arXiv:1207.0319 [hep-ex].
  
\bibitem{CMSmh} CMS Collaboration, 
{\tt http://cdsweb.cern.ch/record/1429928/files/HIG-12-008-pas.pdf}

\bibitem{Erler:2012uu} J.~Erler, arXiv:1201.0695 [hep-ph]

\bibitem{Barate:2003sz} R.~Barate \etal\ (LEP Working Group for Higgs Boson Searches and 
ALEPH, DELPHI, L3 and OPAL Collaborations), \Journal{\PLB}{565}{61}{2003}

\bibitem{TEVNPH:2012ab} Tevatron New-Phenomena and Higgs Working Group and 
CDF and D\O\ Collaborations, arXiv:1203.3774 [hep-ex]

\bibitem{Marciano:1990dp} W.~J.~Marciano and J.~L.~Rosner, \Journal{\PRL}{65}{2963}{1990}

\bibitem{Kennedy:1990ib} D.~C.~Kennedy and P.~Langacker, \Journal{\PRL}{65}{2967}{1990}

\bibitem{Altarelli:1991fk}  G.~Altarelli, R.~Barbieri and S.~Jadach, \Journal{\NPB}{369}{3}{1992}
  
\bibitem{Peskin:1991sw}  M.~E.~Peskin and T.~Takeuchi, \Journal{\PRD}{46}{381}{1992}
  
\bibitem{Veltman:1977kh}  M.~J.~G.~Veltman, \Journal{\NPB}{123}{89}{1977}

\bibitem{Maksymyk:1993zm}  I.~Maksymyk, C.~P.~Burgess and D.~London,
\Journal{\PRD}{50}{529}{1994}

\bibitem{Davoudiasl:2012ag} H.~Davoudiasl, H.~S.~Lee and W.~J.~Marciano,
\Journal{\PRD}{85}{115019}{2012}

\bibitem{Anthony:2005pm} P.~L.~Anthony \etal\ (SLAC E158 Collaboration), 
\Journal{\PRL}{95}{081601}{2005}

\bibitem{Young:2007zs} R.~D.~Young, R.~D.~Carlini, A.~W.~Thomas and J.~Roche,
\Journal{\PRL}{99}{122003}{2007}

\bibitem{Albright:2000xi} C.~Albright \etal, arXiv:hep-ex/0008064

\bibitem{Panman:1996} J.~Panman, {\em Neutrino-electron scattering}, 
pp.~504--544 of Ref.~\cite{Langacker:1996qb}

\bibitem{Hasert:1973cr} F.~J.~Hasert \etal\ (Gargamelle Collaboration), 
\Journal{\PLB}{46}{121}{1973}

\bibitem{Sarantakos:1982bp} S.~Sarantakos, A.~Sirlin and W.~J.~Marciano, 
\Journal{\NPB}{217}{84}{1983}

\bibitem{Dorenbosch:1988is} J.~Dorenbosch \etal\ (CHARM Collaboration),
\Journal{\ZPC}{41}{567}{1989}

\bibitem{Vilain:1994qy} P.~Vilain \etal\ (CHARM~II Collaboration), \Journal{\PLB}{335}{246}{1994}
  
\bibitem{Ahrens:1990fp} L.~A.~Ahrens \etal\ (CALO Collaboration), \Journal{\PRD}{41}{3297}{1990}
  
\bibitem{Allen:1992qe} R.~C.~Allen \etal\ (CNTR Collaboration), \Journal{\PRD}{47}{11}{1993}
  
\bibitem{Auerbach:2001wg} L.~B.~Auerbach \etal\ (LSND Collaboration), 
\Journal{\PRD}{63}{112001}{2001}
  
\bibitem{Deniz:2009mu} M.~Deniz \etal\ (TEXONO Collaboration), 
\Journal{\PRD}{81}{072001}{2010}

\bibitem{Perrier:1996qg} F.~Perrier, 
{\em The Measurement of electroweak parameters from deep inelastic neutrino scattering}, 
pp.~385--490 of Ref.~\cite{Langacker:1996qb}

\bibitem{Conrad:1997ne} J.~M.~Conrad, M.~H.~Shaevitz and T.~Bolton, 
\Journal{\RMP}{70}{1341}{1998}

\bibitem{Llewellyn Smith:1983ie} C.~H.~Llewellyn Smith, \Journal{\NPB}{228}{205}{1983}

\bibitem{Paschos:1972kj} E.~A.~Paschos and L.~Wolfenstein, \Journal{\PRD}{7}{91}{1973}

\bibitem{Dobrescu:2003ta} B.~A.~Dobrescu and R.~K.~Ellis, \Journal{\PRD}{69}{114014}{2004}
    
\bibitem{Callan:1969uq} C.~G.~Callan, Jr. and D.~J.~Gross, \Journal{\PRL}{22}{156}{1969}
    
\bibitem{Gluck:2005xh} M.~Gl\"uck, P.~Jimenez-Delgado and E.~Reya, 
\Journal{\PRL}{95}{022002}{2005}
  
\bibitem{Londergan:2003ij} J.~T.~Londergan and A.~W.~Thomas, 
\Journal{\PRD}{67}{111901}{2003}

\bibitem{Cloet:2009qs} I.~C.~Clo\"et, W.~Bentz and A.~W.~Thomas, 
\Journal{\PRL}{102}{252301}{2009}
  
\bibitem{Diener:2003ss} K.~P.~O.~Diener, S.~Dittmaier and W.~Hollik, 
\Journal{\PRD}{69}{073005}{2004}
  
\bibitem{Arbuzov:2004zr} A.~B.~Arbuzov, D.~Y.~Bardin and L.~V.~Kalinovskaya, 
\Journal{\em JHEP}{0506}{078}{2005}

\bibitem{Allaby:1987vr}  J.~V.~Allaby \etal\ (CHARM Collaboration), \Journal{\ZPC}{36}{611}{1987}

\bibitem{Blondel:1989ev} A.~Blondel \etal\ (CDHS Collaboration),  \Journal{\ZPC}{45}{361}{1990}
  
\bibitem{McFarland:1997wx} K.~S.~McFarland \etal\ (CCFR Collaborations),
\Journal{\EPJC}{1}{509}{1998}

\bibitem{Zeller:2001hh} G.~P.~Zeller \etal\ (NuTeV Collaboration), 
\Journal{\PRL}{88}{091802}{2002} 

\bibitem{Amaldi:1987fu} U.~Amaldi \etal, \Journal{\PRD}{36}{1385}{1987}

\bibitem{Mann:1996qh} A.~K.~Mann, {\em Neutrino proton elastic scattering}, 
pp.~491--503 of Ref.~\cite{Langacker:1996qb}

\bibitem{Leinweber:2004tc} D.~B.~Leinweber \etal, \Journal{\PRL}{94}{212001}{2005}
  
\bibitem{Leinweber:2006ug} D.~B.~Leinweber \etal, \Journal{\PRL}{97}{022001}{2006}

\bibitem{Ahmed:2011vp} Z.~Ahmed \etal\ (HAPPEX Collaboration), 
\Journal{\PRL}{108}{102001}{2012}

\bibitem{Baunack:2009gy} S.~Baunack \etal\ (PVA4 Collaboration),
\Journal{\PRL}{102}{151803}{2009}

\bibitem{Androic:2009aa} D.~Androi\'c \etal\ (G\O\ Collaboration), 
\Journal{\PRL}{104}{012001}{2010}
  
\bibitem{GonzalezJimenez:2011fq} R.~Gonzalez-Jimenez, J.~A.~Caballero and T.~W.~Donnelly,  
arXiv:1111.6918 [nucl-th]

\bibitem{Armstrong:2012bi} D.~S.~Armstrong and R.~D.~McKeown,
\Journal{\ARNPS}{62}{337}{2012}
  
\bibitem{Weinberg:1958ut} S.~Weinberg, \Journal{\PREV}{112}{1375}{1958}
  
\bibitem{Horstkotte:1981ne} J.~Horstkotte \etal\ (BNL--E--613 Collaboration),
\Journal{\PRD}{25}{2743}{1982}
  
\bibitem{Ahrens:1986xe} L.~A.~Ahrens \etal\ (BNL--E--734 Collaboration), 
\Journal{\PRL}{35}{785}{1987}

\bibitem{Zucchelli:2002sa} P.~Zucchelli, \Journal{\PLB}{532}{166}{2002}

\bibitem{Antonelli:2007eb} V.~Antonelli, G.~Battistoni, P.~Ferrario and S.~Forte,
\Journal{\NPPS}{168}{192}{2007}

\bibitem{Kullenberg:2009pu} C.~T.~Kullenberg \etal\ (NOMAD Collaboration), 
\Journal{\PLB}{682}{177}{2009}
  
\bibitem{Wu:2007ab} Q.~Wu \etal\ (NOMAD Collaboration), \Journal{\PLB}{660}{19}{2008}

\bibitem{Rein:1982pf} D.~Rein and L.~M.~Sehgal, \Journal{\NPB}{223}{29}{1983}
     
\bibitem{Grabosch:1985mt}  H.~J.~Grabosch \etal\ (SKAT Collaboration), 
\Journal{\ZPC}{31}{203}{1986}

\bibitem{Bergsma:1985qy} F.~Bergsma  \etal\ (CHARM Collaboration), 
\Journal{\PLB}{157}{469}{1985}

\bibitem{Vilain:1993sf} P.~Vilain \etal\ (CHARM~II Collaboration), \Journal{\PLB}{313}{267}{1993}

\bibitem{Marciano:1980pb} W.~J.~Marciano and A.~Sirlin, \Journal{\PRD}{22}{2695}{1980}
   
\bibitem{Derman:1979zc} E.~Derman and W.~J.~Marciano, 
\Journal{\em Annals Phys.}{121}{147}{1979}

\bibitem{Zykunov:2004nk} V.~A.~Zykunov, \Journal{\PAN}{67}{1342}{2004}

\bibitem{Zykunov:2005md} V.~A.~Zykunov, J.~Suarez, B.~A.~Tweedie and Y.~G.~Kolomensky,
arXiv:hep-ph/0507287

\bibitem{Czarnecki:1995fw} A.~Czarnecki and W.~J.~Marciano, \Journal{\PRD}{53}{1066}{1996}

\bibitem{Erler:2003yk} J.~Erler, A.~Kurylov and M.~J.~Ramsey-Musolf,
\Journal{\PRD}{68}{016006}{2003}

\bibitem{Mammei:2012ph} J.~Mammei \etal\ (MOLLER Collaboration), 
\Journal{\NCC}{035N04}{203}{2012}

\bibitem{Bouchiat:1986gc} M.~A.~Bouchiat and L.~Pottier, \Journal{\em Science}{234}{1203}{1986}

\bibitem{Masterson:1996qi} B.~P.~Masterson and C.~E.~Wieman, 
{\em Atomic parity nonconservation experiments}, 
pp.~545--576 of Ref.~\cite{Langacker:1996qb}

\bibitem{Wood:1997zq} C.~S.~Wood \etal, \Journal{\em Science}{275}{1759}{1997}

\bibitem{Guena:2004sq} J.~Gu\'ena, M.~Lintz and M.~A.~Bouchiat, arXiv:physics/0412017
  
\bibitem{Edwards:1995zz} N.~H.~Edwards, S.~J.~Phipp, P.~E.~G.~Baird and S.~Nakayama,
\Journal{\PRL}{74}{2654}{1995}
 
\bibitem{Vetter:1995vf} P.~A.~Vetter \etal, \Journal{\PRL}{74}{2658}{1995}

\bibitem{Blundell:1996qj} S.~A.~Blundell, W.~R.~Johnson and J.~R.~Sapirstein,
{\em The Theory of atomic parity violation}, 
pp.~577--598 of Ref.~\cite{Langacker:1996qb}
  
\bibitem{Ginges:2003qt} J.~S.~M.~Ginges and V.~V.~Flambaum, \Journal{\PREP} {397}{63}{2004}

\bibitem{Porsev:2009pr} S.~G.~Porsev, K.~Beloy and A.~Derevianko, 
\Journal{\PRL}{102}{181601}{2009}

\bibitem{Dzuba:2012kx} V.~A.~Dzuba, J.~C.~Berengut, V.~V.~Flambaum and B.~Roberts,
arXiv:1207.5864 [hep-ph]

\bibitem{Zeldovich:1958} Ya.~B.~ZelÕdovich, \Journal{\em Sov.\ Phys.\ JETP}{6}{1184}{1958}
  
\bibitem{Haxton:2001ay} W.~C.~Haxton and C.~E.~Wieman, \Journal{\ARNPS}{51}{261}{2001}

\bibitem{Bennett:1999pd} S.~C.~Bennett and C.~E.~Wieman, \Journal{\PRL}{82}{2484}{1999}
  
\bibitem{Bouchiat:1988} M. A. Bouchiat and J. Gu\'ena, 
\Journal{\em J.\ Phys.\ France}{49}{2037}{1988}

\bibitem{Dzuba:1987px} V.~A.~Dzuba, V.~V.~Flambaum, P.~G.~Silvestrov and O.~P.~Sushkov,
\Journal{\JPB}{20}{3297}{1987}
 
\bibitem{Rosner:1995aj} J.~L.~Rosner, \Journal{\PRD}{53}{2724}{1996}
  
\bibitem{Pollock:1992mv} S.~J.~Pollock, E.~N.~Fortson and L.~Wilets,
\Journal{\PRC}{46}{2587}{1992}
  
\bibitem{Chen:1993fw} B.~Q.~Chen and P.~Vogel, \Journal{\PRC}{48}{1392}{1993}

\bibitem{Brown:2008ib} B.~A.~Brown, A.~Derevianko and V.~V.~Flambaum,
\Journal{\PRC}{79}{035501}{2009}

\bibitem{Abrahamyan:2012gp} S.~Abrahamyan \etal\ (PREX Collaboration), 
\Journal{\PRL}{108}{112502}{2012}

\bibitem{RamseyMusolf:1999qk} M.~J.~Ramsey-Musolf, \Journal{\PRC}{60}{015501}{1999}

\bibitem{Dunford:2007df} R.~W.~Dunford and R.~J.~Holt, \Journal{\JPG}{34}{2099}{2007}
  
\bibitem{Behr:2008at} J.~A.~Behr and G.~Gwinner, \Journal{\JPG}{36}{033101}{2009}
    
\bibitem{Wansbeek:2008} L.~W.~Wansbeek \etal, \Journal{\PRA}{78}{050501(R)}{2008}
  
\bibitem{Armstrong:2012ps} D.~S.~Armstrong \etal\ (Qweak Collaboration), 
arXiv:1202.1255 [physics.ins-det]
  
\bibitem{Gorchtein:2008px} M.~Gorchtein and C.~J.~Horowitz, \Journal{\PRL}{102}{091806}{2009}
  
\bibitem{Baunack:2012} D.~Becker, S.~Baunack, and F.~E.~Maas, 
\Journal{\em Hyperfine Interact.}{214}{141}{2013}
  
\bibitem{Souder:1990ia} P.~A.~Souder \etal, \Journal{\PRL}{65}{694}{1990}

\bibitem{Souder:1996qk} P.~A.~Souder, 
{\em Charged lepton hadron asymmetries in fixed target experiments}, 
pp.~599--625 of Ref.~\cite{Langacker:1996qb} 
  
\bibitem{Prescott:1979dh} C.~Y.~Prescott \etal,  \Journal{\PLB}{84}{524}{1979}

\bibitem{Zheng:2012vf} X.~Zheng \etal\ (Jefferson Lab Hall A Collaboration), 
\Journal{\NCC}{035N04}{72}{2012}

\bibitem{Souder:2011zz} P.~A.~Souder (SOLID Collaboration), \Journal{\em AIP Conf.\ Proc.}{1441}{123}{2012}

\bibitem{Argento:1982tq} A.~Argento \etal, \Journal{\PLB}{120}{245}{1983}

\bibitem{Zhu:2000gn} S.~L.~Zhu, S.~J.~Puglia, B.~R.~Holstein and M.~J.~Ramsey-Musolf,
\Journal{\PRD}{62}{033008}{2000}

\bibitem{Spayde:1999qg} D.~T.~Spayde \etal\ (SAMPLE Collaboration), 
\Journal{\PRL}{84}{1106}{2000}

\bibitem{Hasty:2001ep} R.~Hasty \etal\ (SAMPLE Collaboration), 
\Journal{\em Science}{290}{2117}{2000}

\bibitem{Heil:1989dz} W.~Heil \etal, \Journal{\NPB}{327}{1}{1989}
  
\bibitem{Marciano:1982mm} W.~J.~Marciano and A.~Sirlin, \Journal{\PRD}{27}{552}{1983}

\bibitem{Marciano:1993ep} W.~J.~Marciano, 
{\em Radiative corrections to neutral current processes}, 
pp.~170--200 of Ref.~\cite{Langacker:1996qb}

\bibitem{Marciano:1983ss} W.~J.~Marciano and A.~Sirlin, \Journal{\PRD}{29}{75}{1984}

\bibitem{Blunden:2012ty}  P.~G.~Blunden, W.~Melnitchouk and A.~W.~Thomas, 
\Journal{\PRL}{109}{262301}{2012}

\bibitem{Blunden:2011rd} P.~G.~Blunden, W.~Melnitchouk and A.~W.~Thomas, 
\Journal{\PRL}{107}{081801}{2011}

\bibitem{Sibirtsev:2010zg} A.~Sibirtsev, P.~G.~Blunden, W.~Melnitchouk and A.~W.~Thomas,
\Journal{\PRD}{82}{013011}{2010}

\bibitem{Rislow:2010vi} B.~C.~Rislow and C.~E.~Carlson,  \Journal{\PRD}{83}{113007}{2011}

\bibitem{Gorchtein:2011mz} M.~Gorchtein, C.~J.~Horowitz and M.~J.~Ramsey-Musolf, 
\Journal{\PRC}{84}{015502}{2011}

\bibitem{RamseyMusolf:2006vr} M.~J.~Ramsey-Musolf and S.~Su, \Journal{\PREP}{456}{1}{2008}

\bibitem{'tHooft:1979bh} G.~'t Hooft, 
\Journal{\em NATO Adv.\ Study Inst.\ Ser.\ B Phys.}{59}{135}{1980}

\bibitem{Haber:1984rc} H.~E.~Haber and G.~L.~Kane, \Journal{\PREP}{117}{75}{1985}

\bibitem{Martin:1997ns} S.~P.~Martin, arXiv:hep-ph/9709356

\bibitem{ArkaniHamed:1998} N.~Arkani-Hamed, S.~Dimopoulos and G.~R.~Dvali, 
\Journal{\PLB}{429}{263}{1998}

\bibitem{Antoniadis:1998} I.~Antoniadis, N.~Arkani-Hamed, S.~Dimopoulos and G.~R.~Dvali, 
\Journal{\PLB}{436}{257}{1998}
  
\bibitem{Randall:1999} L.~Randall and R.~Sundrum, 
\Journal{\PRL}{83}{4690}{1999}and \Journal{\ibid}{83}{3370}{1999}

\bibitem{ArkaniHamed:2001nc} N.~Arkani-Hamed, A.~G.~Cohen and H.~Georgi,
\Journal{\PLB}{513}{232}{2001}

\bibitem{ArkaniHamed:2002} N.~Arkani-Hamed \etal, \Journal{\em JHEP}{0208}{021}{2002}

\bibitem{Harnik:2003rs}   R.~Harnik, G.~D.~Kribs, D.~T.~Larson and H.~Murayama,
\Journal{\PRD}{70}{015002}{2004}

\bibitem{Csaki:2003zu} C.~Csaki, C.~Grojean, L.~Pilo and J.~Terning,
\Journal{\PRL}{92}{101802}{2004}

\bibitem{Kurylov:2003zh} A.~Kurylov, M.~J.~Ramsey-Musolf and S.~Su,
\Journal{\PRD}{68}{035008}{2003}

\bibitem{Aad:2012pxa} G.~Aad \etal\ (ATLAS Collaboration), \Journal{\PLB}{718}{879}{2013}

\bibitem{Chatrchyan:2013lya} S.~Chatrchyan \etal\ (CMS Collaboration), arXiv:1303.2985 [hep-ex]

\bibitem{Kurylov:2003by} A.~Kurylov, M.~J.~Ramsey-Musolf and S.~Su,
\Journal{\NPB}{667}{321}{2003}

\bibitem{Kurylov:2003xa} A.~Kurylov, M.~J.~Ramsey-Musolf and S.~Su, 
\Journal{\PLB}{582}{222}{2004}
 
\bibitem{Davidson:2001ji} S.~Davidson, S.~Forte, P.~Gambino, N.~Rius and A.~Strumia,
\Journal{\em JHEP}{0202}{037}{2002}

\bibitem{RamseyMusolf:2000qn} M.~J.~Ramsey-Musolf, \Journal{\PRD}{62}{056009}{2000}

\bibitem{Hardy:2008gy} J.~C.~Hardy and I.~S.~Towner, \Journal{\PRC}{79}{055502}{2009}

\bibitem{Britton:1992pg} D. I. Britton \etal, \Journal{\PRL}{68}{3000}{1992}

\bibitem{Czapek:1993kc} G.~Czapek {\em et al.}, \Journal{\PRL}{70}{17}{1993}

\bibitem{Cirigliano:2007ga} V.~Cirigliano and I.~Rosell, \Journal{\em JHEP}{0710}{005}{2007}
   
\bibitem{Bauman:2012fx} S.~Bauman, J.~Erler and M.~J.~Ramsey-Musolf, 
\Journal{\PRD}{87}{035012}{2013}

\bibitem{Chang:2009yw} W.~F.~Chang, J.~N.~Ng and J.~M.~S.~Wu, 
\Journal{\PRD}{79}{055016}{2009}

\bibitem{Erler:2009jh} J.~Erler, P.~Langacker, S.~Munir and E.~Rojas,
\Journal{\em JHEP}{0908}{017}{2009}
 
\bibitem{Li:2009xh} Y.~Li, F.~Petriello and S.~Quackenbush, \Journal{\PRD}{80}{055018}{2009}

\bibitem{Erler:2011ud} J.~Erler, P.~Langacker, S.~Munir and E.~Rojas,
\Journal{\em JHEP}{1111}{076}{2011}
 
\bibitem{Diener:2011jt} R.~Diener, S.~Godfrey and I.~Turan, \Journal{\PRD}{86}{115017}{2012}
  
\bibitem{Buckley:2012tc} M.~R.~Buckley and M.~J.~Ramsey-Musolf, \Journal{\PLB}{712}{261}{2012}

\bibitem{GonzalezAlonso:2012jb} M.~Gonzalez-Alonso and M.~J.~Ramsey-Musolf,
arXiv:1211.4581 [hep-ph]

\bibitem{Herczeg:2003ag} P.~Herczeg, \Journal{\PRD}{68}{116004}{2003}
  
\end{thebibliography}
\end{document}